\documentclass[11pt,a4paper,oneside]{article}
\pdfoutput=1
\usepackage{jheppub}

\usepackage{verbatim}
\usepackage{mathrsfs}
\usepackage{appendix}
\usepackage{caption}
\usepackage{float}
\usepackage{subfig}
\usepackage[vcentermath]{youngtab}
\usepackage{alltt}
\usepackage{multirow}
\usepackage{arydshln}
\usepackage{lscape}
\usepackage{tabu}
\usepackage{amssymb,amsmath,amsfonts,mathrsfs,enumerate,wrapfig,latexsym,wasysym}

\usepackage{graphicx}
\usepackage{adjustbox}
\usepackage{xcolor}
\usepackage{slashed}    
\usepackage{url}
\usepackage{hyperref}
\hypersetup{
	colorlinks = true,
	linkcolor = blue,
	anchorcolor = blue,
	citecolor = blue,
	filecolor = blue,
	urlcolor = magenta,
	bookmarks=true,
	linktocpage=true
}
\usepackage{framed}
\usepackage[numbers,sort&compress]{natbib}
\usepackage{fancyvrb}
\allowdisplaybreaks
\usepackage{multirow}
\usepackage{framed}
\RequirePackage{flushend}
\usepackage{makecell}
\usepackage{color}
\usepackage{pifont}
\usepackage{cancel}
\usepackage{import}
\usepackage{subfiles}
\usepackage{filecontents}
\usepackage{rotating}
\usepackage{threeparttable,tablefootnote}\usepackage{mathtools}
\usepackage{pbox}
\usepackage{setspace}
\usepackage{lscape}
\usepackage{color, colortbl}
\usepackage{pdflscape}
\usepackage{afterpage}
\usepackage{fontawesome}

\usepackage{threeparttable}
\usepackage{slashed}
\usepackage[normalem]{ulem}

\definecolor{darkgreen}{cmyk}{1,0,1,0.4}

\definecolor{pink}{cmyk}{0.4,1,0.3,0}

\def\com2#1{\textcolor{red}{\it{#1}}}

\graphicspath{{plots/}}

\newcommand{\cmark}{\ding{51}}
\newcommand{\xmark}{\ding{55}}
\definecolor{Gray}{gray}{0.9}

\title{A Step Toward Model Comparison: Connecting Electroweak-Scale Observables to BSM through EFT and Bayesian Statistics}

\author[1]{Anisha, Supratim Das Bakshi, Joydeep Chakrabortty,}
\author[2]{and Sunando Kumar Patra} \author{\\}

\affiliation[1]{Department of Physics, Indian Institute of Technology, Kanpur-208016, India}

\affiliation[2]{Department of Physics, Bangabasi Evening College, 19 Rajkumar Chakraborty Sarani, Kolkata 700009, West Bengal, India}

\emailAdd{anisha, supratim, joydeep@iitk.ac.in, sunando.patra@gmail.com }

\abstract{
	Recognizing the potential of effective field theories to posit multiple BSM scenarios  in similar footing, with a possibility to compare them, we inspect the effects of 11 single scalar-multiplet extensions of the SM on the combined set of electroweak precision observables and Higgs signal strength data, by systematically integrating out the heavy multiplets  and computing the resulting SMEFT operators and  Wilson coefficients (WCs) up to one-loop level.
	Noting that multiple BSM models give rise to a degenerate set of WCs, we then perform Bayesian statistical inference both directly on the BSM parameters and on the associated set of independent WCs. 
	Using the posteriors of the BSM parameters, we infer the respective (correlated) WC-distributions and compare both the model independent and dependent analyses by overlaying the 2-D marginal WC-posteriors from both processes, thus laying the ground for a data-driven attempt to compare diverse BSM theories of different origins, and hopefully, a possible way to approach the intractable inverse problem. We also demonstrate, with an example model, the crucial role of theoretical constraints to rule out large chunks of BSM parameter spaces. The entirety of numerical results is available in GitHub \href{https://github.com/effExTeam/SMEFT-EWPO-Higgs}{\faGithub}.}

\begin{document}
	\maketitle
	\section{Introduction}
	In spite of the immense success of the Standard Model (SM), it is still inadequate to explain a plethora of phenomena in the high energy physics spectrum.
	There has been no direct evidence of any beyond Standard Model (BSM) physics after the discovery of the Higgs. We thus need to refer to indirect evidence hinting towards BSM scenarios. 
	Among the observables with the potential to constrain BSM physics and thus to act as indirect evidence for new physics (NP), Electroweak Precision Observables (EWPO) and those from Higgs decays play an important role. To effectively use these observables to constrain BSM parameter-space, we need to bridge the gap between any BSM physics residing at a high scale and the observables lying at low energy. The Standard Model Effective Field Theory (SMEFT) \cite{Buchmuller:1985jz,Grzadkowski:2010es} links the BSM theories to the low energy observables using the higher dimension operators originating from integrating out the heavy degrees of freedom (DOFs). The SMEFT effective operators, for a given mass dimension and defined using particle content and  symmetry of the SM, offer additional contributions to the SM predictions of the low energy observables. These modifications are recast in terms of the Wilson coefficients (WCs) that carry the footprints of the unknown new physics (NP).

	There have been numerous works over the years to constrain the SMEFT Wilson coefficients (WCs) of dimension-6 operators in a model-independent manner. The general strategy has been to perform statistical inference on these WCs using the available data, either taking one of them at a time or all of the pertinent WCs (to which the data are sensitive) together. The SMEFT operators are frequently discussed in Warsaw~\cite{Grzadkowski:2010es} and SILH~\cite{Elias-Miro:2013mua,delAguila:2011zs,Giudice:2007fh} bases. The inferences are drawn mostly in a frequentist framework  \cite{Ciuchini_2013, Ellis:2014jta, Falkowski_2015, Cirigliano_2016, Jana_2018, Berthier:2015oma, Murphy_2018, Ciuchini:2014dea, Haller:2018nnx, Ellis_2018, de_Blas_2016,Ellis:2018gqa,Almeida:2018cld}, though some Bayesian analysis have been done as well \cite{Dumont_2013,van_Beek_2019,Sato:2013ika,Castro:2016jjv}. The main idea here is that once a BSM theory is matched to SMEFT, the bounds on the WCs can be converted to that of the BSM parameters. 
	
	A lot of work has been done to match various BSM theories to the SMEFT (upto one-loop order) \cite{Henning:2014wua, Drozd:2015rsp, Henning:2016lyp, Ellis:2016enq,  Fuentes-Martin:2016uol, Zhang:2016pja, Ellis:2017jns, de_Blas_2018, Kramer:2019fwz, Haisch_2020}, enabling the community to express the SMEFT Wilson coefficients, and in turn, the low energy observables (the EWPO, the Higgs signal strengths, etc.) in terms of the BSM parameters. Some (model-dependent) global fits have also been done to constrain specific BSM parameters from these matching results \cite{Gorbahn:2015gxa,Dawson:2020oco,Ellis:2014jta,Ellis_2018,Drozd:2015kva,DasBakshi:2020ejz}. 
	The main caveats of this yet-accepted-process are two-fold: firstly, not all WCs are modified within the scope of a specific model and even for those which are affected, the effects are not of the same degree, i.e., not all WCs are similarly sensitive to all model parameters. This set of pertinent WCs also varies with chosen BSM models. Secondly, though the model-independent inferences performed on WCs can point to a conservative estimate of the BSM parameter space, in reality, they are often highly non-linear functions of these parameters and the actual parameter-space (obtained from a direct inference on the parameters themselves) may differ a lot from the model-independent estimates.

	The motivation of this work is thus to probe into the relative capacity of these model-independent analyses to predict the BSM parameter-spaces, in comparison with direct inference done on the parameters themselves. In this article, we work with the Warsaw basis (a complete basis) of dimension-6 SMEFT operators, of which 18 operators affect the EWPO and Higgs-decays considered in this analysis. The main challenge in this endeavor is to obtain the SMEFT WCs in terms of the BSM parameters. We use the \emph{Mathematica\textsuperscript \textregistered} package CoDEx \cite{Bakshi:2018ics}, to this end. Given the BSM lagrangian, CoDEx can provide the list of the different dimension-6 operators and their corresponding WCs at one-loop, in terms of different BSM model parameters. For statistical inference, we choose the Bayesian framework and all required analyses are performed using the \emph{Mathematica\textsuperscript \textregistered} package OptEx \cite{OptEx}.

	The work is organized as follows: Section~\ref{sec:obs} introduces the observables relevant to the present analysis; in Section~\ref{sec:model_ind}, we discuss SMEFT contributions to these observables and perform a model-independent analysis using the relevant WCs; in Section~\ref{sec:model_dep}, we introduce 11 BSM scenarios with the potential to affect the observables in this analysis and obtain individual statistical inferences on each of them; in Section~\ref{sec:comparison}, we compare the model-independent and dependent results obtained in the previous two sections by inspecting the WC-space populated by these results.
	
	Numerical results of the entire analysis (including those not included in the draft) are available in the GitHub repository \cite{Githubeffex} \href{https://github.com/effExTeam/SMEFT-EWPO-Higgs}{\faGithub} associated with this work. 
	
\section{The Observables ($\mathcal{O}_i's$)}\label{sec:obs}
	\begin{table}[htb]
	\centering
	\renewcommand{\arraystretch}{1.8}
	\caption{\small The Higgs signal strengths from both ATLAS and CMS.}
	\begin{adjustbox}{width=\textwidth}
		\label{tab:HiggsSignalStrength}
		\begin{tabular}{|c|c|c|}
			\hline
			\multicolumn{2}{|c|}{Higgs signal strengths}& References\\
			\hline
			\multirow{3}{*}{7 and 8 TeV } & Combined ATLAS \& CMS measurements  &  table 8 of ref.~\cite{Khachatryan:2016vau}\\
			\cline{2-3}
			\multirow{3}{*}{Run-I data }& Combined ATLAS \& CMS measurement of $\mu_{pp}^{\mu \mu}$ & table 13 of ref.~\cite{Khachatryan:2016vau} \\
			\cline{2-3} & ATLAS measurement of $\mu_{pp}^{Z \gamma}$ &  Figure~1 of ref.~\cite{Aad:2015gba} \\
			\hline
			\multirow{7}{*}{13 TeV ATLAS} &  $H \rightarrow ZZ^*$ at 139 $fb^{-1}$   &  table 8 of ref.~\cite{Aad:2020mkp}\\
			\cline{2-3}
			\multirow{7}{*}{Run-II data} &  Measurement of $\mu_{pp}^{Z \gamma}$ at 139 $fb^{-1}$ &  ref.~\cite{Aad:2020plj} \\
			\cline{2-3}
			 & Measurement of $\mu_{pp}^{\mu \mu}$ at 139 $fb^{-1}$ &  ref.~\cite{ATLAS:2019ain} \\
			\cline{2-3}
			 & $VH$ $\rightarrow$ $H$ $\rightarrow$ $b\bar{b}$ at 139 $fb^{-1}$ &  ref.~\cite{ATLAS:2020udg} \\
			\cline{2-3}
			 & Measurements for Higgs production through  & Figure 5 of ref.~\cite{Aad:2019mbh}\\
			& gluon and vector boson fusions at 80 $fb^{-1}$ & [Correlations in Figure~6]\\
			\cline{2-3}
			 &  Associated production of Higgs with $t\bar{t}$  &  refs.~\cite{Aaboud:2017jvq,Aaboud:2018urx}\\
			\cline{2-3}
			 & $VH$ $\rightarrow$ $H$ $\rightarrow$ $WW^{*}$ at 36.1 $fb^{-1}$ & ref.~\cite{Aad:2019lpq}\\
			\hline
			\multirow{2}{*}{13 TeV CMS} &  \multirow{2}{*}{Signal strengths data up to 35.9 $fb^{-1}$}   &  table 3 of ref.~\cite{Sirunyan:2018koj}\\
			& & [Correlations in auxiliary material]\\
			\cline{2-3}
			Run-II data & Measurements of $\mu_{ZH}^{cc}$  and $\mu_{WH}^{cc}$ & ref.~\cite{Sirunyan:2019qia}\\
			\hline
		\end{tabular}
	\end{adjustbox}
\end{table}

	As mentioned before, the chosen set of observables for both model dependent and independent analyses in the present work are the EWPO and Higgs signal strengths. We summarize both the experimental inputs and the SM expressions of the observables in this section.
		
	\subsection{Electroweak Precision Observables} \label{subsection:EWPO_SM}
	The  EWPO under consideration for our analysis include the higher-order radiative corrections which are parametrized in terms of the five SM parameters: $Z$-boson mass ($m_{Z}$), Higgs mass ($ m_{H}$), top quark mass ($m_{t}$), strong coupling constant ($\alpha_{s}(m^{2}_{Z})$), and hadronic contributions to the running of $\alpha$ ($\Delta\alpha^{(5)}_{had}(m^{2}_{Z})$). As experimental inputs, we have used: (i) EWPO  measured at the $Z$-mass pole \cite{Haller:2018nnx} and their correlations \cite{ALEPH:2005ab}, (ii) mass and decay width of $W$ \cite{Tanabashi:2018oca}.	
	Some more details on these corrections to the EWPO are listed below: 
	\begin{itemize}
		\item  \underline{ \text{sin}$^2\theta_{\text{eff}}^{l}$}: receives up to full two-loop electroweak,  partial three-loop and four-loop QCD corrections, see ref.~\cite{Awramik:2006uz,Haller:2018nnx}. The  missing higher-order corrections is estimated to be $4.7 \times 10^{-5}$, included as theoretical uncertainty in the computation.
		
		\item \underline{Partial decay width ratios and Hadronic peak cross-section of $Z$}:  receives up to  full two-loop fermionic corrections, see ref.~\cite{Freitas:2014hra,Haller:2018nnx}.
		
		\item \underline{$Z$ pole asymmetry observables}:
	estimated using $ \sin ^2\theta _ {\text{eff}}^f$\footnote{$\theta _ {\text{eff}}^f$, the effective Weinberg mixing angle, receives the corrections from fermions only.}, see ref.~\cite{Flacher:2008zq,Haller:2018nnx}.
		
		\item \underline{Mass of $W$ boson}: receives up to two-loop complete and four-loop QCD corrections, see ref.~\cite{Awramik:2003rn,Haller:2018nnx}.
	
		\item \underline{Decay width of $W$ boson}: receives up to one-loop electroweak corrections, see ref.~\cite{Cho:2011rk,Haller:2018nnx}.
	\end{itemize}
	
	\subsection{Higgs signal strengths}\label{subsection:HiggsSignalStrength}
	The Higgs signal strengths, used in our analysis, contain the latest Run-I and -II LHC data. The details of the relevant experimental inputs  are tabulated in table \ref{tab:HiggsSignalStrength}.

\section{Model Independent Analysis}\label{sec:model_ind}

	\subsection{SMEFT contributions to the observables}\label{section:np_corrections}
\begin{table}[ht]
	\caption{\small These are the 18 dimension-6 effective operators (Warsaw basis) offer additional contributions to the EWPO and Higgs signal strengths. Here, $\tau^{I}$ are normalized Pauli matrices; $I=1,2,3$. }
	\centering
	\renewcommand{\arraystretch}{2.0}
	{\small\begin{tabular}{|c|c|c|c|c|c|}
			\hline
			$Q_H$  &  $\left(H^{\dagger }H )^3\right.$ & $Q_{HG}$  &  $\left(H^{\dagger }H \right)G_{\mu \nu }{}^aG^{a,\mu \nu }$ & $Q_{He}$  &  $\left(H^{\dagger }\it{ i }\overleftrightarrow{\mathcal{D}   }_{\mu }\it{ H }\right) \text{(}\bar{e} \gamma ^{\mu }\it{e} \text{ )}$ \\
			\hline
			$Q_{H\square }$  &  $\left(H^{\dagger }H \text{)$\square $(}H^{\dagger }H \right)$ & $Q_{Hl}^{(1)}$  &  $\left(H^{\dagger }\it{ i }\overleftrightarrow{\mathcal{D}   }_{\mu }\it{ H }\right) \text{(}\bar{l} \gamma ^{\mu }\it{l} \text{ )}$ & $Q_{Hu}$  &  $\left(H^{\dagger }\it{ i }\overleftrightarrow{\mathcal{D}   }_{\mu }\it{ H }\right) \text{(}\bar{u} \gamma ^{\mu }\it{u} \text{ )}$  \\ 
			\hline
			$Q_{HD}$  &  $\left(H^{\dagger }\mathcal{D}_{\mu }H )^*\right(H^{\dagger }\mathcal{D}^{\mu }H )$ & $Q_{Hl}^{(3)}$  &  $\left(H^{\dagger }\it{ i }\tau ^I \overleftrightarrow{\mathcal{D}   }_{\mu }\it{ H }\right) \text{(}\bar{l} \tau ^I \gamma ^{\mu }\it{l} \text{ )}$ & $Q_{Hd}$  &  $\left(H^{\dagger }\it{ i }\overleftrightarrow{\mathcal{D}   }_{\mu }\it{ H }\right) \text{(}\bar{d} \gamma ^{\mu }\it{d} \text{ )}$  \\
			\hline
			$Q_{HB}$  &  $\left(H^{\dagger }H \right)B_{\mu \nu }B^{\mu \nu }$ & $Q_{Hq}^{(1)}$  &  $\left(H^{\dagger }\it{ i }\overleftrightarrow{\mathcal{D}   }_{\mu }\it{ H }\right) \text{(}\bar{q} \gamma ^{\mu }\it{q} \text{ )}$ & $Q_{eH}$  &  $\left(  H^{\dagger }H \right) \text{(}\bar{l}\text{ e }H \text{)+h.c.}$  \\
			\hline
			$Q_{HW}$  &  $\left(H^{\dagger }H \right)W_{\mu \nu }{}^I W^{I,\mu \nu }$ & $Q_{Hq}^{(3)}$  &  $\left(H^{\dagger }\it{ i }\tau^I\overleftrightarrow{\mathcal{D}   }_{\mu }\it{ H }\right) \text{(}\bar{q} \tau^I \gamma ^{\mu }\it{q} \text{ )}$ & $Q_{uH}$  &  $\left(H^{\dagger }H \right) \text{(}\bar{q} \ \it{ u } \ \tilde{H}\text{)+h.c.}$  \\
			\hline
			$Q_{HWB}$  &  $\left(H^{\dagger }\tau^I H \right)W_{\mu \nu }{}^I B^{\mu \nu }$ & $Q_{ll}$  &  $\left(\bar{l} \gamma _{\mu }\it{l}  \right) \left(\bar{l} \gamma ^{\mu }\it{l} \text{ )}\right.$ & $Q_{dH}$  &  $\left(H^{\dagger }H \text{)(}\bar{q} \ \it{d} \ H \text{)+h.c.}\right.$     \\
			\hline
	\end{tabular}}
	\label{table:operators}
\end{table}

	The SMEFT induces corrections to the fit observables capturing the new physics lying beyond the cut-off scale ($\Lambda$) of the EFT. We consider the SMEFT contributions to these observables from dimension-6 effective operators (in Warsaw basis) \cite{Grzadkowski:2010es}. The EWPO and the Higgs observables can be expressed in terms of the associated WCs ($\mathcal{C}_{i}$) as:
\begin{align}\label{eq:obs_np}
\mathcal{O}_{\text{NP}}=\mathcal{O}_{\text{SM}}+\sum_{i}\frac{\mathcal{A}_{i}}{\Lambda^2} \mathcal{C}_{i} \,,
\end{align}
	where, $\mathcal{O}_{\text{NP}}$ represents the expressions of observables after including the SMEFT dimension-6 operator corrections and $\mathcal{O}_{\text{SM}}$ represents the SM expressions for the observables discussed in Section~\ref{sec:obs}. The  $\mathcal{A}_{i}$'s are functions of the five SM parameters (see Subsection~\ref{subsection:EWPO_SM}), $ i $ runs over the number of dimension-6 operators pertinent to the observables in question (18 in this work, see table~\ref{table:operators}).  The WCs encapsulate the effect of the NP on top of the SM estimates. Contributions from the SMEFT operators to the observables are discussed below.

	The SM expressions of the EWPO are modified by the effective operators: $Q_H$, $Q_{HD}$, $Q_{H\square }$, $Q_{HWB}$,  $Q_{Hl}^{(1)}$,  $Q_{Hl}^{(3)}$, $Q_{He}$, $Q_{Hq}^{(1)}$,  $Q_{Hq}^{(3)}$, $Q_{Hu}$, $Q_{Hd}$, and $Q_{ll}$. The details about these corrections are summarized in Appendix~\ref{sec:dim6correction}. The modification of the theoretical predictions of the Higgs boson production and decay rates due to SMEFT operators (in SILH-like basis) are discussed in ref.~\cite{Murphy:2017omb} that we rewrite in terms of dimension-6 operators in  Warsaw basis (for operator basis translation, see \cite{Alonso:2013hga}). Following ref.~\cite{Murphy:2017omb}, we only consider the contributions from the 3rd generation of fermions for the operators $Q_{eH},~Q_{uH}$, and $Q_{dH}$.

\begin{figure}[ht]
	\centering
	\includegraphics[ width=\textwidth, height=20cm]{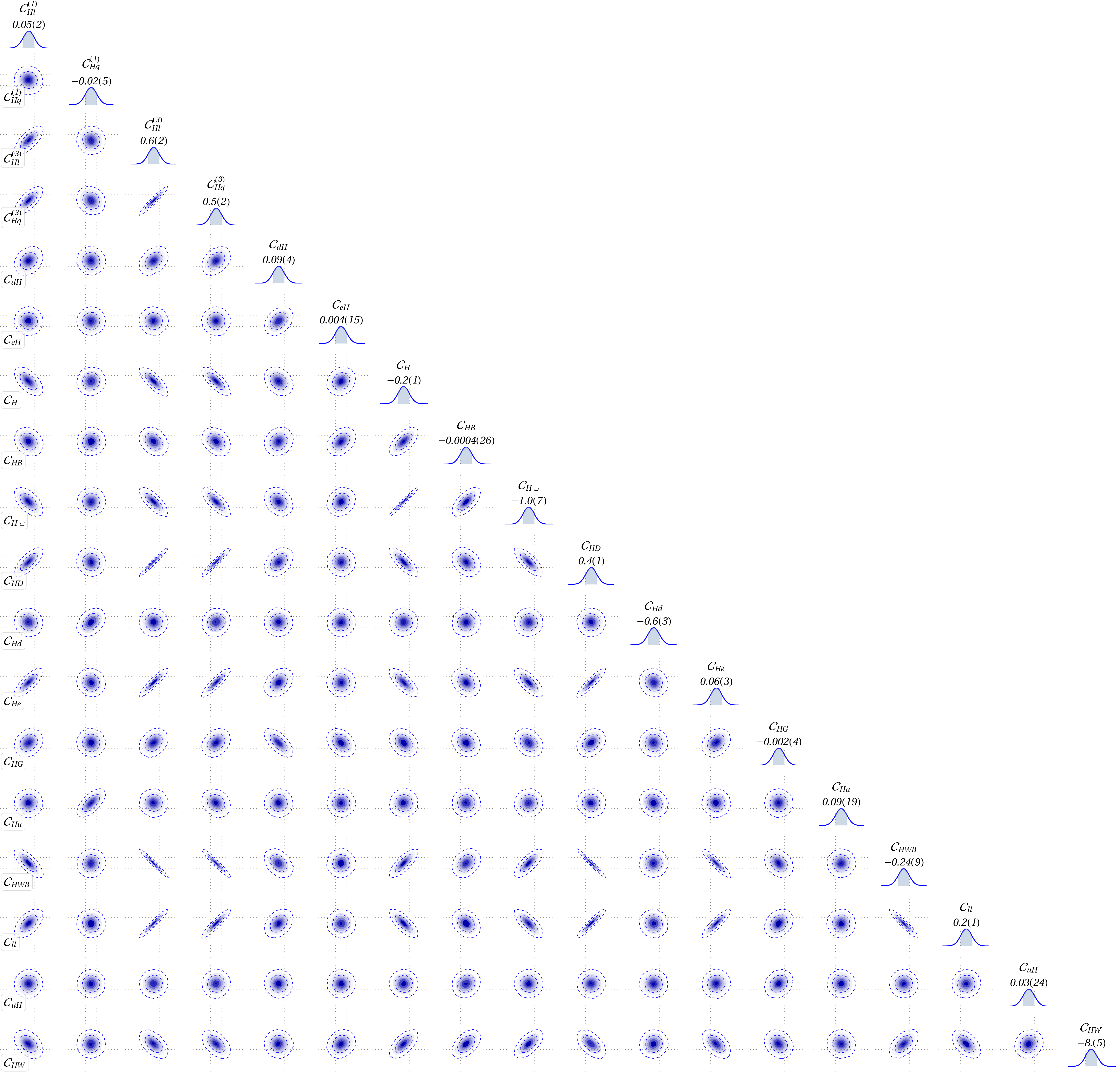}
	\caption{\small The marginalized one- and two-dimensional posteriors of the SMEFT WCs showing the correlations among them.}
	\label{fig:wctriplot}
\end{figure}

\begin{table}[ht]
	\large
	\centering
	\renewcommand{\arraystretch}{3.2}
	\caption{\small `Model-independent' fit results: 18  WCs  ($\in \mathcal{O}_i$'s): in presence of all (column I), single (column III); 10  WCs ($\in$ 18) $\in$ our adopted BSMs (column IV).  `Model-dependent' fit results: WCs ($\subset$ 10)  $\in$ individual BSM and functions of respective model parameters, see  table \ref{table:models} (columns V - X). We set the cut-off scale $\Lambda$ = $1$ TeV.}
	\begin{adjustbox}{width=\textwidth}
		{\large\begin{tabular}{|c|c|c|c|c|c|c|c|c|c|c|c|c|c|c|}
		\hline
		\multirow{2}{*}{WCs} & \multicolumn{3}{c|}{Model Independent Fits} & \multicolumn{11}{c|}{Fits with WCs $\in$ single BSM} \\
		\cline{2-15}
		 &  18 WCs $\in$ $\mathcal{O}_i$'s  & Individual WC Fit  & 10 WCs $\in$ BSMs & $SM+\mathcal{S}$ & $SM+\mathcal{S}_{2}$ & $SM+\Delta$ & \multicolumn{1}{c|}{SM+$\mathcal{H}_{2}$} & \multicolumn{1}{c|}{SM+$\Delta_{1}$} & \multicolumn{1}{c|}{SM+$\Sigma$}& \multicolumn{1}{c|}{SM+$\varphi_{1}$} & \multicolumn{1}{c|}{SM+$\varphi_{2}$} &   \multicolumn{1}{c|}{SM+$\Theta_{1}$}  & \multicolumn{1}{c|}{SM+$\Theta_{2}$} & \multicolumn{1}{c|}{SM+$\Omega$} \\
		\hline \hline
		$\mathcal{C}_H$  & $ -0.19(12)$ &  $ -0.01(64)\times 10^{-2}$ & $0.060(76)$ &$0.050(45)$ &$0.075(66)$& $0.026(54)$ & \multicolumn{3}{c|}{$0.051(75)$} &\multicolumn{2}{c|} {$0.076(67)$} &\multicolumn{3}{c|} {$0.066(73)$} \\
		\hline
		$\mathcal{C}_{H\square }$  &  $ -1.01(68)$ & $ 0.006(37)$ & $0.35(44)$ & $0.29(26)$ & $0.44(38)$ & $0.15(31)$ & \multicolumn{3}{c|}{$0.30(44)$}&\multicolumn{2}{c|}{$0.44(39)$} &\multicolumn{3}{c|} {$0.39(42)$}\\ 
		\hline
		$\mathcal{C}_{HD}$  &  $ 0.40(15)$ & $-0.09(19)\times 10^{-2} $ & $0.48(73)\times 10^{-2}$ & \xmark & \xmark & $-0.08(19)\times 10^{-2}$ & \multicolumn{3}{c|}{$ 0.46(72)\times 10^{-2}$} & \multicolumn{2}{c|}{\xmark} &\multicolumn{3}{c|} {\xmark}\\
		\hline
		$\mathcal{C}_{HB}$  &  $ -0.04(26)\times 10^{-2}$ & $-0.06(14)\times 10^{-2}$ & $0.16(25)\times 10^{-2}$ &\xmark &$0.11(21)\times 10^{-2}$& \xmark & \multicolumn{3}{c|}{$0.13(25)\times 10^{-2}$}& \multicolumn{2}{c|}{$0.10(21)\times 10^{-2}$} &\multicolumn{3}{c|} {$0.07(23)\times 10^{-2}$} \\
		\hline
		$\mathcal{C}_{HW}$  &  $ -7.82(496)$ & $-2.74(363)$ & $-0.52(419)$ & \xmark &\xmark & $-1.60(380)$ & \multicolumn{3}{c|}{$-0.82(419)$}& \multicolumn{2}{c|}{\xmark} &\multicolumn{3}{c|} {$-1.43(402)$}\\
		\hline
		$\mathcal{C}_{HWB}$  &  $ -0.24(9)$ & $-0.07(10)\times 10^{-2}$ & $-0.31(39)\times 10^{-2}$ & \xmark &\xmark & \xmark  & \multicolumn{3}{c|}{$-0.29(38)\times 10^{-2}$} & \multicolumn{2}{c|}{\xmark} &\multicolumn{3}{c|} {\xmark} \\
		\hline
		$\mathcal{C}_{HG}$  &  $ -0.18(38)\times 10^{-2}$  & $0.06(16)\times 10^{-2}$ & $-0.48(37)\times 10^{-2}$ &\xmark &\xmark & \xmark &\multicolumn{3}{c|}{\xmark} & \multicolumn{2}{c|}{$-0.02(19)\times 10^{-2}$} &\multicolumn{3}{c|} {$-0.03(19)\times 10^{-2}$}  \\
		\hline
		$\mathcal{C}_{eH}$  &  $ 0.004(15)$   & $ -0.006(13)$ & $0.004(15)$ &\xmark & \xmark  & $0.005(13)$ & \multicolumn{3}{c|}{$-0.002(14)$} & \multicolumn{2}{c|}{\xmark}&\multicolumn{3}{c|} {\xmark} \\
		\hline
		$\mathcal{C}_{uH}$  &  $ 0.03(24)$ & $0.09(21)$ & $0.02(23)$ &\xmark & \xmark & $0.09(22)$ & \multicolumn{3}{c|}{$0.07(23)$}& \multicolumn{2}{c|}{\xmark} &\multicolumn{3}{c|} {\xmark}\\
		\hline
		$\mathcal{C}_{dH}$  &  $ 0.087(37)$  & $ 0.016(14)$ & $0.053(35)$ & \xmark &\xmark & $0.011(18)$ & \multicolumn{3}{c|}{$0.015(19)$} & \multicolumn{2}{c|}{\xmark} &\multicolumn{3}{c|} {\xmark}\\
		\hline \hline
		$\mathcal{C}_{Hl}^{(1)}$  &  $ 0.048(18)$ & $0.40(47)\times 10^{-2}$ &  \multicolumn{12}{c|}{\xmark} \\
		\hline
		$\mathcal{C}_{Hl}^{(3)}$  &  $ 0.57(23)$ & $-0.32(46)\times 10^{-2}$ &  \multicolumn{12}{c|}{\xmark} \\
		\hline
		$\mathcal{C}_{Hq}^{(1)}$  &  $ -0.016(50)$ & $0.013(21)$ &  \multicolumn{12}{c|}{\xmark} \\
		\hline
		$\mathcal{C}_{Hq}^{(3)}$  &  $ 0.52(23)$ &  $0.23(85)\times 10^{-2} $ &  \multicolumn{12}{c|}{\xmark} \\ 
		\hline
		$\mathcal{C}_{He}$  &  $ 0.059(29)$ & $-0.90(66)\times 10^{-2}$&  \multicolumn{12}{c|}{\xmark} \\
		\hline
		$\mathcal{C}_{Hu}$  &  $ 0.09(19)$ & $ 0.005(55)$ & \multicolumn{12}{c|}{\xmark} \\
		\hline
		$\mathcal{C}_{Hd}$  &  $ -0.59(28)$  & $ -0.073(75)$ & \multicolumn{12}{c|}{\xmark} \\
		\hline
		$\mathcal{C}_{ll}$  &  $ 0.23(11)$ & $ 0.55(75)\times 10^{-2}$& \multicolumn{12}{c|}{\xmark} \\
		\hline
	\end{tabular}}
	\end{adjustbox}
	\label{table:wcfitval}
\end{table}	 

	\subsection{Statistical Inference}\label{sec:modind_stat}
	Adopting the Bayesian framework, all inference throughout this work is obtained by sampling the un-normalized posterior distribution using a Markov Chain Monte Carlo (MCMC). The algorithm used is Metropolis-Hastings \cite{hastings1970} and instead of using multiple walkers for assuring convergence, we depend on a single long chain. Ensuring that the random variable samples are independent and identically distributed ($iid$) and that the chain is converged to desired quantiles is done by diagnostic checks and sequential runs, following the prescriptions of Raftery and Lewis \cite{raftery1992}. As sanity checks, all corresponding frequentist maximum likelihood estimations (MLE) are also obtained for comparison of the best-fit results, goodness-of-fit tests, and outlier estimation. Using pulls and Cook's distances \cite{CookDist1,CookDist2,CookCutoff}, we have ensured that none of the observables are both an outlier and a disproportionately influential point in our analysis. The MLE and parameter-uncertainty estimation using Hessians enables us to quickly select the initial points and proposal spreads of the MCMC chains as well. 

	For the first part of our analysis, we perform a model-independent statistical inference from a total of 88 observables mentioned in Section~\ref{sec:obs} in terms of 18 SMEFT dimension-6 WCs as free parameters and 5 SM parameters: $m_{Z}$, $ m_{H}$, $m_{t}$, $\alpha_{s}(m^{2}_{Z})$, and $\Delta\alpha^{(5)}_{had}(m^{2}_{Z})$ as nuisance parameters. To obtain the nuisance-priors, we have performed an SM fit of the EWPO, details of which are given in Appendix~\ref{appendix:SMfit}.

	Priors for the SM parameters are introduced as a multi-normal distribution, following the result of the fit mentioned above. For the SMEFT model-independent analysis, uniform, uninformative priors are taken for all free parameters (WCs). We have found that the range $\{-10,10\}$ is good enough for all WCs except $\mathcal{C}_{HW}$ and $\mathcal{C}_{uH}$, for which the range $\{-50,50\}$ is chosen. We perform two types of fits at this stage: taking all relevant WCs together, and taking one WC at a time. It is expected that the inferred parameter-space of any one WC will be smaller for the fit with only that WC, whereas, in presence of all other WCs, it may have a considerably larger parameter space. The results of these fits are tabulated for comparison in table~\ref{table:wcfitval}. The first column lists the WCs, the second column is the result of the fit with all observables taken together, and in the third one, values in each row show the result of the fit with only the corresponding WC. The fourth (last under model independent fits) column lists the fit results of the maximum set of 10 operators, relevant to the model-dependent part of this analysis, taken together. More on these in Section~\ref{section:models}. All fits are done with the cut-off set as $\Lambda$ = 1 TeV. For the fit with all the parameters taken together, the one and two-dimensional marginal probability distributions for the SMEFT WCs are shown in Figure~\ref{fig:wctriplot}. 
	
\section{Model Dependent Analysis}\label{sec:model_dep}
	
	\subsection{Realizing BSMs in terms of SMEFT operators }\label{section:models}
	\begin{table}[h!]
	\centering
	\renewcommand{\arraystretch}{1.5}
	\caption{\small The nomenclature and quantum numbers of the SM ``fields" relevant for this work. The BSM Lagrangians are defined in terms of these SM and heavy fields (table \ref{table:models}).}
	\begin{threeparttable}
		\begin{tabular}{|c|c|c|c|c|}
			\hline
			\multirow{2}{*}{SM fields} & \multirow{2}{*}{Spin} &\multicolumn{3}{c|}{SM quantum numbers}\\
			\cline{3-5}
			&&$SU(3)_{_{C}}$&$SU(2)_{_{L}}$&$U(1)_{_Y}$\tnote{1}\\
			\hline \hline
			$q_{_L}$ & $\frac{1}{2}$ & 3 & 2 & $\frac{1}{6}$\\
			\hline
			$l_{_L}$ & $\frac{1}{2}$ & 1 & 2 & -$\frac{1}{2}$\\
			\hline
			$u_{_R}$ & $\frac{1}{2}$ & 3 & 1 & $\frac{2}{3}$\\
			\hline
			$d_{_R}$ & $\frac{1}{2}$ & 3 & 1 & -$\frac{1}{3}$\\
			\hline
			$e_{_R}$ & $\frac{1}{2}$ & 1 & 1 & -1\\
			\hline
			$H$ 	 & 0 			 & 1 & 2 & $\frac{1}{2}$\\
				\hline
			$B_{\mu\nu}$ 	 & 1 			 & 1 & 1 & 0\\
				\hline
			$W_{\mu\nu}$ 	 & 1 			 & 1 & 3 & 0\\
				\hline
		$B_{\mu\nu}$ 	 & 1 			 & 8 & 1 & 0\\
			\hline
		\end{tabular}
		\begin{tablenotes}
			\item[1] \footnotesize{Hypercharge convention: $Q_{\text{em}}$ = $T_{3}$ + Y, where $Q_{\text{em}}$, $T_{3}$ and Y are electro-magnetic charge, third component of isospin quantum number and hypercharge respectively.}
		\end{tablenotes}
	\end{threeparttable}\label{tab:SMfieldconvention}
\end{table}

	We consider the BSM scenarios which are single-heavy-multiplet-extensions of the SM. Once the massive particle(s) are integrated out, their impacts are captured through the higher dimensional effective operators made up of SM DOFs. Then the renormalizable BSM theories can be realized as effective ones and are expressed as as: 
\begin{align}\label{eq:bsmlagr}
	\mathcal{L}_{_\text{eff}} = \mathcal{L}_{_\text{SM}}^{d\leq 4} +  \mathcal{L}_{_\text{SM}}^{\text{EFT}}. 
\end{align}
	Here, $\mathcal{L}_{_\text{SM}}^{\text{EFT}}$ can be expressed in a compact form as  $\sum\limits_{d=5,...} \sum\limits_{i} \left(\mathcal{C}^{(d)}_{i} / \Lambda^{d-4}\right) Q_{i}^{(d)}$,
	where, $\mathcal{C}^{(d)}_{i}$'s  and $Q_{i}^{(d)}$'s are the WCs and the SM-invariant effective operators of mass dimension-$d$, respectively. Here, $ i$ runs over the number of independent effective operators, i.e., the dimension of the operator basis for a given mass-dimension. In this work, we restrict our study to dimension-6 effective operators in Warsaw basis only. The SM field-convention is in table~\ref{tab:SMfieldconvention}. The SM Lagrangian is defined as:
	\begin{align}\label{eq:smlagr}
		\mathcal{L}_{\text{SM}}^{d\leq 4} &= -\frac{1}{4} G^a_{\mu\nu}G^{a,{\mu\nu}}-\frac{1}{4} W^I_{\mu\nu}W^{I,{\mu\nu}}-\frac{1}{4}B_{\mu\nu}B^{\mu\nu}+ |D_{\mu}H|^2 - \mu_{_H}^2 |H|^2 - \lambda_{H} |H|^4 \nonumber\\
		&+\bar{l}_{_L}i \slashed{D} \,l_{_L} + \bar{q}_{_L}i \slashed{D} \,q_{_L} + \bar{e}_{_R} i \slashed{D} \, e_{_R} + \bar{u}_{_R}\, i \slashed{D} \, u_{_R} + \bar{d}_{_R} \, i \slashed{D}\, d_{_R}\nonumber\\
		&-\left\lbrace Y_{\text{SM}}^{(e)}  H^\dagger \bar{e}_{_R} \, l_{_L}		 \, +  Y_{\text{SM}}^{(u)} \widetilde{H}^\dagger \bar{u}_{_R}  \, q_{_L} \,  + \, Y_{\text{SM}}^{(d)} H^\dagger	 \bar{d}_{_R}  \, q_{_L}		 \, + \,\text{h.c.} \right\rbrace ,
	\end{align}
	where, $B_{\mu\nu}, W^I_{\mu\nu}, \text{ and } G^a_{\mu\nu}$ are the field strength tensors of the SM gauge groups  $U(1)_{_Y}$, $SU(2)_{_L}$, and $SU(3)_{_C}$ respectively with $a=1,\cdots,8$. The $D$'s are the covariant derivatives. $Y_{\text{SM}}$'s are the SM Yukawa couplings, and $\widetilde{H} = i  \,  \sigma_2 \, H^\ast$.  
	
\begin{table}[h!]
	\centering
	\caption{\small SM gauge quantum numbers for the heavy BSM scalars are in the 2nd column. Relevant effective operators are in columns III - XII. Ticks (\cmark) and crosses (\xmark) represent whether that operator (columns) is generated from the heavy fields (rows) or not. 6 different classes with identical set of operators are separated by triple-lines.}
	\renewcommand*{\arraystretch}{2}
	\begin{adjustbox}{width=0.8\textwidth}
		\begin{tabular}{|p{25pt}|p{68pt}|c|c|c|c|c|c|c|c|c|c|}
			\hline
			\small Heavy\newline BSM \newline fields&\small The SM Gauge\newline quantum nos. \newline(Color, Isospin, \newline Hypercharge)& $Q_H$ & $Q_{H\square}$ & $Q_{HD}$ & $Q_{HB}$ & $Q_{HW}$ & $Q_{HWB}$ & $Q_{HG}$ &$Q_{eH}$& $Q_{uH}$ & $Q_{dH}$ \\
			\hline \hline 
			$\mathcal{S}$&(1,1,0) & \cmark & \cmark & \xmark & \xmark & \xmark & \xmark & \xmark &\xmark &\xmark &\xmark \\
			\hline \hline \hline
			$\mathcal{S}_{2}$&(1,1,2) & \cmark & \cmark & \xmark & \cmark & \xmark & \xmark & \xmark &\xmark &\xmark &\xmark \\
			\hline \hline \hline
			$\Delta$&(1,3,0) & \cmark & \cmark & \cmark & \xmark & \cmark & \xmark & \xmark &\cmark &\cmark &\cmark \\
			\hline \hline \hline 
			$\mathcal{H}_{2}$&(1,2,${-\frac{1}{2}}$) & \cmark & \cmark & \cmark & \cmark & \cmark & \cmark & \xmark &\cmark &\cmark &\cmark \\
			\hline
			$\Delta_{1}$&(1,3,1) & \cmark & \cmark & \cmark & \cmark & \cmark & \cmark & \xmark &\cmark &\cmark &\cmark \\
			\hline
			$\Sigma$&(1,4,${\frac{1}{2}}$) & \cmark & \cmark & \cmark & \cmark & \cmark & \cmark & \xmark &\cmark &\cmark &\cmark \\ 
			\hline \hline \hline
			$\varphi_{1}$&(3,1,${-\frac{1}{3}}$) & \cmark & \cmark & \xmark & \cmark & \xmark & \xmark & \cmark &\xmark &\xmark &\xmark \\
			\hline
			$\varphi_{2}$&(3,1,$-{\frac{4}{3}}$) & \cmark & \cmark & \xmark & \cmark & \xmark & \xmark & \cmark &\xmark &\xmark &\xmark \\
			\hline \hline \hline
			$\Theta_{1}$&(3,2,$\frac{1}{6}$) & \cmark & \cmark & \xmark & \cmark & \cmark & \xmark & \cmark &\xmark &\xmark &\xmark \\
			\hline
			$\Theta_{2}$&(3,2,$\frac{7}{6}$) & \cmark & \cmark & \xmark & \cmark & \cmark & \xmark & \cmark &\xmark &\xmark &\xmark \\
			\hline
			$\Omega$&(3,3,-$\frac{1}{3}$) & \cmark & \cmark & \xmark & \cmark & \cmark & \xmark & \cmark &\xmark &\xmark &\xmark \\
			\hline
		\end{tabular}
	\end{adjustbox}
	\label{table:models}
\end{table}
	
	In the present analysis, we consider 11 BSM scenarios which are single heavy field extensions of the SM. We choose these models carefully to encompass the variety of phenomenological features. To proceed further,  we integrate out the heavy fields belonging to the adopted BSMs and compute the effective operators and their respective WCs. It is important to note here, that the WCs are the functions of BSM parameters and are thus not entirely independent\footnote{It is worthy to mention that this is very much unlike the usual SMEFT approach where all the WCs are assumed to be independent and free parameters.}. Here, we present the analytical structures of the WCs which are computed up to one-loop level, considering only heavy field propagators in the loop (pure heavy-loop\footnote{We have ignored the contributions from heavy-light mixing in the loop.}), using the Mathematica package CoDEx \cite{Bakshi:2018ics}.  We summarize the characteristics of the {\it to-be integrated out} BSM fields corresponding to our adopted scenarios and encapsulate the respective effective operators in table~\ref{table:models}.

	In principle, though one must write down the most general gauge-invariant theories involving these heavy fields and integrate them out to compute all possible effective operators, our  present analysis is driven by a set of chosen observables that encompass a very specific set of effective operators. Any other operator, that does not belong to that set, remain unconstrained in our analysis, and are irrelevant for the purpose of this work. This is why we only note down those interactions involving the BSM fields, which can lead to the desired operators. As an example, Yukawa-type interaction terms are viable in the case of certain BSM scenarios (with the scalar as the only heavy field), and these terms generate WCs of four-fermionic effective operators that do not contribute to our set of observables. These types of interactions are ignored here, without loss of any generality. To check the complete list of WCs and their expressions in terms of BSM parameters, please go to the GitHub repository \cite{Githubeffex}, where we make the full workflow and results (BSM theory implementation and the resulting effective operators and their WCs) available in Mathematica notebook files. In this work, we have ignored the Renormalization Group (RG) evolution of the effective operators, which are computed up to one-loop level\footnote{As we have checked for some of the cases, incorporation of RG evolution of the operators from NP to electroweak scale does not alter our conclusion significantly and hence, is of less practical relevance.}. Thus, the WCs, computed at the matching scale, are assumed to be unaltered at the electroweak scale.

	Considering the points made above, we further note that out of the 18 relevant WCs (Figure~\ref{fig:wctriplot}) for the present set of observables, a maximum of 10 operators can be exhausted (table~\ref{table:models}), in presence of one or more BSM scenarios of the 11 considered here. This justifies the absence of $Q_{Hl}^{(1)}$,  $Q_{Hl}^{(3)}$, $Q_{Hq}^{(1)}$, $Q_{Hq}^{(3)}$, $Q_{He}$, $Q_{Hu}$, $Q_{Hd}$, and $Q_{ll}$ operators in the model-dependent part of our analysis. 

	In the following subsections, we first introduce the relevant part of the BSM interactions, and then tabulate the effective operators and respective WCs as functions of BSM parameters. Here, we consider the mass of the heavy fields to be the same as the cut-off scale. Thus the all the dimension-6 operators are suppressed by mass-squared terms of the integrated out heavy fields.
	
	\subsection{Real Singlet Scalar}\label{sec:class5}
	This is the first of three BSM scenarios with a unique WCs-space. The SM is extended by a real singlet heavy scalar ($\mathcal{S}$) and the relevant part of  interactions involving $ \mathcal{S}$ is given as\footnote{As the SM Lagrangian is always there for all BSMs, we are not quoting that part of the Lagrangian explicitly for each model.} \cite{Zhang:2016pja,Jiang:2018pbd,Dawson:2017vgm,Haisch:2020ahr}:
\begin{align}\label{eq:model-S}
	\mathcal{L}_{\mathcal{S}} &\supset  \frac{1}{2} \, D_{\mu}\mathcal{S} \, D^{\mu}\mathcal{S} -\frac{1}{2} m_{_\mathcal{S}}^2 \, \mathcal{S}^2 - c_{_{\mathcal{S},a}} |H|^2 \mathcal{S} - \frac{\kappa_{_\mathcal{S}}}{2} |H|^2 \mathcal{S}^2 - \frac{\mu_{_\mathcal{S}}}{3!}  \mathcal{S}^3 - \frac{\lambda_{_\mathcal{S}}}{4!}  \mathcal{S}^4,
\end{align}
	where, $D_{\mu}$ is the covariant derivative\footnote{In this case $D_{\mu}=\partial_{\mu}$ for a real singlet scalar, but for rest of the scenarios, $D_{\mu}$ possesses non-trivial structures. As its explicit form is not required for this discussion, we will not mention it further in detail.} and $ m_{_{\mathcal{S}}} $ is the mass of the heavy scalar. We list the effective operators and the respective WCs, generated after $ {\mathcal{S}}$ is integrated out, in table~\ref{tab:realsingletWCs}. Note that the WCs are functions of BSM parameters depicted in eq.~\ref{eq:model-S}.
	
\begin{table*}[h!]
	\caption{\small The relevant effective operators (in Warsaw basis) and the associated WCs, generated once the heavy real SM singlet scalar ($\mathcal{S}$) is integrated out up to one-loop level, are tabulated. The WCs are expressed as functions of BSM parameters, see eq.~\ref{eq:model-S}.}
	\label{tab:realsingletWCs}
	\centering
	\renewcommand{\arraystretch}{2.2}
	\begin{tabular}{|*{2}{c|}}
		\hline 
		Effective operator & Wilson coefficient (SM + $\mathcal{S}$)\\
		\hline \hline
		\multirow{2}{*}{$Q_H$}  &  $-\frac{c_{\text{$\mathcal{S}$,a}}^2 \kappa _S \lambda _{\mathcal{S}}}{32 \pi ^2 m_S^4}+\frac{c_{\text{$\mathcal{S}$,a}}^2 \kappa _S \mu _{\mathcal{S}}^2}{32 \pi ^2 m_S^6}-\frac{c_{\text{$\mathcal{S}$,a}} \kappa _S^2 \mu _{\mathcal{S}}}{64 \pi ^2 m_S^4}+\frac{c_{\text{$\mathcal{S}$,a}}^3 \lambda _{\mathcal{S}} \mu _{\mathcal{S}}}{48 \pi ^2 m_S^6}$  \\
		&$-\frac{c_{\text{$\mathcal{S}$,a}}^3 \mu _{\mathcal{S}}^3}{96 \pi ^2 m_S^8}+\frac{c_{\text{$\mathcal{S}$,a}}^3 \mu _{\mathcal{S}}}{6 m_S^6}-\frac{c_{\text{$\mathcal{S}$,a}}^2 \kappa _S}{2 m_S^4}-\frac{\kappa _S^3}{192 \pi ^2 m_S^2}$\\
		\hline
		$Q_{H\square }$  &  $-\frac{5 c_{\text{$\mathcal{S}$,a}} \kappa _S \mu _{\mathcal{S}}}{192 \pi ^2 m_S^4}-\frac{c_{\text{$\mathcal{S}$,a}}^2 \lambda _{\mathcal{S}}}{32 \pi ^2 m_S^4}+\frac{11 c_{\text{$\mathcal{S}$,a}}^2 \mu _{\mathcal{S}}^2}{384 \pi ^2 m_S^6}-\frac{c_{\text{$\mathcal{S}$,a}}^2}{2 m_S^4}-\frac{\kappa _S^2}{384 \pi ^2 m_S^2}$  \\
		\hline
	\end{tabular}
\end{table*}

	\subsection{Complex Singlet Scalar}\label{sec:class6}
	In the next BSM scenario with a unique choice of WCs, the SM is extended by a complex singlet heavy scalar ($\mathcal{S}_{2}$) with hypercharge Y=$2$. The relevant part of interactions involving $ \mathcal{S}_2 $ is noted as \cite{deBlas:2014mba,deBlas:2017xtg}:
\begin{align}\label{eq:model-S2}
	\mathcal{L}_{\mathcal{S}_{_2}} &\supset   \left(D_{\mu}\mathcal{S}_{_2}\right)^\dagger \, \left(D^{\mu}\mathcal{S}_{_2}\right) - m_{\mathcal{S}_{_2}}^2 \mathcal{S}_{_2}^\dagger \mathcal{S}_{_2}  - \frac{\eta_{_{\mathcal{S}_{_2}}}}{2} |H|^2  |\mathcal{S}_{_2}|^2  - \lambda_{_{\mathcal{S}_{_2}}} |\mathcal{S}_{_2}|^4,
\end{align}
	where $ m_{_{\mathcal{S}_{_2}}} $ is the mass of heavy scalar ($\mathcal{S}_{_2}$), which gets integrated out, leading to the effective operators and respective WCs depicted in table~\ref{tab:complexsingletWCs}. The WCs are functions of BSM parameters, see eq.~\ref{eq:model-S2}.

\begin{table}[h!]
	\caption{\small  WCs (similar to table \ref{tab:realsingletWCs}) after integrating out the heavy complex SM singlet scalar ($\mathcal{S}_{2}$); see eq.~\ref{eq:model-S2}.}
	\label{tab:complexsingletWCs}
	\centering
	\renewcommand{\arraystretch}{2.2}
	\begin{tabular}{|*{2}{c|}}
		\hline 
		Effective operator & Wilson coefficient (SM + $\mathcal{S}_2$) \\
		\hline \hline
		$Q_H$  &  $\frac{\eta _{\mathcal{S}_2}^3}{96 \pi ^2 m_{\mathcal{S}_2}^2}$  \\
		\hline
		$Q_{HB}$  &  $-\frac{g_Y^2 \eta _{\mathcal{S}_2}}{48 \pi ^2 m_{\mathcal{S}_2}^2}$  \\
		\hline
		$Q_{H\square }$  &  $-\frac{\eta _{\mathcal{S}_2}^2}{192 \pi ^2 m_{\mathcal{S}_2}^2}$  \\
		\hline
	\end{tabular}
\end{table}
	
	\subsection{Isospin-Triplet Real Scalar}\label{sec:class4}
	In the third and the final BSM scenario with a unique choice of WCs, the SM is extended with a color-singlet, isospin-triplet heavy scalar ($\Delta$) with hypercharge Y=$0$. We write the relevant interactions involving $\Delta$ as \cite{Henning:2014wua,Ellis:2017jns}:
\begin{align}\label{eq:model-Delta}
	\mathcal{L}_{\Delta} &\supset  \frac{1}{2} \left(D_{\mu}\Delta\right)^I \, \left(D^{\mu}\Delta\right)^I - \frac{1}{2} m_{_\Delta}^2 \, \Delta^I \Delta^I + 2 \, \kappa_{_\Delta} \, H^\dagger \tau^I H \, \Delta^a - \eta_{_\Delta} |H|^2 \Delta^I \Delta^I - \frac{\lambda_{_\Delta}}{4} \left(\Delta^I \Delta^I\right)^2, 
\end{align}
	where $ m_{_{\Delta}} $ is the mass of $\Delta$. In table~\ref{tab:realtripletWCs} we list the effective operators and WCs generated after $ {\Delta}$ is integrated out. The WCs are functions of BSM parameters mentioned in eq.~\ref{eq:model-Delta}. 
	
\begin{table*}[h!]
	\caption{\small WCs (similar to table \ref{tab:realsingletWCs}) after integrating out the heavy real triplet scalar ($\Delta$); see eq.~\ref{eq:model-Delta}.}
	\label{tab:realtripletWCs}
	\small
	\centering
	\renewcommand{\arraystretch}{2.2}
	\begin{tabular}{|*{2}{c|}}
		\hline 
		Effective operator & Wilson coefficient (SM + $\Delta$) \\
		\hline \hline
		$Q_H$  &  $-\frac{5 \eta _{\Delta } \kappa _{\Delta }^2 \lambda _{\Delta }}{8 \pi ^2 m_{\Delta }^4}-\frac{\eta _{\Delta } \kappa _{\Delta }^2}{m_{\Delta }^4}-\frac{\eta _{\Delta }^3}{8 \pi ^2 m_{\Delta }^2} +\frac{5 \lambda _H \kappa _{\Delta }^2 \lambda _{\Delta }}{2 \pi ^2 m_{\Delta }^4}+\frac{4 \lambda _H \kappa _{\Delta }^2}{m_{\Delta }^4}$\\
		\hline
		$Q_{H\square }$  &  $-\frac{\eta _{\Delta }^2}{32 \pi ^2 m_{\Delta }^2}+\frac{5 \kappa _{\Delta }^2 \lambda _{\Delta }}{8 \pi ^2 m_{\Delta }^4}+\frac{\kappa _{\Delta }^2}{m_{\Delta }^4}$  \\
		\hline
		$Q_{HD}$  &  $-\frac{5 \kappa _{\Delta }^2 \lambda _{\Delta }}{4 \pi ^2 m_{\Delta }^4}-\frac{2 \kappa _{\Delta }^2}{m_{\Delta }^4}$  \\
		\hline
		$Q_{HW}$  &  $\frac{\eta _{\Delta } g_W^2}{96 \pi ^2 m_{\Delta }^2}$  \\
		\hline
		$Q_{eH}$  &  $\frac{5 Y_{\text{SM}}^{(e)} \kappa _{\Delta }^2 \lambda _{\Delta }}{8 \pi ^2 m_{\Delta }^4}+\frac{Y_{\text{SM}}^{(e)} \kappa _{\Delta }^2}{m_{\Delta }^4}$  \\
		\hline
		$Q_{uH}$  &  $\frac{5 Y_{\text{SM}}^{(u)} \kappa _{\Delta }^2 \lambda _{\Delta }}{8 \pi ^2 m_{\Delta }^4}+\frac{Y_{\text{SM}}^{(u)} \kappa _{\Delta }^2}{m_{\Delta }^4}$  \\
		\hline
		$Q_{dH}$  &  $\frac{5 Y_{\text{SM}}^{(d)} \kappa _{\Delta }^2 \lambda _{\Delta }}{8 \pi ^2 m_{\Delta }^4}+\frac{Y_{\text{SM}}^{(d)} \kappa _{\Delta }^2}{m_{\Delta }^4}$  \\
		\hline
	\end{tabular}
\end{table*}
	
	\subsection{Color-Singlet Isospin-Multiplet Complex Scalars}\label{sec:class1}
	Next, we discuss the class of BSM theories, which are extensions of the SM with $SU(3)_{_C}$ singlet but isospin non-singlet complex heavy scalar multiplets ($\mathcal{H}_{2}$, $\Delta_{1}$, $\Sigma$). 

	\subsubsection*{{\em SM+$\mathcal{H}_{2}$}}
	The SM is extended by a heavy $SU(2)_L$ complex doublet scalar ($\mathcal{H}_{2}$) with hypercharge  $Y=-\frac{1}{2}$ of mass $ m_{_{\mathcal{H}_{2}}} $. The  relevant $ Z_2 $ invariant interactions of $\mathcal{H}_{2}$ are noted as \cite{Deshpande:1977rw,Henning:2014wua,Nie_1999,Branchina:2018qlf,Pilaftsis:1999qt}:
\begin{align}\label{eq:2hdmlag}
	\mathcal{L}_{_{\mathcal{H}_{2}}}&\supset \left|D_{\mu}{\mathcal{H}_{2}}\right|^{2}-m_{_{\mathcal{H}_{2}}}^{2}|{\mathcal{H}_{2}}|^{2}-\frac{\lambda_{_{\mathcal{H}_{2}}}}{4}|{\mathcal{H}_{2}}|^4 -\lambda_{_{\mathcal{H}_{2},1}}|\widetilde{H}|^2|{\mathcal{H}_{2}}|^2-\lambda_{_{\mathcal{H}_{2},2}}|\widetilde{H}^{\dagger}{\mathcal{H}_{2}}|^2\nonumber\\
	&-\lambda_{_{\mathcal{H}_{2},3}}[(\widetilde{H}^{\dagger} {\mathcal{H}_{2}})^2+({\mathcal{H}_{2}}^{\dagger}\widetilde{H})^2].
\end{align}
	In table~\ref{tab:class1WCs}, we list the effective operators and their respective the WCs, functions of BSM parameters in eq.~\ref{eq:2hdmlag}, which are generated after $ \mathcal{H}_{2}$ is integrated out.
	
\begin{table*}[h!]
	\caption{\small WCs (similar to table \ref{tab:realsingletWCs}) after integrating out the heavy complex scalars, $\mathcal{H}_2$, $\Delta_1$, and $\Sigma$; see eqs. \ref{eq:2hdmlag}~-~\ref{eq:quartetlag}.}
	\label{tab:class1WCs}
	\centering
	\renewcommand{\arraystretch}{2.2}
	\small{
		\begin{tabular}{|c|c|c|c|}
			\hline
			Effective & \multicolumn{3}{c|}{Wilson coefficients}\\
			\cline{2-4}
			operators& SM + $\mathcal{H}_2$ & SM + $\Delta_1$ & SM + $\Sigma$ \\
			\hline \hline
			\multirow{5}{*}{$Q_H$}  &  $\frac{\lambda _H \left(\lambda _{_{\mathcal{H}_2,2}}\right){}^2}{48 \pi ^2 m_{\mathcal{H}_2}{}^2}+\frac{\lambda _H \left(\lambda _{\mathcal{H}_2,3}\right){}^2}{12 \pi ^2 m_{\mathcal{H}_2}{}^2}$  &  $\frac{8 \lambda _H \mu _{\Delta _1}^2}{m_{_{\Delta_1}}4}-\frac{4 \left(\lambda _{\Delta _1,1}\right) \mu _{\Delta _1}^2}{m_{_{\Delta_1}}4}$  &  $-\frac{5 \zeta _1 \zeta _2^2}{128 \pi ^2 m_{\Sigma }{}^2}-\frac{5 \zeta _1 \mu _{\Sigma }^2}{36 \pi ^2 m_{\Sigma }{}^2}$\\
			&$-\frac{\left(\lambda _{\mathcal{H}_2,1}\right){}^3}{48 \pi ^2 m_{\mathcal{H}_2}{}^2}-\frac{\left(\lambda _{\mathcal{H}_2,1}\right){}^2 \lambda _{_{\mathcal{H}_2,2}}}{32 \pi ^2 m_{\mathcal{H}_2}{}^2}$&$-\frac{12 \left(\lambda _{\Delta _1,1}\right) \left(\lambda _{\Delta _1,3}\right) \mu _{\Delta _1}^2}{\pi ^2 m_{_{\Delta_1}}4}-\frac{4 \left(\lambda _{\Delta _1,4}\right) \mu _{\Delta _1}^2}{m_{_{\Delta_1}}4}$&$-\frac{\zeta _1^3}{24 \pi ^2 m_{\Sigma }{}^2}-\frac{5 \zeta _2 \mu _{\Sigma }^2}{144 \pi ^2 m_{\Sigma }{}^2}$\\
			&$-\frac{\lambda _{\mathcal{H}_2,1} \left(\lambda _{_{\mathcal{H}_2,2}}\right){}^2}{32 \pi ^2 m_{\mathcal{H}_2}{}^2}-\frac{\lambda _{\mathcal{H}_2,1} \left(\lambda _{\mathcal{H}_2,3}\right){}^2}{8 \pi ^2 m_{\mathcal{H}_2}{}^2}$&$+\frac{\lambda _H \left(\lambda _{\Delta _1,4}\right){}^2}{12 \pi ^2 m_{_{\Delta_1}}2}-\frac{\left(\lambda _{\Delta _1,1}\right){}^3}{4 \pi ^2 m_{_{\Delta_1}}2}$&$+\frac{5 \lambda _H \zeta _2^2}{96 \pi ^2 m_{\Sigma }{}^2}+\frac{5 \lambda _H \mu _{\Sigma }^2}{54 \pi ^2 m_{\Sigma }{}^2}$\\
			&$-\frac{\left(\lambda _{_{\mathcal{H}_2,2}}\right){}^3}{96 \pi ^2 m_{\mathcal{H}_2}{}^2}-\frac{\lambda _{_{\mathcal{H}_2,2}} \left(\lambda _{\mathcal{H}_2,3}\right){}^2}{8 \pi ^2 m_{\mathcal{H}_2}{}^2}$&$-\frac{5 \left(\lambda _{\Delta _1,1}\right) \left(\lambda _{\Delta _1,4}\right){}^2}{16 \pi ^2 m_{_{\Delta_1}}2}-\frac{3 \left(\lambda _{\Delta _1,4}\right){}^3}{32 \pi ^2 m_{_{\Delta_1}}2}$&\\
			&&$-\frac{16 \left(\lambda _{\Delta _1,1}\right) \left(\lambda _{\Delta _1,2}\right) \mu _{\Delta _1}^2}{\pi ^2 m_{_{\Delta_1}}4}-\frac{3 \left(\lambda _{\Delta _1,1}\right){}^2 \left(\lambda _{\Delta _1,4}\right)}{8 \pi ^2 m_{_{\Delta_1}}2}$&\\
			\hline
			\multirow{2}{*}{$Q_{H\square }$}  &  $-\frac{\left(\lambda _{\mathcal{H}_2,1}\right){}^2}{96 \pi ^2 m_{\mathcal{H}_2}{}^2}-\frac{\lambda _{\mathcal{H}_2,1} \lambda _{_{\mathcal{H}_2,2}}}{96 \pi ^2 m_{\mathcal{H}_2}{}^2}$  &  $\frac{\mu _{\Delta _1}^2}{m_{_{\Delta_1}}4}-\frac{\left(\lambda _{\Delta _1,1}\right){}^2}{16 \pi ^2 m_{_{\Delta_1}}2}$  &  $-\frac{\zeta _1^2}{48 \pi ^2 m_{\Sigma }{}^2}+\frac{5 \zeta _2^2}{384 \pi ^2 m_{\Sigma }{}^2}$  \\
			&$+\frac{\left(\lambda _{_{\mathcal{H}_2,2}}\right){}^2}{384 \pi ^2 m_{\mathcal{H}_2}{}^2}+\frac{\left(\lambda _{\mathcal{H}_2,3}\right){}^2}{96 \pi ^2 m_{\mathcal{H}_2}{}^2}$&$-\frac{\left(\lambda _{\Delta _1,1}\right) \left(\lambda _{\Delta _1,4}\right)}{16 \pi ^2 m_{_{\Delta_1}}2}+\frac{\left(\lambda _{\Delta _1,4}\right){}^2}{192 \pi ^2 m_{_{\Delta_1}}2}$&$+\frac{5 \mu _{\Sigma }^2}{432 \pi ^2 m_{\Sigma }{}^2}$\\
			\hline
			$Q_{HD}$  &  $\frac{\left(\lambda _{\mathcal{H}_2,3}\right){}^2}{24 \pi ^2 m_{\mathcal{H}_2}{}^2}-\frac{\left(\lambda _{_{\mathcal{H}_2,2}}\right){}^2}{96 \pi ^2 m_{\mathcal{H}_2}{}^2}$  &  $\frac{4 \mu _{\Delta _1}^2}{m_{_{\Delta_1}}4}-\frac{\left(\lambda _{\Delta _1,4}\right){}^2}{24 \pi ^2 m_{_{\Delta_1}}2}$  &  $\frac{5 \mu _{\Sigma }^2}{108 \pi ^2 m_{\Sigma }{}^2}-\frac{5 \zeta _2^2}{192 \pi ^2 m_{\Sigma }{}^2}$  \\
			\hline
			$Q_{HW}$  &  $\frac{\lambda _{\mathcal{H}_2,1} g_W^2}{384 \pi ^2 m_{\mathcal{H}_2}{}^2}+\frac{\lambda _{_{\mathcal{H}_2,2}} g_W^2}{768 \pi ^2 m_{\mathcal{H}_2}{}^2}$  &  $\frac{\left(\lambda _{\Delta _1,1}\right) g_W^2}{48 \pi ^2 m_{_{\Delta_1}}2}+\frac{\left(\lambda _{\Delta _1,4}\right) g_W^2}{96 \pi ^2 m_{_{\Delta_1}}2}$  &  $\frac{5 \zeta _1 g_W^2}{192 \pi ^2 m_{\Sigma }{}^2}$  \\
			\hline
			$Q_{HB}$  &  $\frac{\lambda _{\mathcal{H}_2,1} g_Y^2}{384 \pi ^2 m_{\mathcal{H}_2}{}^2}+\frac{\lambda _{_{\mathcal{H}_2,2}} g_Y^2}{768 \pi ^2 m_{\mathcal{H}_2}{}^2}$  &  $\frac{\left(\lambda _{\Delta _1,1}\right) g_Y^2}{32 \pi ^2 m_{_{\Delta_1}}2}+\frac{\left(\lambda _{\Delta _1,4}\right) g_Y^2}{64 \pi ^2 m_{_{\Delta_1}}2}$  &  $\frac{\zeta _1 g_Y^2}{192 \pi ^2 m_{\Sigma }{}^2}$  \\
			\hline
			$Q_{HWB}$  &  $\frac{\lambda _{_{\mathcal{H}_2,2}} g_W g_Y}{192 \pi ^2 m_{\mathcal{H}_2}{}^2}$  &  $-\frac{\left(\lambda _{\Delta _1,4}\right) g_W g_Y}{24 \pi ^2 m_{_{\Delta_1}}2}$  &  $\frac{5 \zeta _2 g_W g_Y}{192 \pi ^2 m_{\Sigma }{}^2}$  \\
			\hline
			$Q_{eH}$  &  $\frac{\left(\lambda _{_{\mathcal{H}_2,2}}\right){}^2 Y_{\text{SM}}^{(e)}}{192 \pi ^2 m_{\mathcal{H}_2}{}^2}+\frac{\left(\lambda _{\mathcal{H}_2,3}\right){}^2 Y_{\text{SM}}^{(e)}}{48 \pi ^2 m_{\mathcal{H}_2}{}^2}$  &  $\frac{2 Y_{\text{SM}}^{(e)} \mu _{\Delta _1}^2}{m_{_{\Delta_1}}4}+\frac{\left(\lambda _{\Delta _1,4}\right){}^2 Y_{\text{SM}}^{(e)}}{48 \pi ^2 m_{_{\Delta_1}}2}$  &  $\frac{5 Y_{\text{SM}}^{(e)} \zeta _2^2}{384 \pi ^2 m_{\Sigma }{}^2}+\frac{5 Y_{\text{SM}}^{(e)} \mu _{\Sigma }^2}{216 \pi ^2 m_{\Sigma }{}^2}$  \\
			\hline
			$Q_{uH}$  &  $\frac{\left(\lambda _{_{\mathcal{H}_2,2}}\right){}^2 Y_{\text{SM}}^{(u)}}{192 \pi ^2 m_{\mathcal{H}_2}{}^2}+\frac{\left(\lambda _{\mathcal{H}_2,3}\right){}^2 Y_{\text{SM}}^{(u)}}{48 \pi ^2 m_{\mathcal{H}_2}{}^2}$  &  $\frac{2 Y_{\text{SM}}^{(u)} \mu _{\Delta _1}^2}{m_{_{\Delta_1}}4}+\frac{\left(\lambda _{\Delta _1,4}\right){}^2 Y_{\text{SM}}^{(u)}}{48 \pi ^2 m_{_{\Delta_1}}2}$  &  $\frac{5 Y_{\text{SM}}^{(u)} \zeta _2^2}{384 \pi ^2 m_{\Sigma }{}^2}+\frac{5 Y_{\text{SM}}^{(u)} \mu _{\Sigma }^2}{216 \pi ^2 m_{\Sigma }{}^2}$  \\
			\hline
			$Q_{dH}$  &  $\frac{\left(\lambda _{_{\mathcal{H}_2,2}}\right){}^2 Y_{\text{SM}}^{(d)}}{192 \pi ^2 m_{\mathcal{H}_2}{}^2}+\frac{\left(\lambda _{\mathcal{H}_2,3}\right){}^2 Y_{\text{SM}}^{(d)}}{48 \pi ^2 m_{\mathcal{H}_2}{}^2}$  &  $\frac{2 Y_{\text{SM}}^{(d)} \mu _{\Delta _1}^2}{m_{_{\Delta_1}}4}+\frac{\left(\lambda _{\Delta _1,4}\right){}^2 Y_{\text{SM}}^{(d)}}{48 \pi ^2 m_{_{\Delta_1}}2}$  &  $\frac{5 Y_{\text{SM}}^{(d)} \zeta _2^2}{384 \pi ^2 m_{\Sigma }{}^2}+\frac{5 Y_{\text{SM}}^{(d)} \mu _{\Sigma }^2}{216 \pi ^2 m_{\Sigma }{}^2}$  \\
			\hline
	\end{tabular}}
\end{table*}
	
	\subsubsection*{{\em SM + $\Delta_1$}}
	Here, we choose the heavy field  to be an  isospin-complex triplet scalar ($\Delta_{1}$) with hypercharge $Y=1$ of mass $ m_{_{\Delta_1}} $. The  interactions involving $\Delta_{1}$ which are relevant fo us are given as  \cite{Arhrib:2011uy}:
\begin{align}\label{eq:seesawlag}
	\mathcal{L}_{\Delta_1}&\supset \text{Tr}[(D_{\mu}\Delta_1)^{\dagger}(D^{\mu}\Delta_1)]-m_{_{\Delta_1}}^{2}\text{Tr}[\Delta_1^{\dagger}\Delta_1]-\left\lbrace \mu_{_{\Delta_1}} (H^{T} i \sigma_{2}\Delta_1^{\dagger}H)+\text{h.c.}\right\rbrace \nonumber \\
	&- \lambda_{_{\Delta_1,1}}(H^{\dagger}H)\text{Tr}(\Delta_1^{\dagger} \Delta_1) -\lambda_{_{\Delta_1,2}}\left[\text{Tr}(\Delta_1^{\dagger} \Delta_1)\right]^2-\lambda_{_{\Delta_1,3}}\text{Tr}\left[(\Delta_1^{\dagger} \Delta_1)^2\right]-\lambda_{_{\Delta_1,4}} \, H^{\dagger}  \Delta_1 \Delta_1^{\dagger}H,
\end{align}
    Once this field is integrated out the same set of effective operators as the previous case are generated, see table~\ref{tab:class1WCs}. But the WCs are now functions of BSM parameters given in eq.~\ref{eq:seesawlag}. 
	
	\subsubsection*{{\em SM + $\Sigma$}}
	The last BSM scenario considered in this class, is the extension of the SM by an isospin-complex quartet heavy scalar ($\Sigma$) with hypercharge $Y=\frac{1}{2}$ of mass $ m_{_{\Sigma}} $. The $ Z_2 $ invariant relevant interactions of $\Sigma$ are given as \cite{Babu:2009aq,Bambhaniya:2013yca,Dawson:2017vgm}:
\begin{align}\label{eq:quartetlag}
	\mathcal{L}_{\Sigma} &\supset   \left(D_{\mu}\Sigma\right)^\dagger \, \left(D^{\mu}\Sigma\right) - m_{_\Sigma}^2 \, \Sigma^\dagger \Sigma - \mu_{_\Sigma} \, \left[\left(\Sigma^\dagger H\right)^2 + \text{h.c.}\right] - \zeta_{_{1}} \left(H^{\dagger}H\right) \left(\Sigma^{\dagger}\Sigma\right)\nonumber\\
	& - \zeta_{_{2}} \left(H^{\dagger} \tau^I  H\right) \left(\Sigma^{\dagger} \, T_{_4}^I \, \Sigma\right) - \lambda_{_{\Sigma,1}} \left(\Sigma^\dagger \Sigma\right)^2  - \lambda_{_{\Sigma,2}} \left(\Sigma^\dagger \, T_{_4}^I \, \Sigma\right)^2.
\end{align}
 Once $ \Sigma $ is integrated out, the same set of effective operators are generated, but the associated WCs have different functional dependence on the BSM parameters depicted in eq.~\ref{eq:quartetlag}, see table~\ref{tab:class1WCs}.
	
	\subsection{Color-Triplet Isospin-Singlet Complex Scalars}\label{sec:class2}
	The class of BSMs we will discuss next, consists of two scenarios, where the SM is extended by $SU(2)_{_L}$ singlet but color triplet heavy complex scalar fields ($\varphi_{1}$ and $\varphi_{2}$) with different hypercharges. 
	
	\subsubsection*{{\em SM + $ \varphi_{_1} $}}
	Our first choice in this category is the heavy color-triplet, isospin-singlet complex scalar ($\varphi_{1}$) with hypercharge $Y=-\frac{1}{3}$ of mass $m_{_{\varphi_{_1}}}$. The interactions of our interest involving $ \varphi_{_1} $ are \cite{Bauer:2015knc,Bandyopadhyay:2016oif}:
\begin{align}\label{eq:varphi1lag}
	\mathcal{L}_{\varphi_1} &\supset  \left(D_{\mu} \varphi_1\right)^\dagger \, \left(D^{\mu} \varphi_1\right) - m_{{\varphi_1}}^2 \, \varphi_1^\dagger \varphi_1- \eta_{\varphi_{_1}} H^\dagger H \, \varphi_{_1}^\dagger \varphi_{_1}  -\lambda_{\varphi_{_1}} \left(\varphi_{_1}^\dagger \varphi_{_1}\right)^2.
\end{align}
	In table~\ref{tab:class2WCs}, we list the WCs along with the effective operators that are generated after $ \varphi_{_1}$ is integrated out. As expected, these WCs are functions of the BSM parameters noted in eq.~\ref{eq:varphi1lag}.

\begin{table*}[h!]
	\caption{\small WCs (similar to table \ref{tab:realsingletWCs}) after integrating out the heavy scalars, $\varphi _1$ and $\varphi _2$; see eqs. \ref{eq:varphi1lag} and \ref{eq:varphi2lag}.}
	\label{tab:class2WCs}
	\centering
	\renewcommand{\arraystretch}{2.2}
	\begin{tabular}{|c|c|c|}
		\hline 
		Effective & \multicolumn{2}{c|}{Wilson coefficients} \\
		\cline{2-3}
		operator& SM + $\varphi_{1}$ & SM + $\varphi_{2}$ \\
		\hline \hline
		$Q_H$  &  $-\frac{\eta _{\varphi _1}^3}{32 \pi ^2 m_{\varphi _1}^2}$  &  $-\frac{\eta _{\varphi _2}^3}{32 \pi ^2 m_{\varphi _2}^2}$  \\
		\hline
		$Q_{H\square }$  &  $-\frac{\eta _{\varphi _1}^2}{64 \pi ^2 m_{\varphi _1}^2}$  &  $-\frac{\eta _{\varphi _2}^2}{64 \pi ^2 m_{\varphi _2}^2}$  \\
		\hline
		$Q_{HB}$  &  $\frac{g_Y^2 \eta _{\varphi _1}}{576 \pi ^2 m_{\varphi _1}^2}$  &  $\frac{g_Y^2 \eta _{\varphi _2}}{36 \pi ^2 m_{\varphi _2}^2}$  \\
		\hline
		$Q_{HG}$  &  $\frac{g_S^2 \eta _{\varphi _1}}{384 \pi ^2 m_{\varphi _1}^2}$  &  $\frac{g_S^2 \eta _{\varphi _2}}{384 \pi ^2 m_{\varphi _2}^2}$  \\
		\hline
	\end{tabular}
\end{table*}
	
	\subsubsection*{{\em SM + $ \varphi_{_2} $}}
	Another scenario belonging to the same class is found when the SM is extended by a color-triplet, isospin-singlet complex scalar ($\varphi_{2}$) of hypercharge Y=$-\frac{4}{3}$ and of mass $ m_{_{\varphi_{_2}}} $. The  interactions of our interest involving $ \varphi_{_2}$ are \cite{Davidson_2010,Arnold:2013cva}:
\begin{align}\label{eq:varphi2lag}
	\mathcal{L}_{\varphi_2} &\supset  \left(D_{\mu} \varphi_2\right)^\dagger \, \left(D^{\mu} \varphi_2\right) - m_{{\varphi_2}}^2 \, \varphi_2^\dagger \varphi_2- \eta_{\varphi_{_2}} H^\dagger H \, \varphi_{_2}^\dagger \varphi_{_2}  -\lambda_{\varphi_{_2}} \left(\varphi_{_2}^\dagger \varphi_{_2}\right)^2.
\end{align}
	We find that the exact set of effective operators are generated, as $ \varphi_{_1}$, once $ \varphi_{_2}$ is integrated out. For this case, the WCs, functions of BSM parameters noted in eq.~\ref{eq:varphi2lag}, are captured in table~\ref{tab:class2WCs}.
	
	\subsection{Color-Triplet Isospin-Multiplet Complex Scalars}\label{sec:class3}
	The last class of BSM scenarios are extensions of the SM with heavy complex scalar fields $\Theta_{1}$, $\Theta_{2}$, and $\Omega$ charged under both color and isospin symmetries. 
	
	\subsubsection*{{\em SM + $ \Theta_{_1} $}}
	The first one is the extension of SM  with a color-triplet, isospin-doublet complex scalar ($\Theta_{1}$) with hypercharge $Y=\frac{1}{6}$ of mass $m_{_{\Theta_{_1}}}$. The relevant part of the interactions involving $\Theta_{_1}$ is \cite{Buchmuller:1986zs,Arnold:2013cva}:
\begin{align}\label{eq:theta1lag}
	\mathcal{L}_{\Theta_1} &\supset  \left(D_{\mu} \Theta_1\right)^\dagger \, \left(D^{\mu} \Theta_1\right) - m_{{\Theta_1}}^2 \, \Theta_1^\dagger \Theta_1- \eta_{_{\Theta_1}} H^\dagger H \, \Theta_1^\dagger \Theta_1  -\lambda_{\Theta_1} \left(\Theta_1^\dagger \Theta_1\right)^2.
\end{align}
	The effective operators and the associated WCs that are generated after integrating out ${\Theta_{_1}}$ are depicted in table~\ref{tab:class3WCs}. The WCs are expressed in terms of the BSM parameters mentioned in eq.~\ref{eq:theta1lag}.

\begin{table*}[h!]
	\caption{\small WCs (similar to table \ref{tab:realsingletWCs}) after integrating out the heavy scalars $\Theta_1$, $\Theta_2$, and $\Omega$; see eqs. \ref{eq:theta1lag}~-~\ref{eq:omegalag}.}
	\label{tab:class3WCs}
	\centering
	\renewcommand{\arraystretch}{2.2}
	\begin{tabular}{|c|c|c|c|}
		\hline
		Effective & \multicolumn{3}{c|}{Wilson coefficients} \\
		\cline{2-4}
		operator& SM + $\Theta_{1}$ & SM + $\Theta_{2}$ & SM + $\Omega$ \\
		\hline \hline
		$Q_H$  &  $-\frac{\eta _{\Theta _1}^3}{16 \pi ^2 m_{\Theta _1}^2}$  &  $-\frac{\eta _{\Theta _2}^3}{16 \pi ^2 m_{\Theta _2}^2}$  &  $-\frac{3 \eta _{\Omega}^3}{32 \pi ^2 m_{\Omega}^2}$  \\
		\hline
		$Q_{H\square }$  &  $-\frac{\eta _{\Theta _1}^2}{32 \pi ^2 m_{\Theta _1}^2}$  & $-\frac{\eta _{\Theta _2}^2}{32 \pi ^2 m_{\Theta _2}^2}$    &  $-\frac{3 \eta _{\Omega}^2}{64 \pi ^2 m_{\Omega}^2}$  \\
		\hline
		$Q_{HB}$  &  $\frac{g_Y^2 \eta _{\Theta _1}}{1152 \pi ^2 m_{\Theta _1}^2}$  &  $\frac{49 g_Y^2 \eta _{\Theta _2}}{1152 \pi ^2 m_{\Theta _2}^2}$  &  $\frac{g_Y^2 \eta _{\Omega}}{192 \pi ^2 m_{\Omega}^2}$  \\
		\hline
		$Q_{HG}$  &  $\frac{g_W^2 \eta _{\Theta _1}}{128 \pi ^2 m_{\Theta _1}^2}$  &  $\frac{g_W^2 \eta _{\Theta _2}}{128 \pi ^2 m_{\Theta _2}^2}$  &  $\frac{g_W^2 \eta _{\Omega}}{32 \pi ^2 m_{\Omega}^2}$  \\
		\hline
		$Q_{HW}$  &  $\frac{g_S^2 \eta _{\Theta _1}}{192 \pi ^2 m_{\Theta _1}^2}$  & $\frac{g_S^2 \eta _{\Theta _2}}{192 \pi ^2 m_{\Theta _2}^2}$   &  $\frac{g_S^2 \eta _{\Omega}}{128 \pi ^2 m_{\Omega}^2}$  \\
		\hline
	\end{tabular}
\end{table*}

	\subsubsection*{{\em SM + $ \Theta_{_2} $}}
	In the second instance, the heavy field ($\Theta_{2}$) with mass $m_{_{\Theta_{_2}}}$ is a color-triplet, isospin-doublet complex scalar with hypercharge $Y=\frac{7}{6}$. The interactions of $\Theta_{_2}$ that are relevant to our analysis are  \cite{Buchmuller:1986zs,Arnold:2013cva,Davidson_2010}:
\begin{align}\label{eq:theta2lag}
	\mathcal{L}_{\Theta_2} &\supset  \left(D_{\mu} \Theta_2\right)^\dagger \, \left(D^{\mu} \Theta_2\right) - m_{{\Theta_2}}^2 \, \Theta_2^\dagger \Theta_2- \eta_{_{\Theta_2}} H^\dagger H \, \Theta_2^\dagger \Theta_2  -\lambda_{\Theta_2} \left(\Theta_2^\dagger \Theta_2\right)^2.
\end{align}
	Similar to the previous case, the exact same operators are generated once ${\Theta_{_2}}$ is integrated out. The WCs, functions of the BSM parameters (eq.~\ref{eq:theta2lag}), are enlisted in table~\ref{tab:class3WCs}.
	
	\subsubsection*{{\em SM + $ \Omega $}}
	In the last example-model under this class, we choose the heavy field to be a color-triplet, isospin-triplet  scalar ($\Omega$) with hypercharge $Y=-\frac{1}{3}$ and mass $ m_{_{\Omega}} $. The  interactions involving $ \Omega $ that are relevant for this work are \cite{Buchmuller:1986zs,Arnold:2013cva}:
\begin{align}\label{eq:omegalag}
	\mathcal{L}_{\Omega} &\supset  \left(D_{\mu} \Omega\right)^\dagger \, \left(D^{\mu} \Omega\right) - m_{{\Omega}}^2 \, \Omega^\dagger \Omega- \eta_{_{\Omega}} H^\dagger H \, \Omega^\dagger \Omega  -\lambda_{\Omega} \left(\Omega^\dagger \Omega\right)^2.
\end{align}
	Once ${\Omega}$ is integrated out the same set of effective operators similar to the previous two scenarios are emerged. 	In table~\ref{tab:class3WCs}, we capture the WC which are functions of  model parameters given in eq.~\ref{eq:omegalag}. 
    
   \subsection{Statistical Inference}\label{sec:moddep_stat}
   To obtain the bounds on the model parameters, first the low energy observables are written in terms of dimension-6 SMEFT WCs and the SM parameters and then are mapped to the effective theory by expressing the WCs as functions of the BSM couplings and the cut-off scale $\Lambda$\footnote{The masses of the heavy scalars are equivalent to the cut off scale $\Lambda$, which is chosen as to be 1 TeV in the entirety of the analysis.}, following Sections~\ref{sec:class1}~-~\ref{sec:class6}.
    
   The statistical methodology is similar to that described in Section~\ref{sec:modind_stat}, with the exception that the free fit parameters here are those of the BSM models. In general, we choose uniform priors of $\{-4\pi,4\pi\}$ for BSM quartic couplings due to perturbativity and $\{-\Lambda, \Lambda\}$ for couplings with mass dimension 1 (since cut-off scale $\Lambda$ = 1 TeV). In certain scenarios, the Bayesian fit is insensitive to certain model parameters, and in most cases, these insensitive parameters are BSM self-quartic couplings.
   
   In the following text, we discuss the fits on the model parameter space of different SM extensions. We first discuss the models with multiple parameters followed by single parameter models.
   \subsubsection{Models with multiple parameters}
\begin{figure}[h!]
	\centering
   	\subfloat[SM+$\mathcal{S}$]
   	{\includegraphics[width=0.26\textwidth]{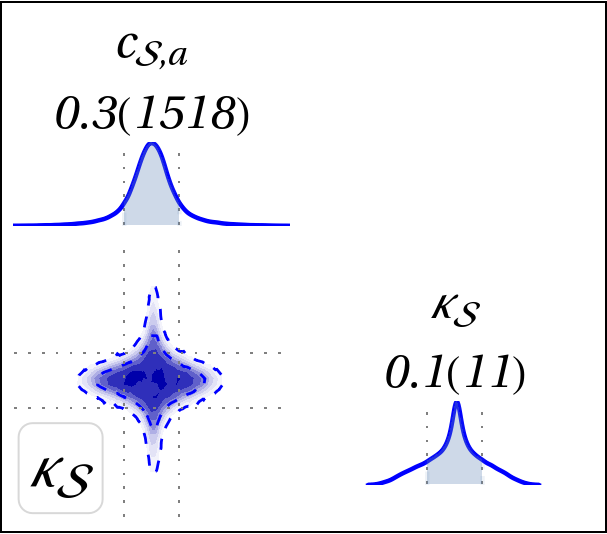}\label{fig:triplot_s}}~
   		\subfloat[SM+$\Delta$]
   	{\includegraphics[width=0.34\textwidth]{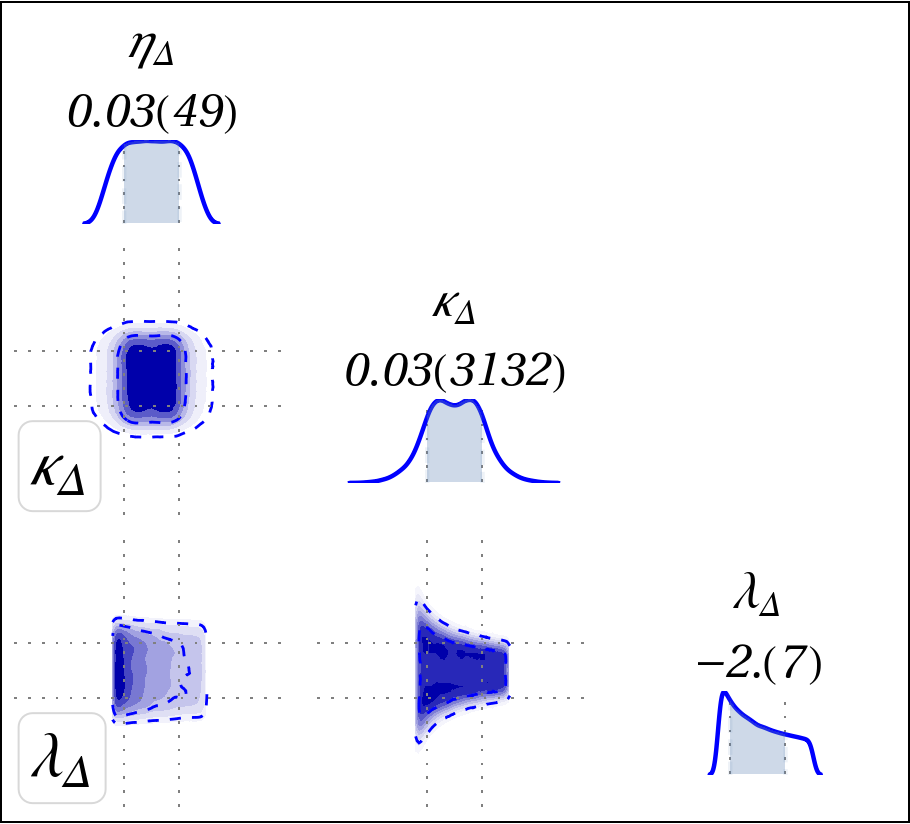}\label{fig:triplot_del}}~
   	\subfloat[SM+$\Delta_{1}$]
   	{\includegraphics[width=0.34\textwidth]{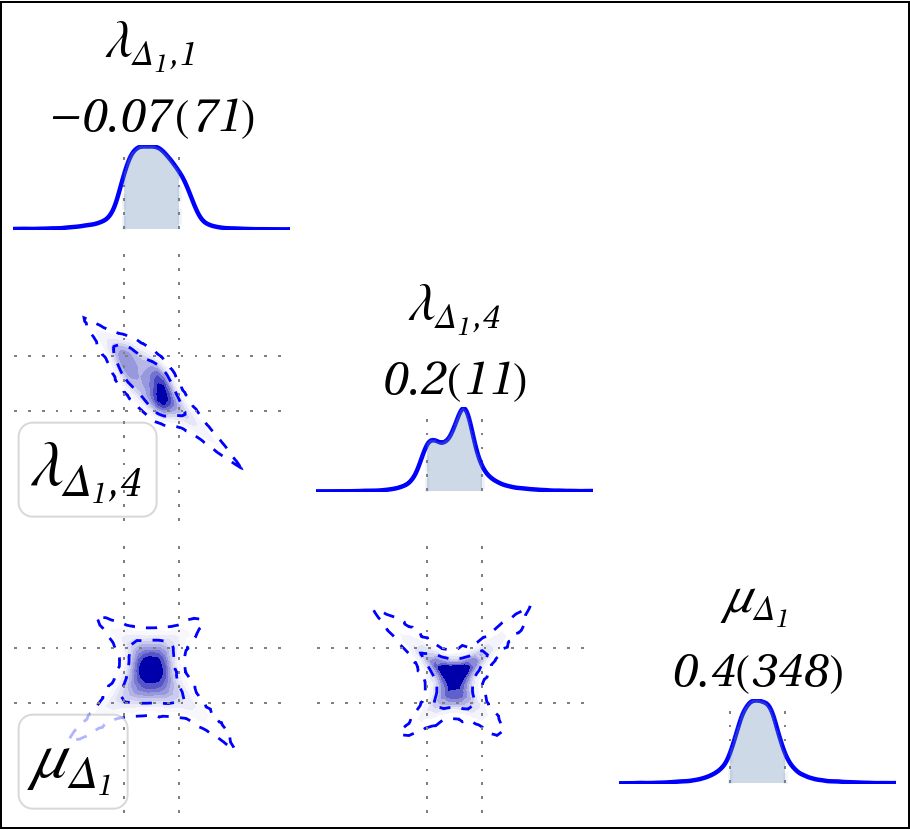}\label{fig:triplot_del1}}\\
   	\subfloat[SM+$\mathcal{H}_{2}$]
   	{\includegraphics[width=0.34\textwidth]{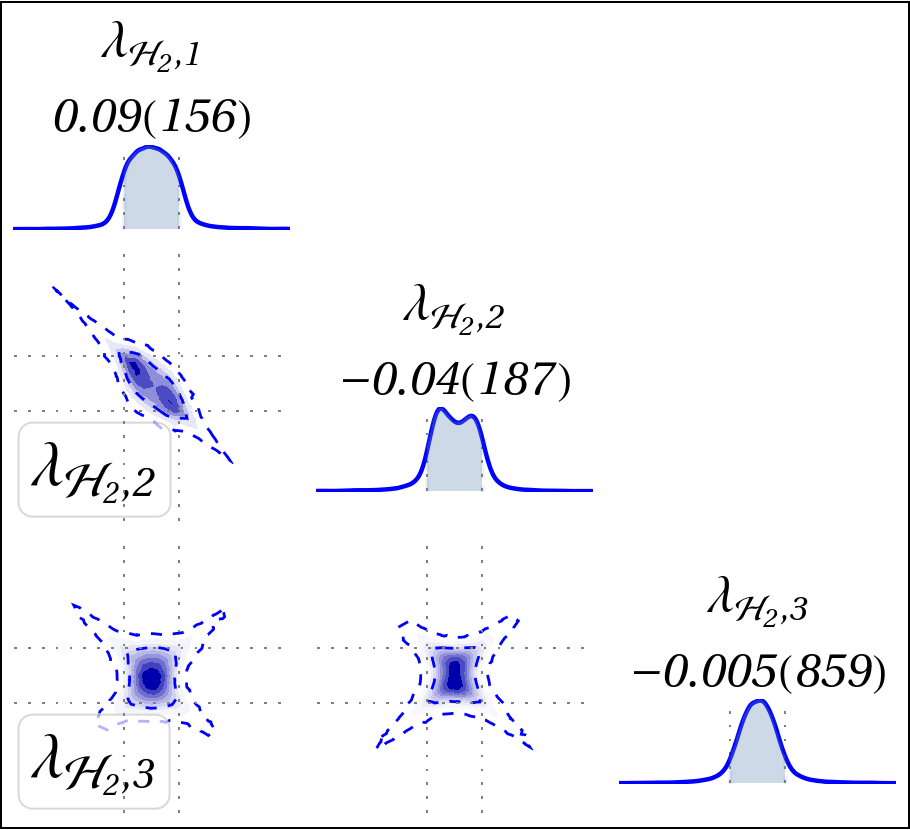}\label{fig:triplot_h2}}~
   	\subfloat[SM+$\Sigma$]
   	{\includegraphics[width=0.34\textwidth]{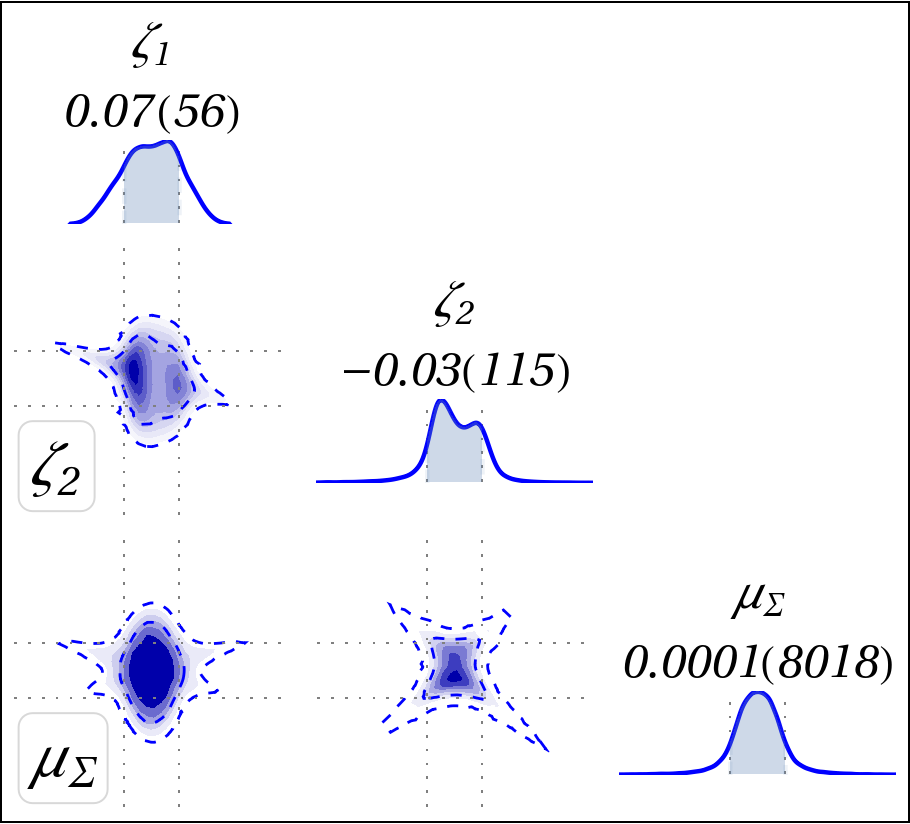}\label{fig:triplot_sig}}
   	\caption{\small The one- and two-dimensional marginalized posteriors for different SM extensions showing the correlations among them. Only the BSM parameters are shown. For full tri-plots like these, including the SM-parameters, please check our GitHub repository \cite{Githubeffex}. Here, the fitted values of $c_{\mathcal{S},a}, \kappa_{_\Delta}$, and $\mu_{\Delta_1}$ are expressed in GeV.}
   	\label{fig:triplot_models}
\end{figure}
   \subsubsection*{{\em SM+$\mathcal{S}$}}\label{subsec:singletfit}
   In this model, we find a set of two WCs, $\mathcal{C}_{H}$ and $\mathcal{C}_{H\square}$ that depend upon the model parameters $c_{_\mathcal{S},a}$, $\kappa_{_\mathcal{S}}$, $\mu_{_\mathcal{S}}$ and $\lambda_{_\mathcal{S}}$ (see column II of table~\ref{tab:realsingletWCs}). We perform the fit assuming the uniform priors of $\{-\Lambda, \Lambda\}$ for $c_{_\mathcal{S},a}$ and $\mu_{_\mathcal{S}}$, and $\{-4\pi,4\pi\}$ for $\kappa_{_\mathcal{S}}$ and $\lambda_{_\mathcal{S}}$. We find that the observables the fit used in this analysis are insensitive to $\mu_{_\mathcal{S}}$ and $\lambda_{_\mathcal{S}}$, i.e., these have negligible effects on the posterior distributions of the other parameters. Therefore, without loss of generality, these  are set to be zero and the fit results of $c_{_\mathcal{S},a}$ and $\lambda_{_\mathcal{S}}$ are obtained along with five SM parameters. The one (showing individual parameter space) and two (showing correlations) dimensional marginal posterior distributions, encapsulating the correlations between them, are shown in Figure~\ref{fig:triplot_s}.
   
      \subsubsection*{\em SM+$\Delta$}
   The 7 WCs in the effective theory resulting from extending the SM with a heavy real $SU(2)_{L}$ triplet scalar ($\Delta$) constrain three model parameters $\eta_{\Delta}$, $\kappa_{\Delta }$, and $\lambda_{\Delta}$ (see second column of table~\ref{tab:complexsingletWCs}). The ranges $\{-4\pi,4\pi\}$ for $\eta_{\Delta}$ and $\lambda_{\Delta}$, and $\{-\Lambda, \Lambda\}$ for $\kappa_{\Delta }$ are set as uniform priors. The marginal posteriors are shown in Figure~\ref{fig:triplot_del}.
   
   \subsubsection*{\em SM+$\Delta_{1}$}\label{subsec:seesawfit}
   For the complex $SU(2)_{L}$ triplet scalar ($\Delta_{1}$) extension, a set of 9 WCs are related to  the five model parameters $\mu_{\Delta_{1}}$, $\lambda_{\Delta_{1},\{1,2,3,4\}}$ (third column of table \ref{tab:class1WCs}). The fit is performed taking uniform priors within range of  $\{-4\pi,4\pi\}$ for the scalar quartic couplings $\lambda_{\Delta_{1},\{1,2,3,4\}}$ and $\{-\Lambda, \Lambda\}$ for $\mu_{\Delta_{1}}$ along with the SM parameters. Just like the model above, the fit is insensitive to  $\lambda_{\Delta_{1},2}$, and $\lambda_{\Delta_{1},3}$ and hence they are set to be zero. The marginal posteriors of $\lambda_{\Delta_{1},1}$, $\lambda_{\Delta_{1},4}$, and $\mu_{\Delta_{1}}$ are shown in Figure~\ref{fig:triplot_del1}. 
   
   \subsubsection*{\em SM+$\mathcal{H}_{2}$}
   We explore the model parameter space of $\mathcal{H}_{2}$ extended BSM scenario utilizing the 9 WCs written as functions of three model parameters listed in the second column of table~\ref{tab:class1WCs}. The BSM quartic couplings $\lambda_{\mathcal{H}_{2},{1}}$, $\lambda_{\mathcal{H}_{2},2}$, and $\lambda_{\mathcal{H}_{2},3}$ are fitted along with the SM parameters assuming uniform priors of $\{-4\pi,4\pi\}$. The one and two-dimensional marginal posterior for the BSM parameters $\lambda_{\mathcal{H}_{2},{1}}$, $\lambda_{\mathcal{H}_{2},2}$, and $\lambda_{\mathcal{H}_{2},3}$ are presented in the Figure~\ref{fig:triplot_h2}.

   \subsubsection*{\em SM+$\Sigma$}
   Analysis, similar to those described above for different BSM extensions, is done for the model parameters of SM+$\Sigma$ using the one-loop matching results. In this $SU(2)_{L}$ quartet scalar extension of $\Sigma$, a list of 9 WCs are expressed as functions of three model parameters $\zeta_{_{1}}$, $\zeta_{_{2}}$, and $\mu_{_{\Sigma}}$ (see the fourth column of table~\ref{tab:class1WCs}). Posterior distributions of these model parameters, along with SM parameters, are sampled assuming uniform priors of $\{-4\pi,4\pi\}$ for $\zeta_{_{1}}$, $\zeta_{_{2}}$, and $\mu_{_{\Sigma}}$. The one dimensional marginal posteriors of $\zeta_{_{1}}$, $\zeta_{_{2}}$, and $\mu_{_{\Sigma}}$, along with their two-dimensional counterparts, are shown in Figure~\ref{fig:triplot_sig}.
   
   \subsubsection{Models with one parameter}
   In case of the rest of the BSM scenarios, containing heavy scalars $\mathcal{S}_{2},\varphi_{1},\varphi_{2},\Omega,\Theta_{1},\Theta_{2}$, the WCs are functions of only one BSM parameter for each model. The fit results are shown in table~\ref{table:1modelparam}. We refrain from showing their posteriors as these are quite imprecise with a large, equi-probable parameter space while being consistent with zero.
\begin{table}[h!]
	\centering
   	\caption{\small Bayesian fit results of BSM theories with a single parameter.}
   	\begin{adjustbox}{width=0.9\textwidth}
   		\renewcommand{\arraystretch}{3.5}
   		\begin{tabular}{|c|c|c|c|c|c|c|}
   			\hline
   			Model & SM+$\mathcal{S}_{2}$ & SM+$\varphi_{1}$ & SM+$\varphi_{2} $ &SM+$\Omega$& SM+$\Theta_{1}$&SM+$\Theta_{2}$   \\
   			\hline
   			Model & $\eta_{\mathcal{S}_{2}} =$ &  $\eta_{\varphi_{1}} =$ & $\eta_{\varphi_{2}} =$ & $\eta_{\Omega} =$ & $\eta_{\Theta_{1}} =$  & $\eta_{\Theta_{2}} =$ \\
   			Parameter & $0.067\pm1.141$ &  $0.076\pm0.794$ & $0.016\pm0.793$ & $0.093\pm0.533$ & $0.095\pm0.619$  & $0.032\pm0.622$ \\
   			\hline
   		\end{tabular}
   	\end{adjustbox}
   	\label{table:1modelparam}
\end{table}
	

\begin{figure}[h!]
	\centering
	\includegraphics[ width=\textwidth]{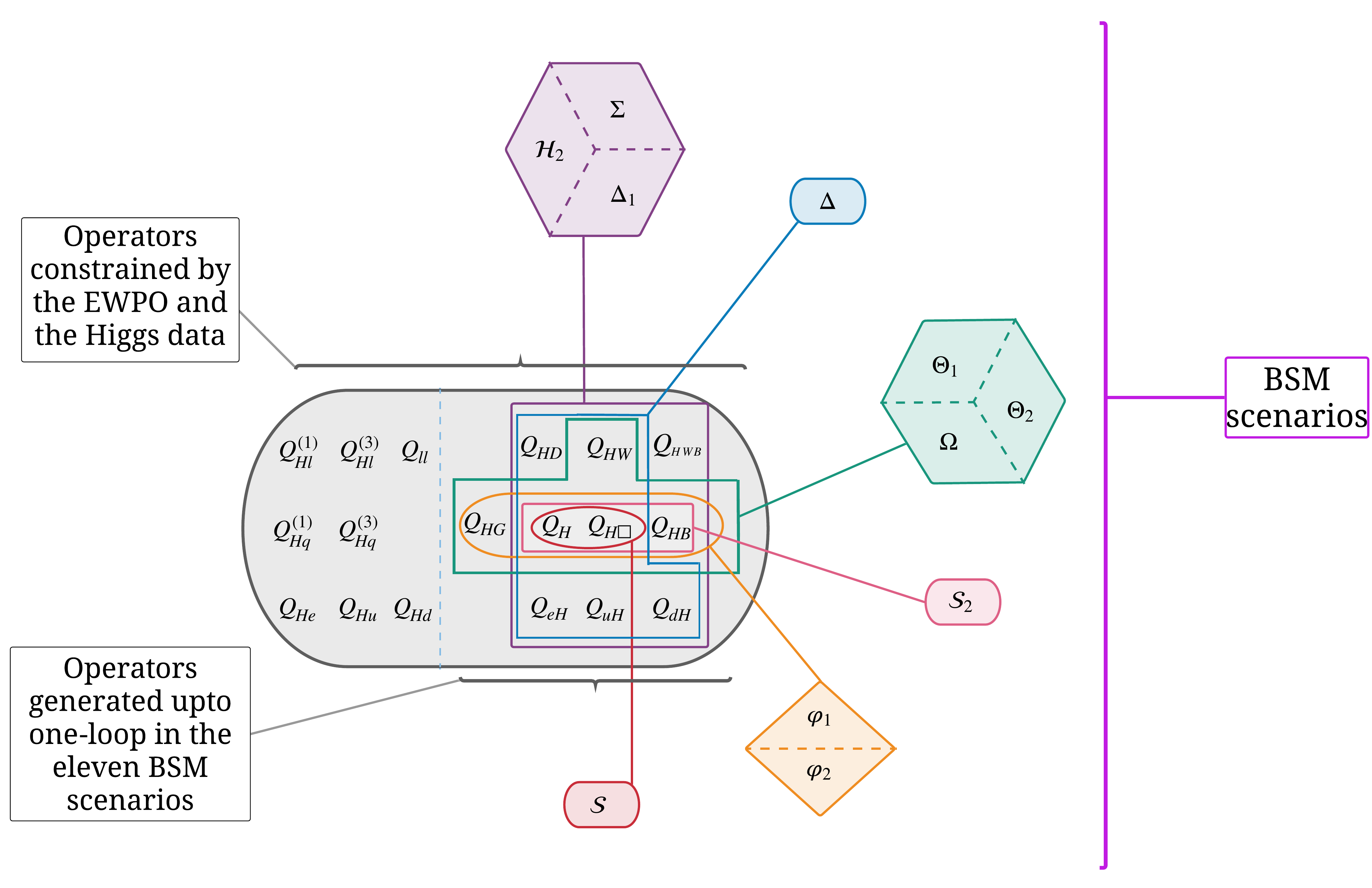}
	\caption{\small This flowchart encapsulates the `model-independent' vs `model-dependent' analysis discussed in Section~\ref{sec:comparison}.}
	\label{fig:flowchart}
\end{figure}


\begin{figure}[t]
	\centering
	\subfloat[$\mathcal{C}_{H}$ - $\mathcal{C}_{H\square}$]
	{\includegraphics[width=0.45\textwidth, height=6.7cm]{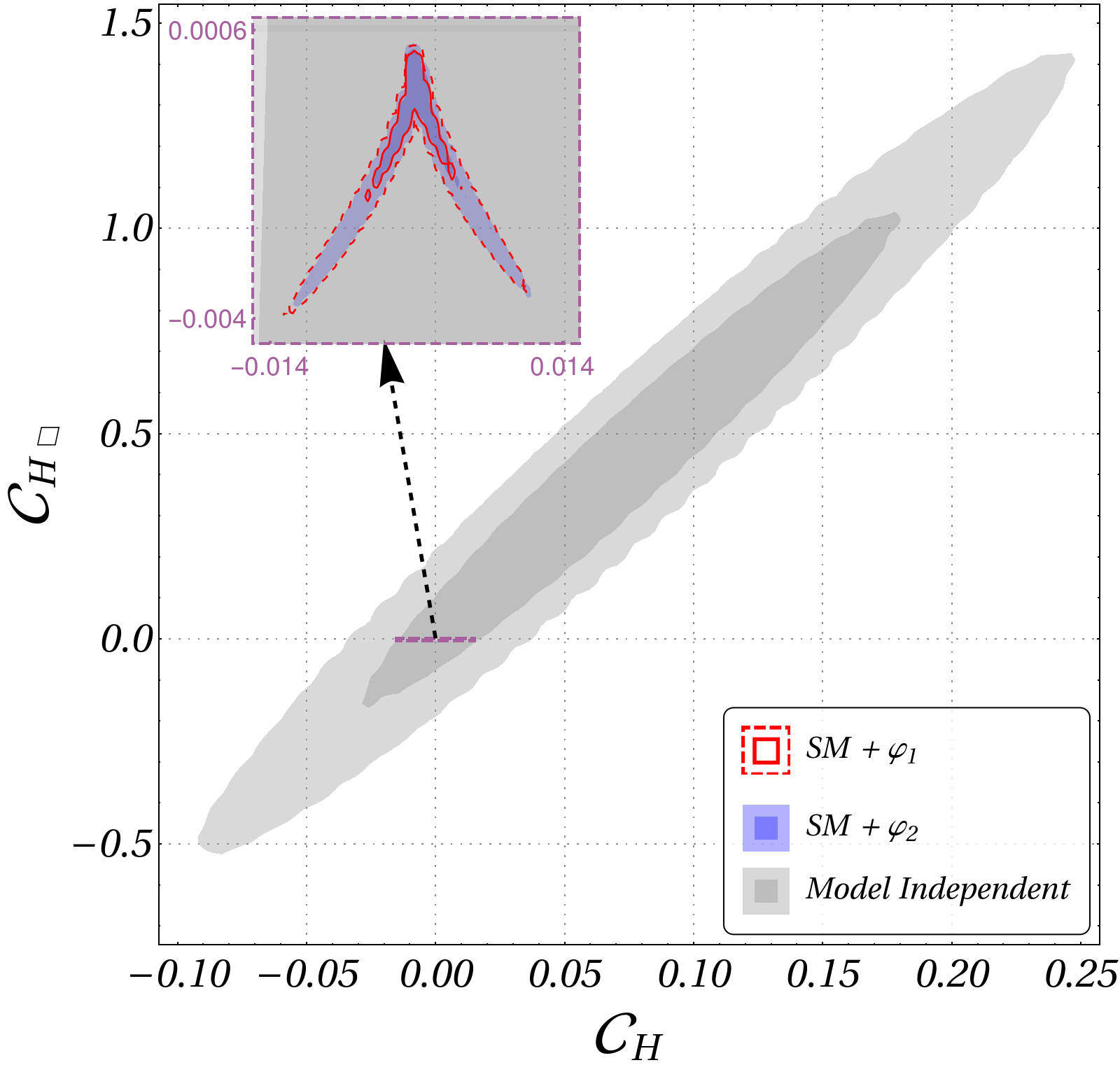}}~
	\subfloat[$\mathcal{C}_{H}$ - $\mathcal{C}_{HB}$]
	{\includegraphics[width=0.45\textwidth, height=6.7cm]{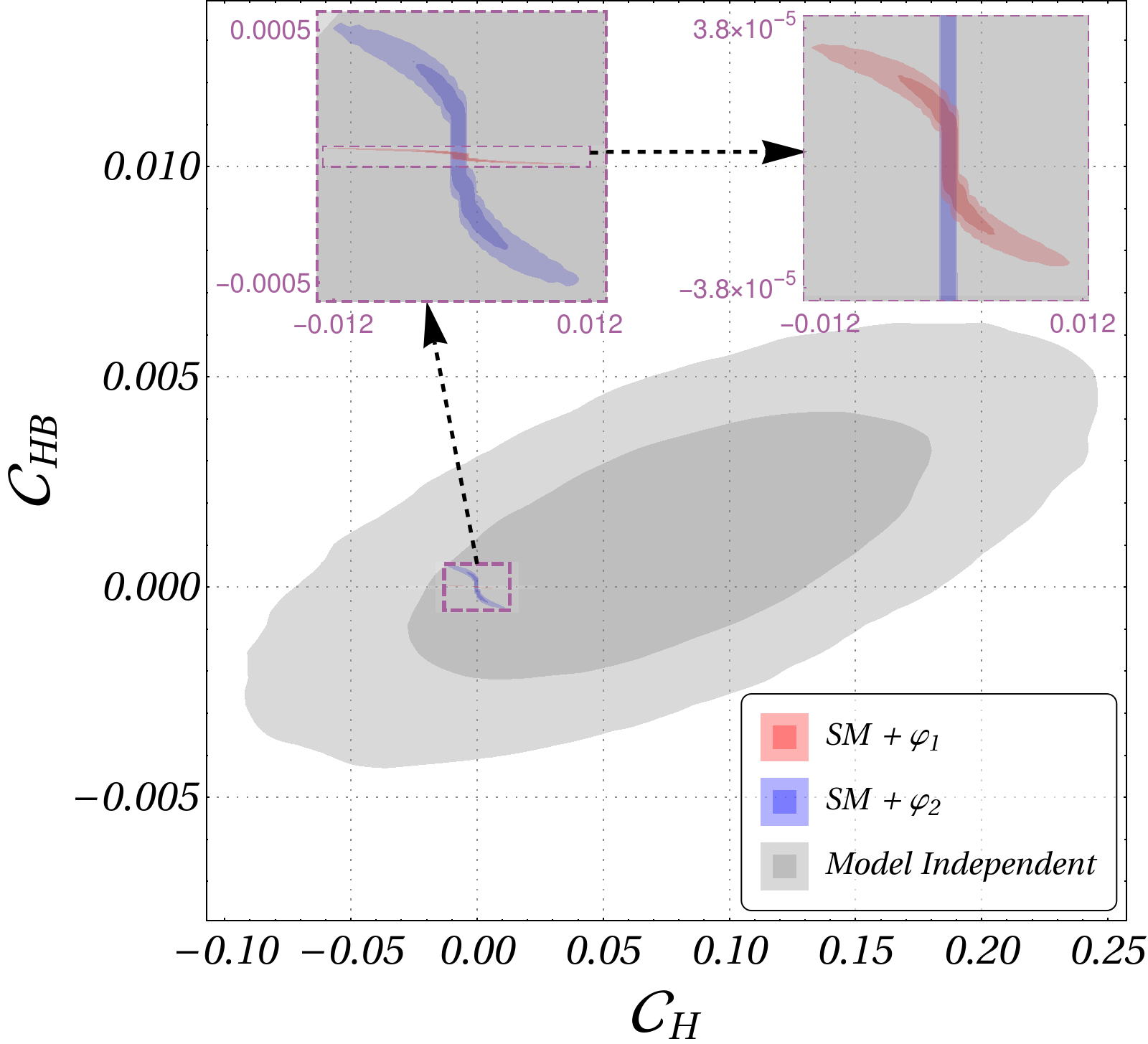}}\\
	\subfloat[$\mathcal{C}_{H}$ - $\mathcal{C}_{HG}$]
	{\includegraphics[width=0.45\textwidth, height=6.7cm]{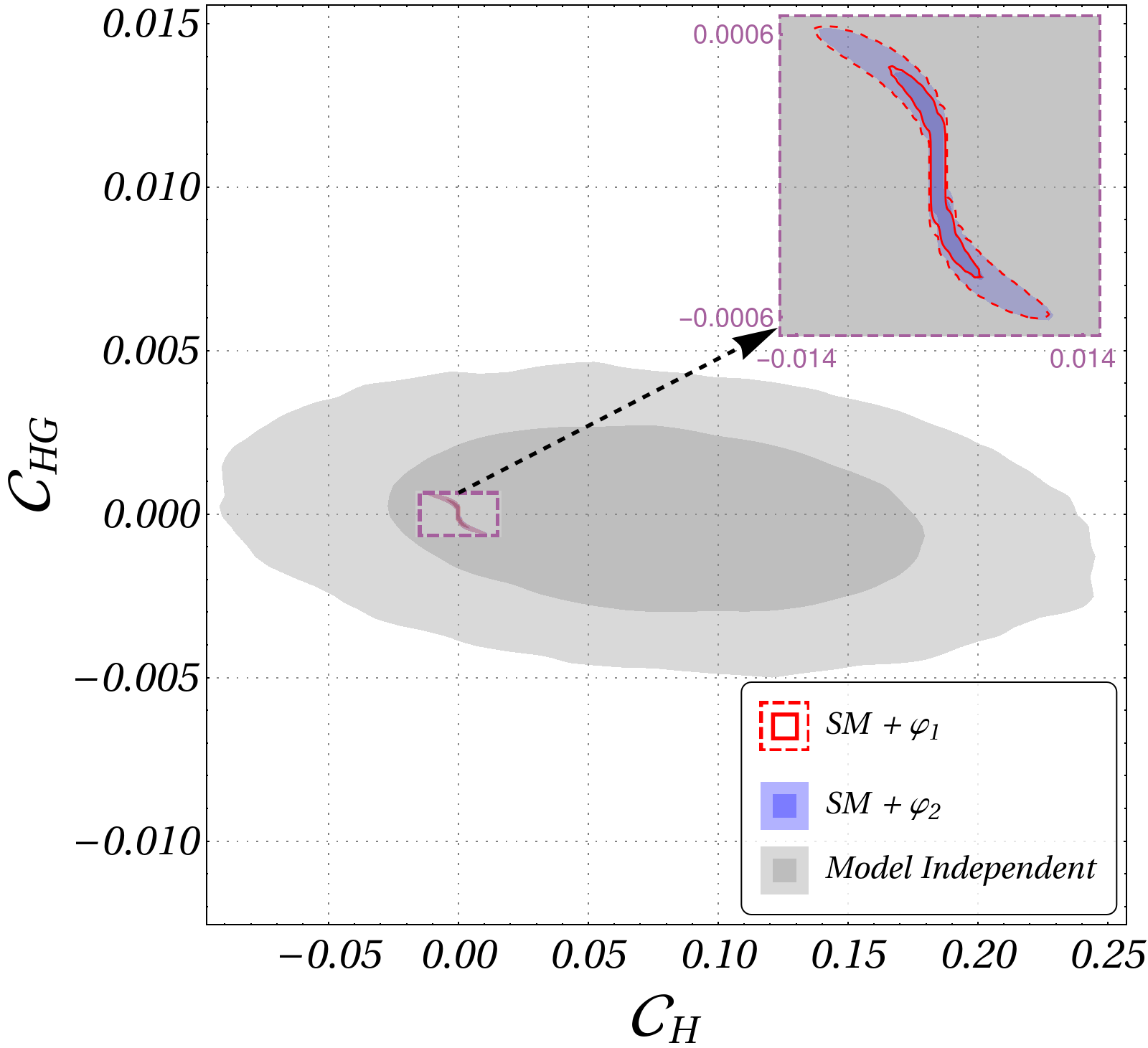}}
	\caption{\small Comparison of the two-dimensional posteriors of model-independent WCs (gray; 68\% (darker) and 95\% (lighter) credible intervals) with those generated from the class of degenerate leptoquark singlet scenarios. The red-dashed-bounded and blue regions correspond to the similar regions obtained from the SM+$\varphi_{1}$ and SM+$\varphi_{2}$ respectively. Zoomed-in spaces are shown inset.}
	\label{fig:leptoquark_singlet_wc}
\end{figure}

\begin{figure}[t]
	\centering
	\subfloat[$\mathcal{C}_{H}$ - $\mathcal{C}_{H\square}$]
	{\includegraphics[width=0.45\textwidth, height=6.7cm]{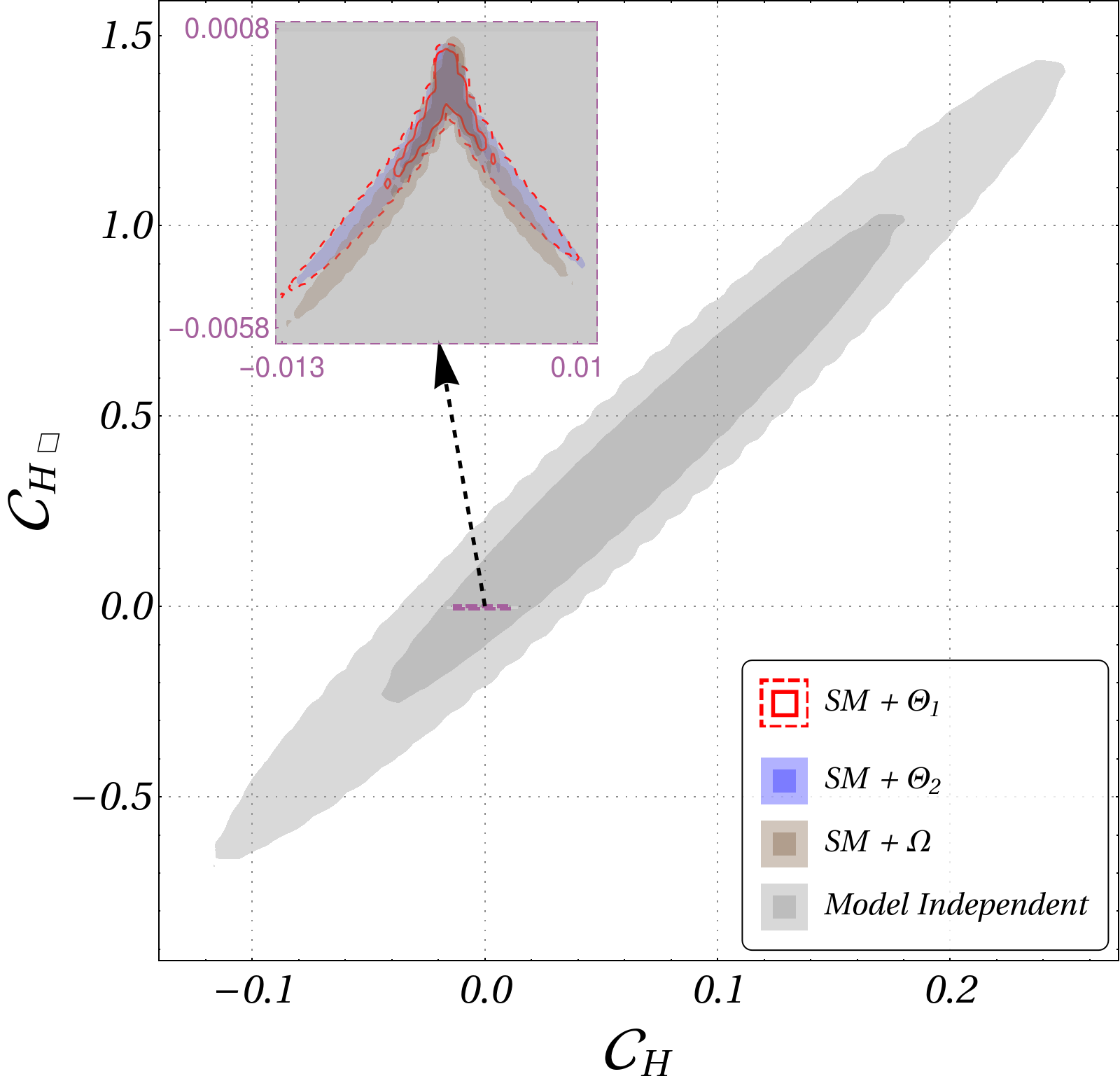}\label{fig:leptoquark_multiplet_qHqHbox}}~
	\subfloat[$\mathcal{C}_{H}$ - $\mathcal{C}_{HB}$]
	{\includegraphics[width=0.45\textwidth, height=6.7cm]{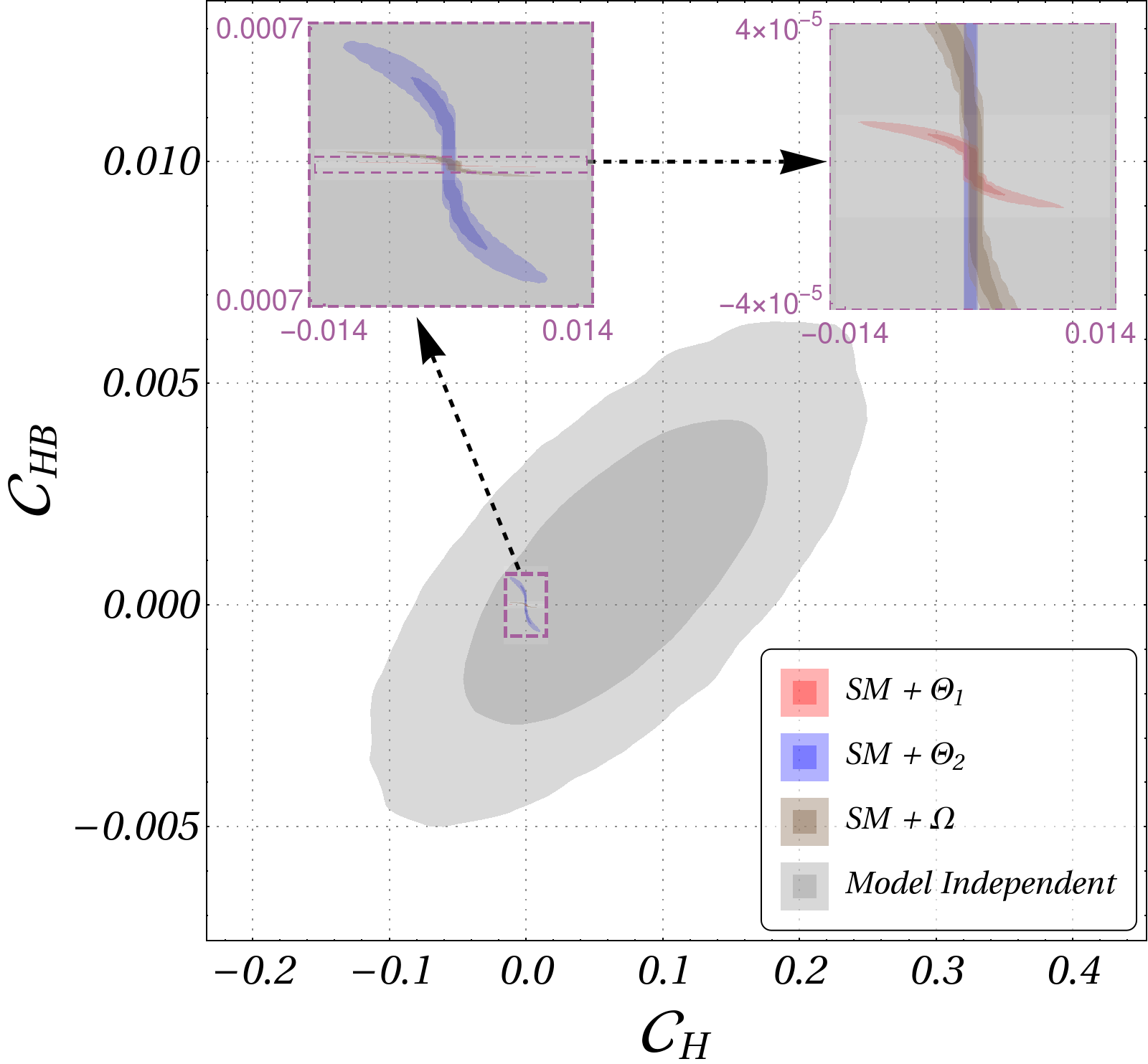}\label{fig:leptoquark_multiplet_qHqHB}}\\
	\subfloat[$\mathcal{C}_{H}$ - $\mathcal{C}_{HW}$]
	{\includegraphics[width=0.45\textwidth, height=6.7cm]{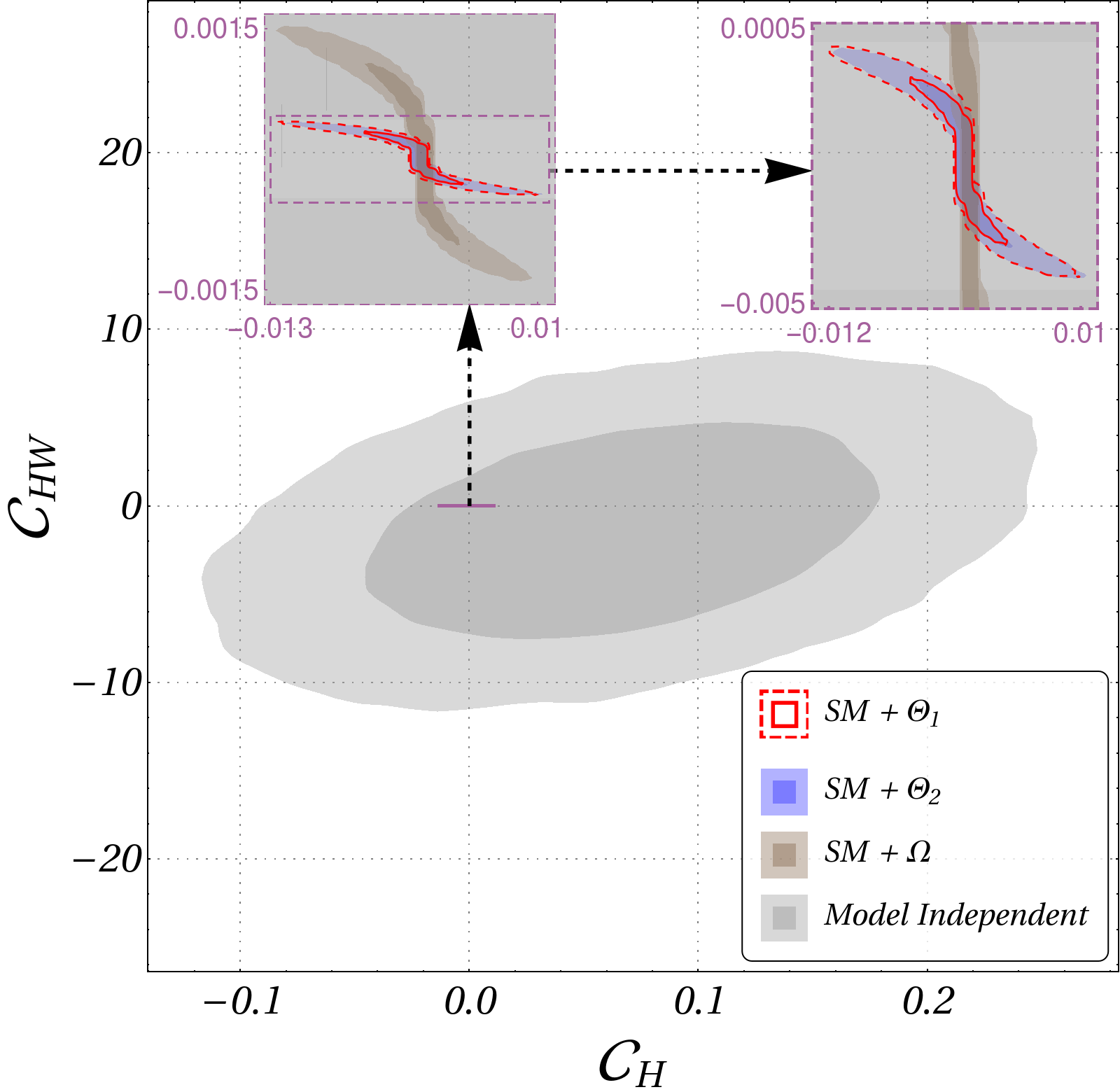}\label{fig:leptoquark_multiplet_qHqHW}}~
	\subfloat[$\mathcal{C}_{H}$ - $\mathcal{C}_{HG}$]
	{\includegraphics[width=0.45\textwidth, height=6.7cm]{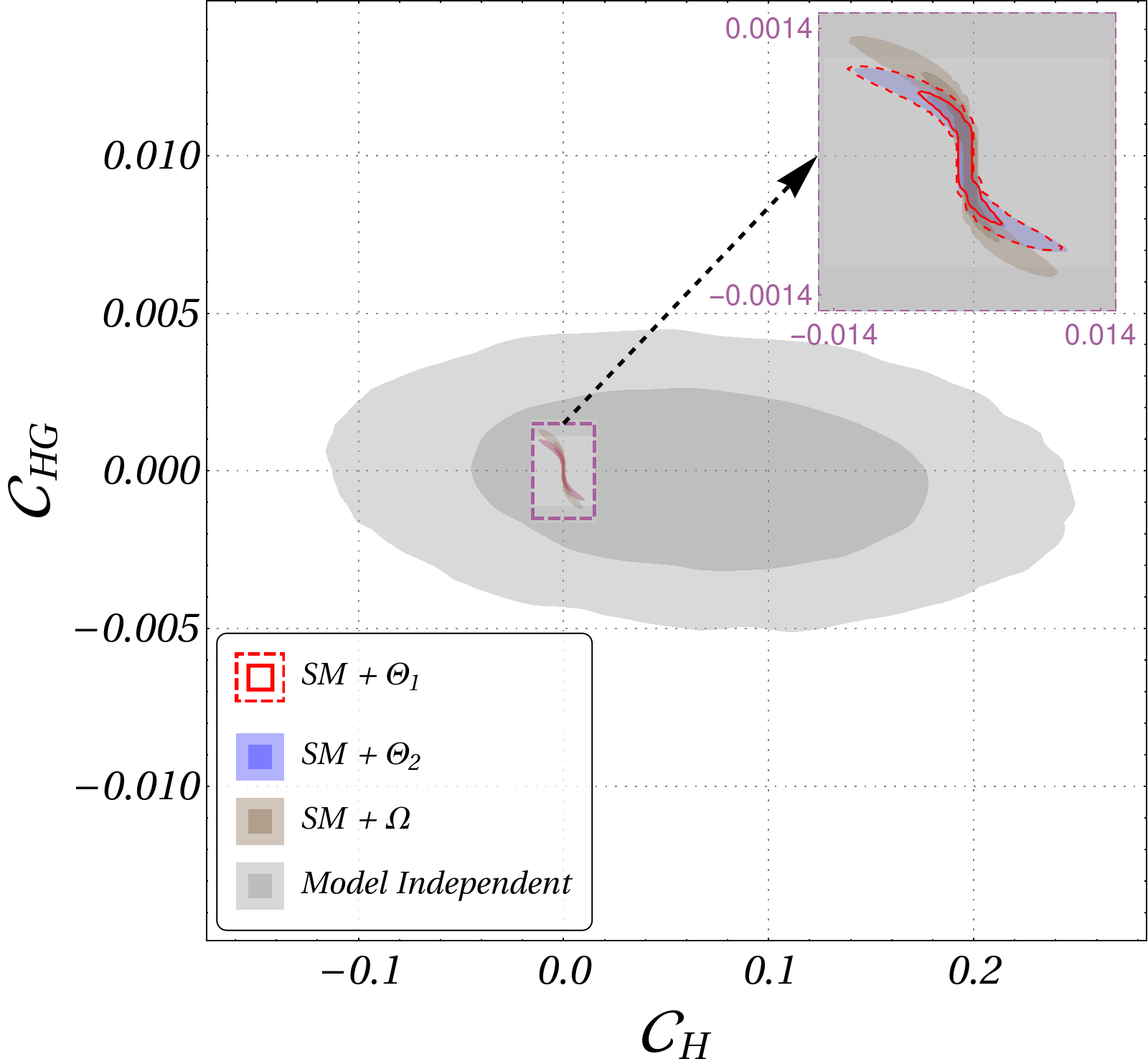}}
	\caption{\small Plots similar to Figure~\ref{fig:leptoquark_singlet_wc}, corresponding to the class of leptoquark-multiplet scenarios, namely, SM+$\Theta_{1}$, SM+$\Theta_{2}$ and SM+$\Omega$.}
	\label{fig:leptoquark_multiplet_wc}
\end{figure}
	
\begin{figure}[t]
	\centering
	\subfloat[$\mathcal{C}_{H}$ - $\mathcal{C}_{H\square}$]
	{\includegraphics[width=0.45\textwidth, height=6.7cm]{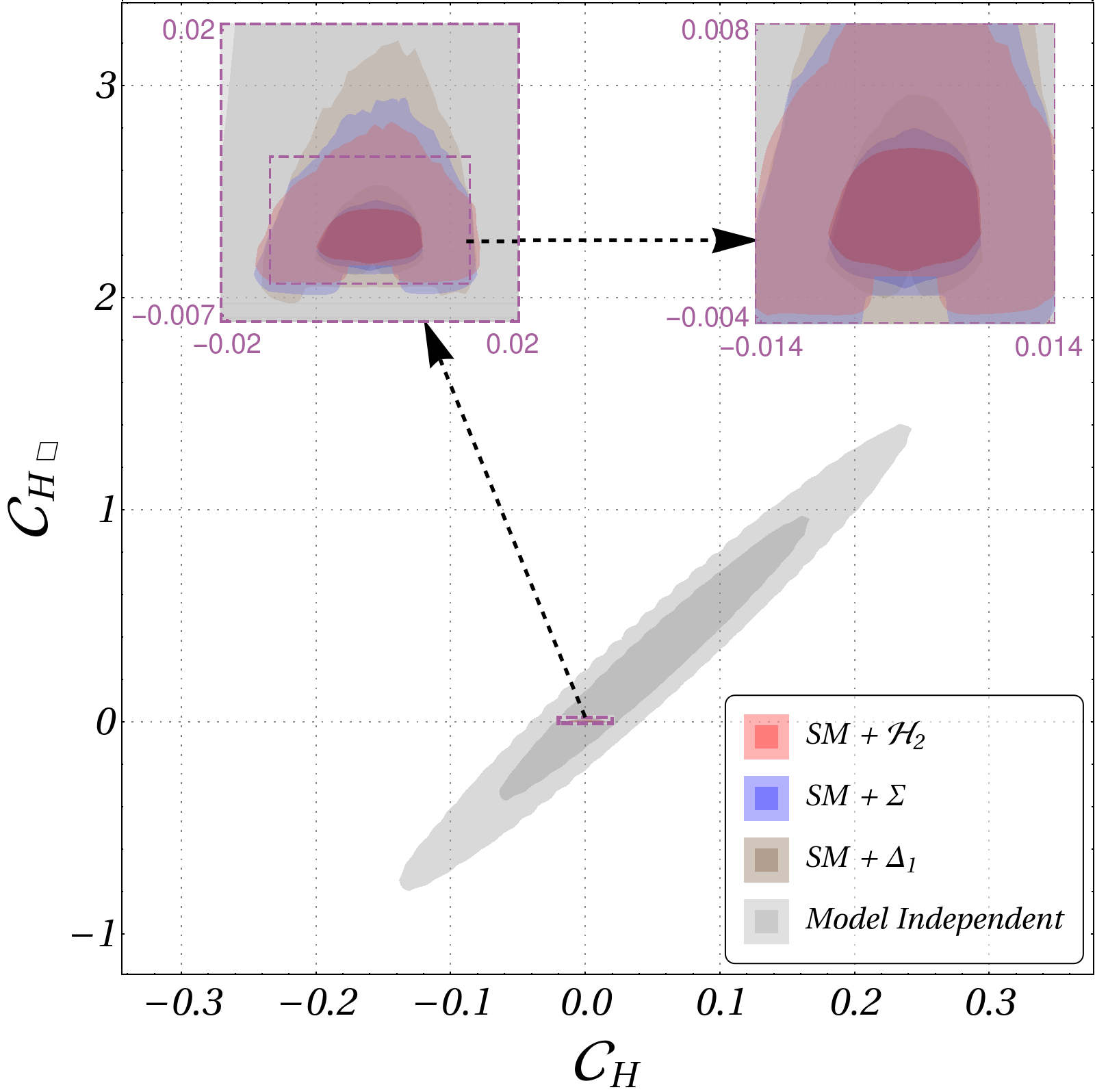}}~
	\subfloat[$\mathcal{C}_{H}$ - $\mathcal{C}_{HD}$]
	{\includegraphics[width=0.45\textwidth, height=6.7cm]{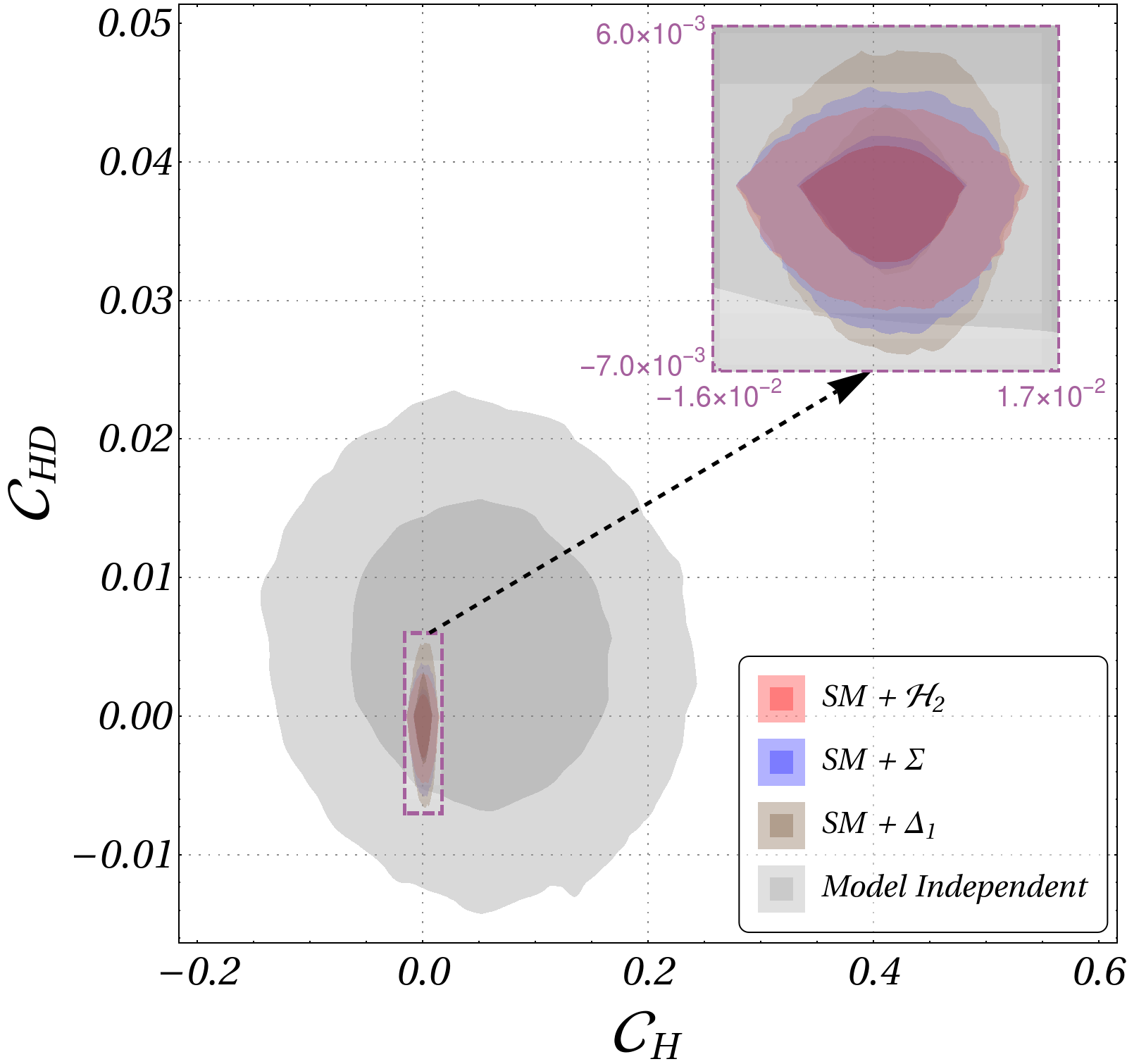}}\\
	\subfloat[$\mathcal{C}_{H}$ - $\mathcal{C}_{HB}$]
	{\includegraphics[width=0.45\textwidth, height=6.7cm]{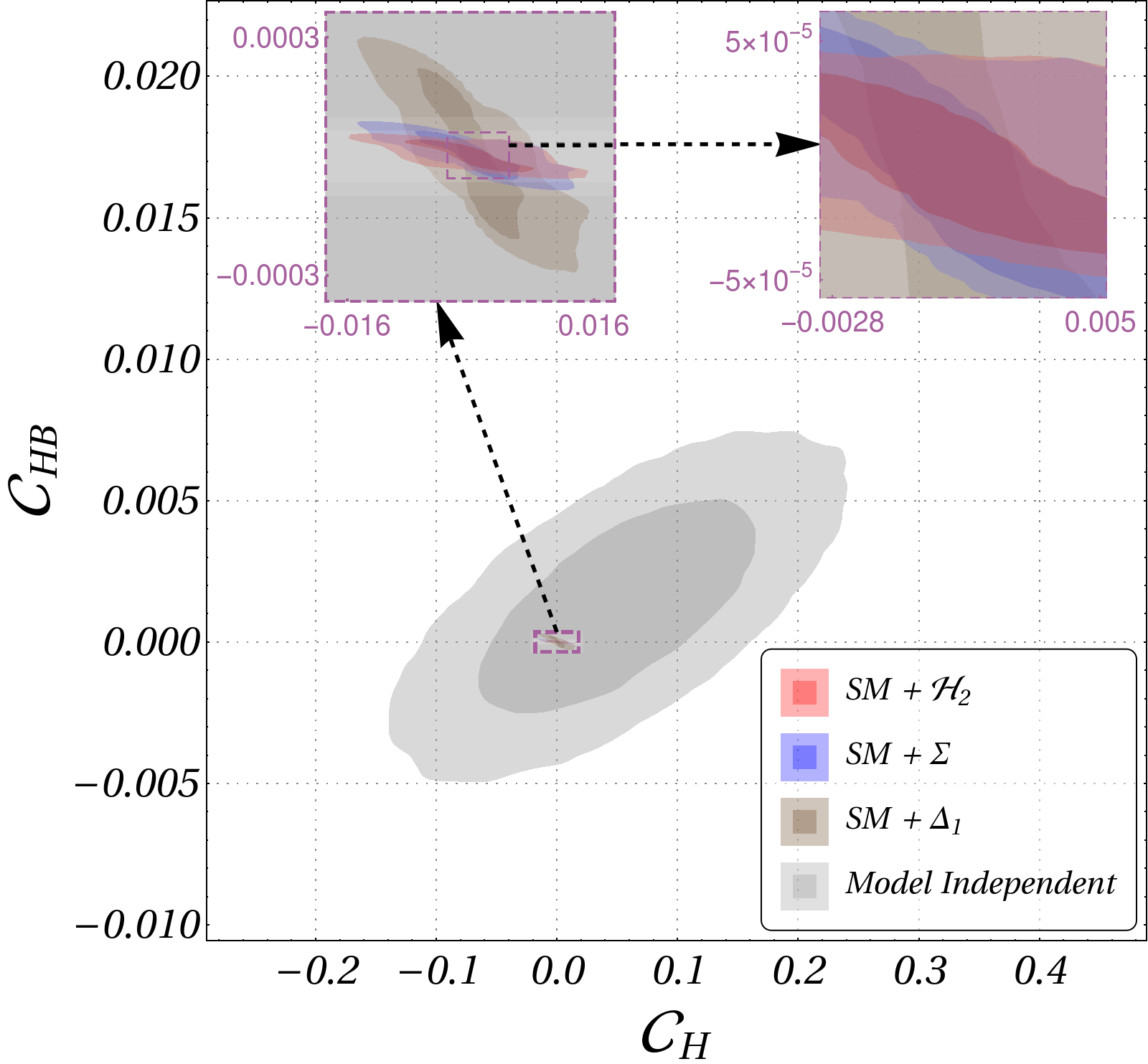}}~
	\subfloat[$\mathcal{C}_{H}$ - $\mathcal{C}_{HWB}$]
	{\includegraphics[width=0.45\textwidth, height=6.7cm]{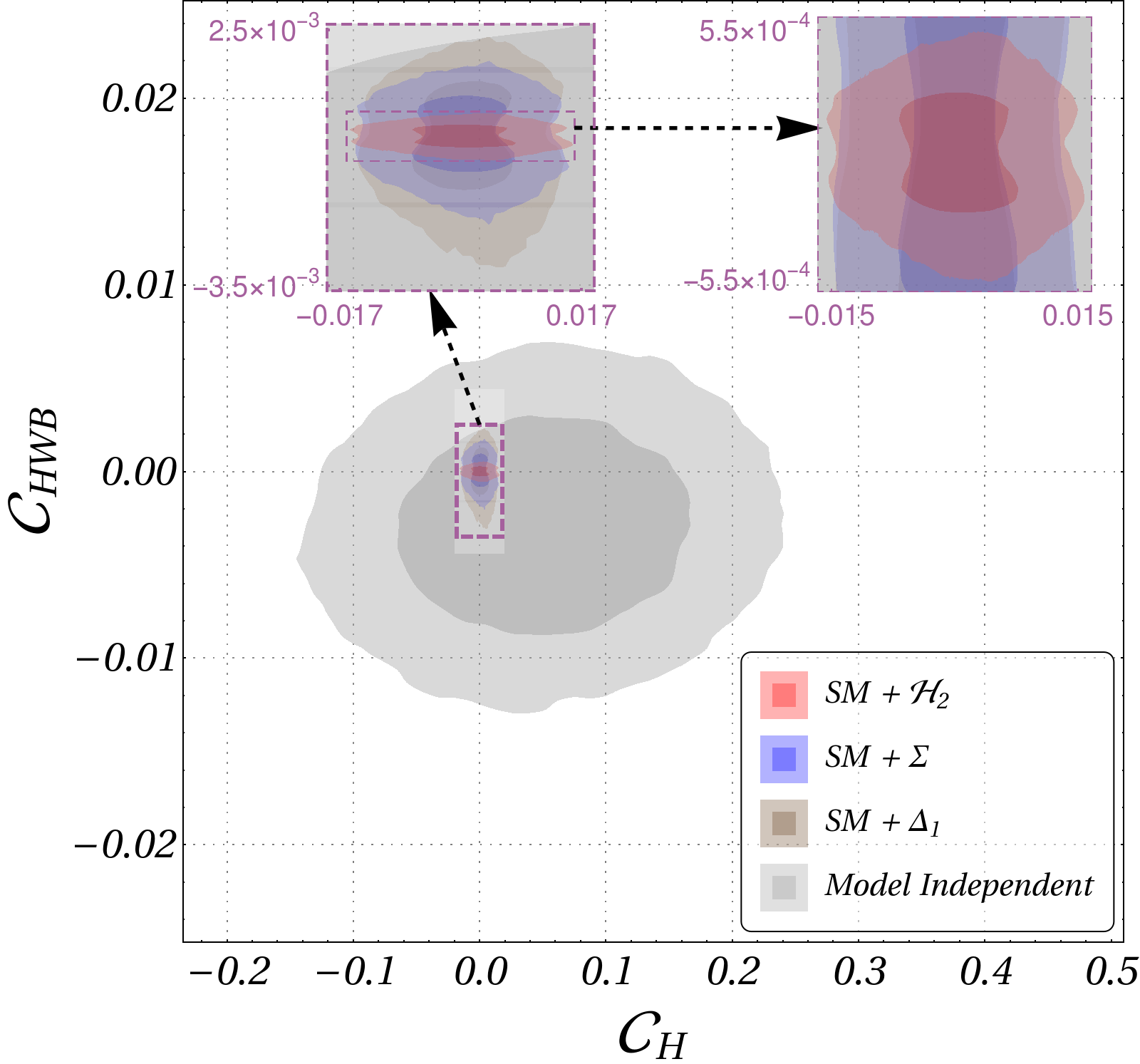}}\\
	\caption{\small Plots similar to Figure~\ref{fig:leptoquark_singlet_wc}, corresponding to the class of electroweak multiplet scalar scenarios, namely, SM+$\Delta_{1}$, SM+$\mathcal{H}_{2}$, and SM+$\Sigma$. Continued to Figure~\ref{fig:scalar_multiplet_wc2}.}
	\label{fig:scalar_multiplet_wc1}
\end{figure}

\section{Model-Independent vs Model-Dependent Analyses}\label{sec:comparison}
\begin{figure}[t]
	\centering
	\subfloat[$\mathcal{C}_{H}$ - $\mathcal{C}_{HW}$]
	{\includegraphics[width=0.45\textwidth, height=6.7cm]{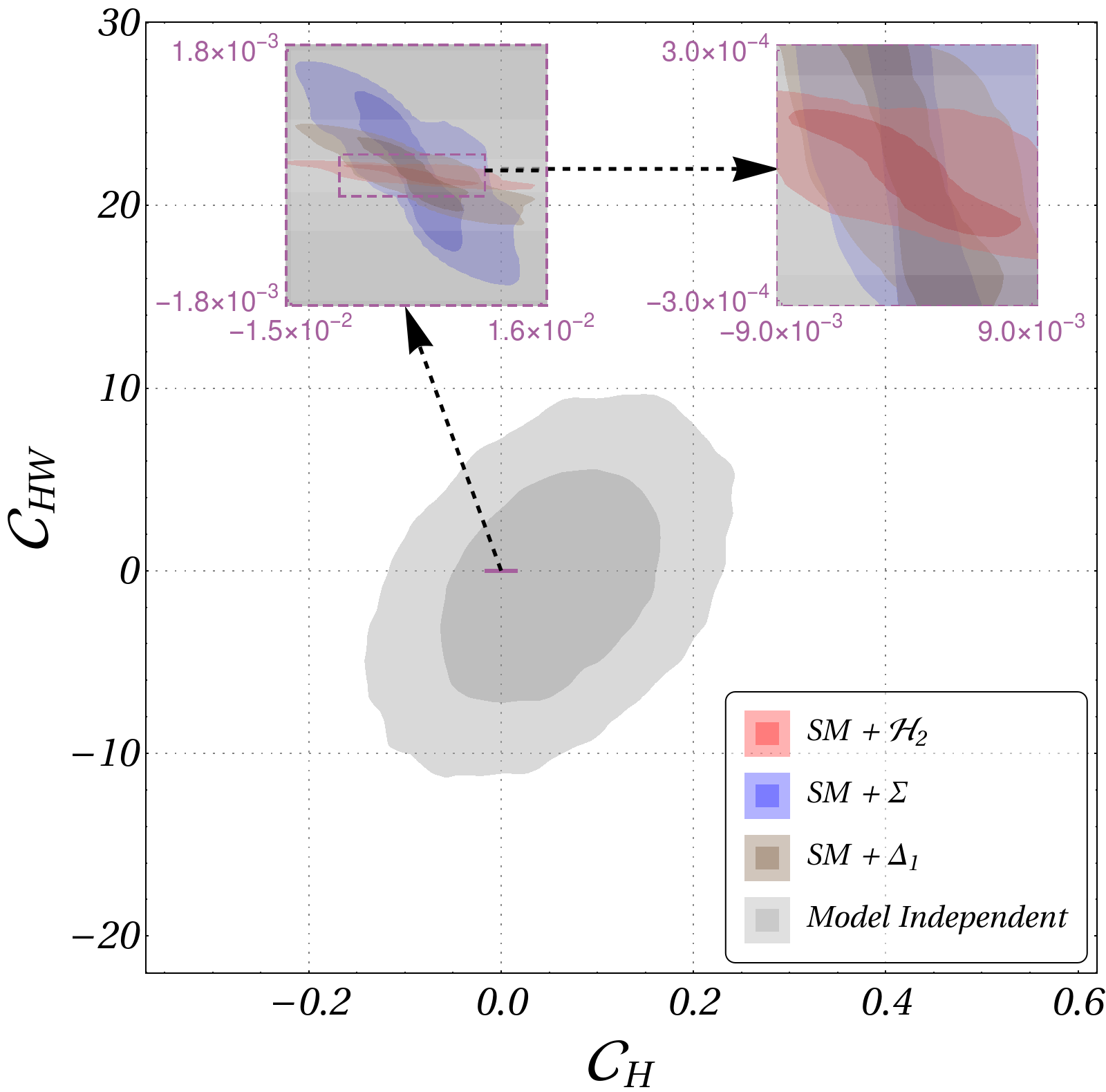}}~
	\subfloat[$\mathcal{C}_{H}$ - $\mathcal{C}_{eH}$]
	{\includegraphics[width=0.45\textwidth, height=6.7cm]{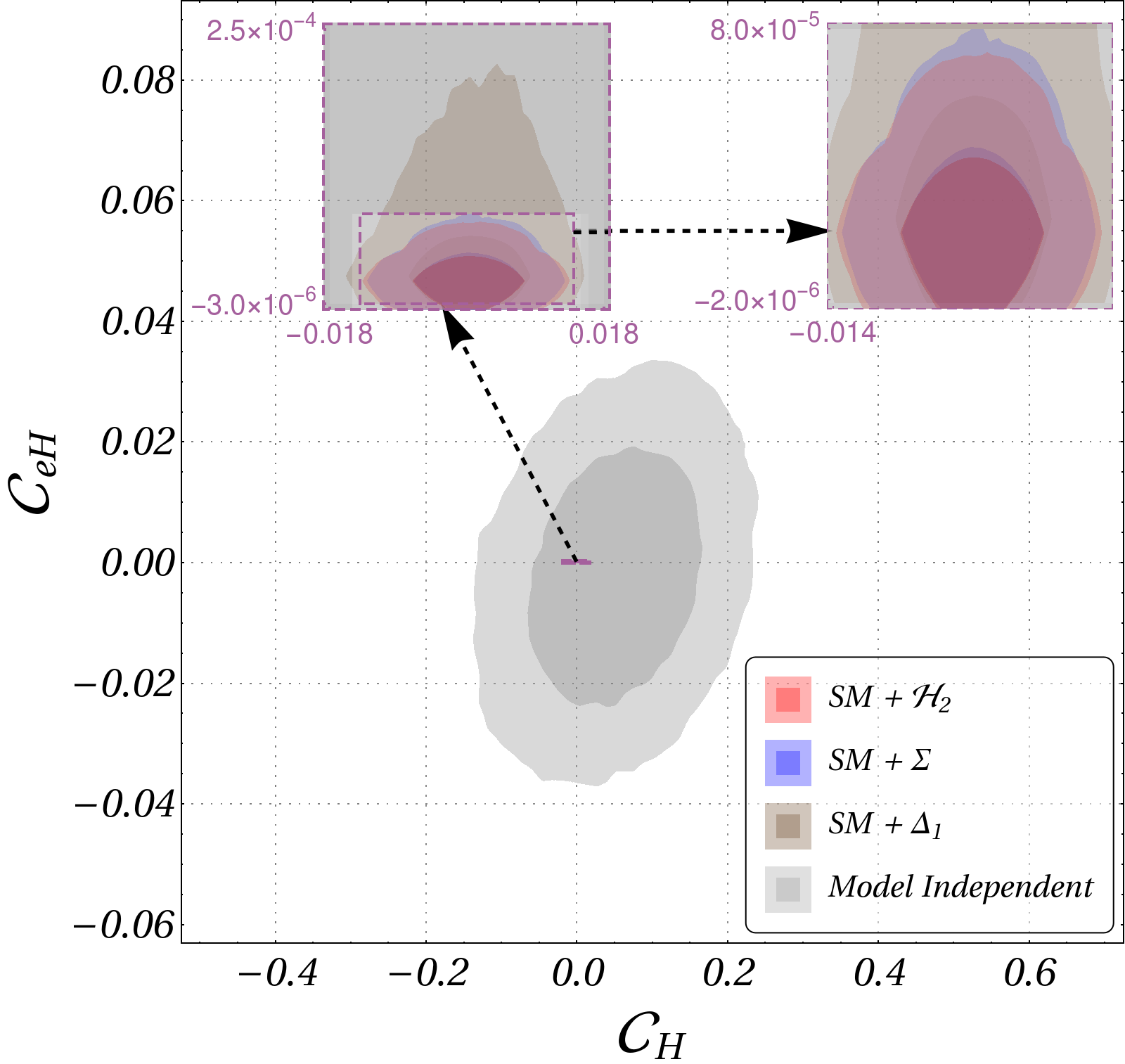}}\\
	\subfloat[$\mathcal{C}_{H}$ - $\mathcal{C}_{uH}$]
	{\includegraphics[width=0.45\textwidth, height=6.7cm]{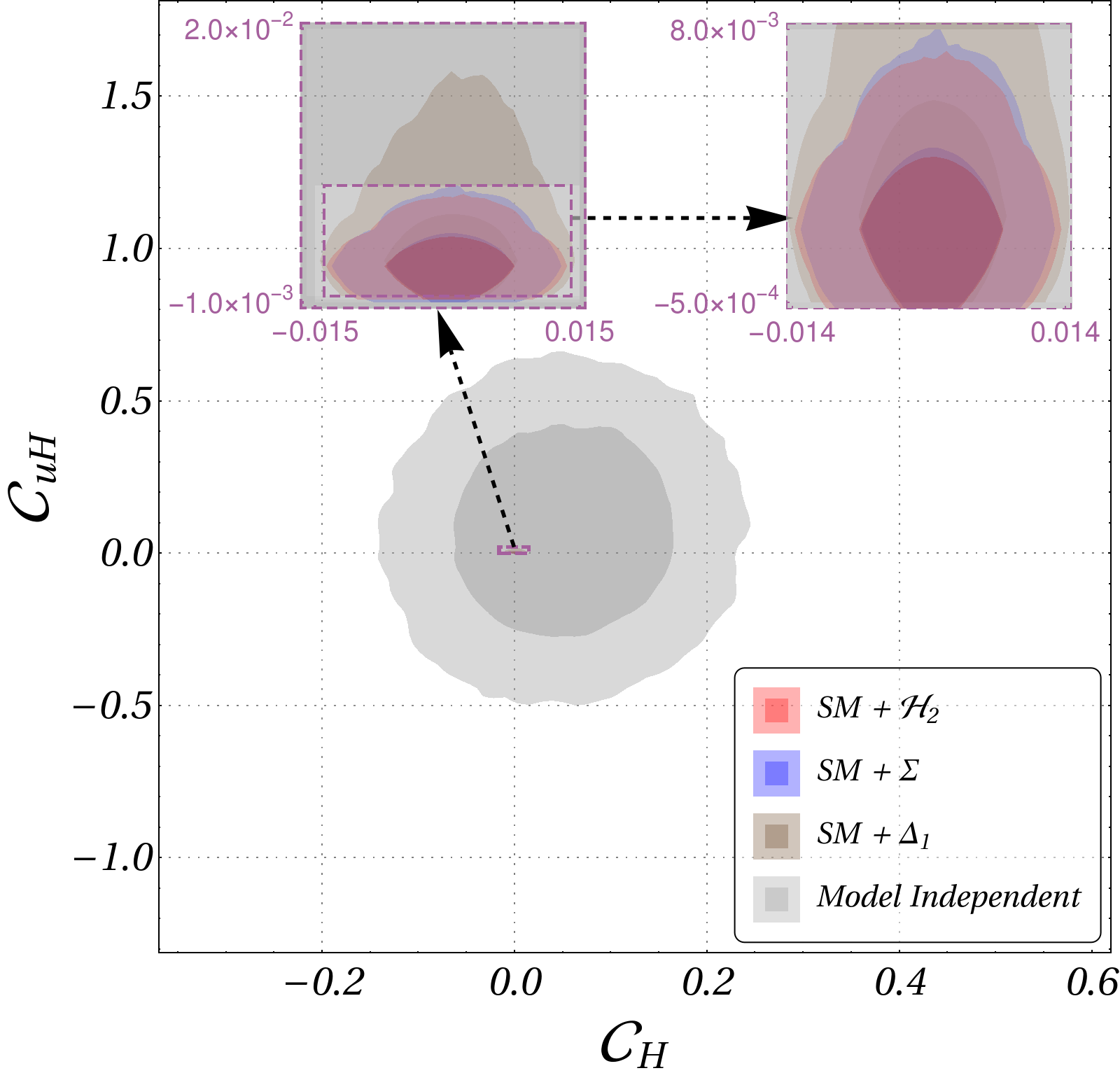}}~
	\subfloat[$\mathcal{C}_{H}$ - $\mathcal{C}_{dH}$]
	{\includegraphics[width=0.45\textwidth, height=6.7cm]{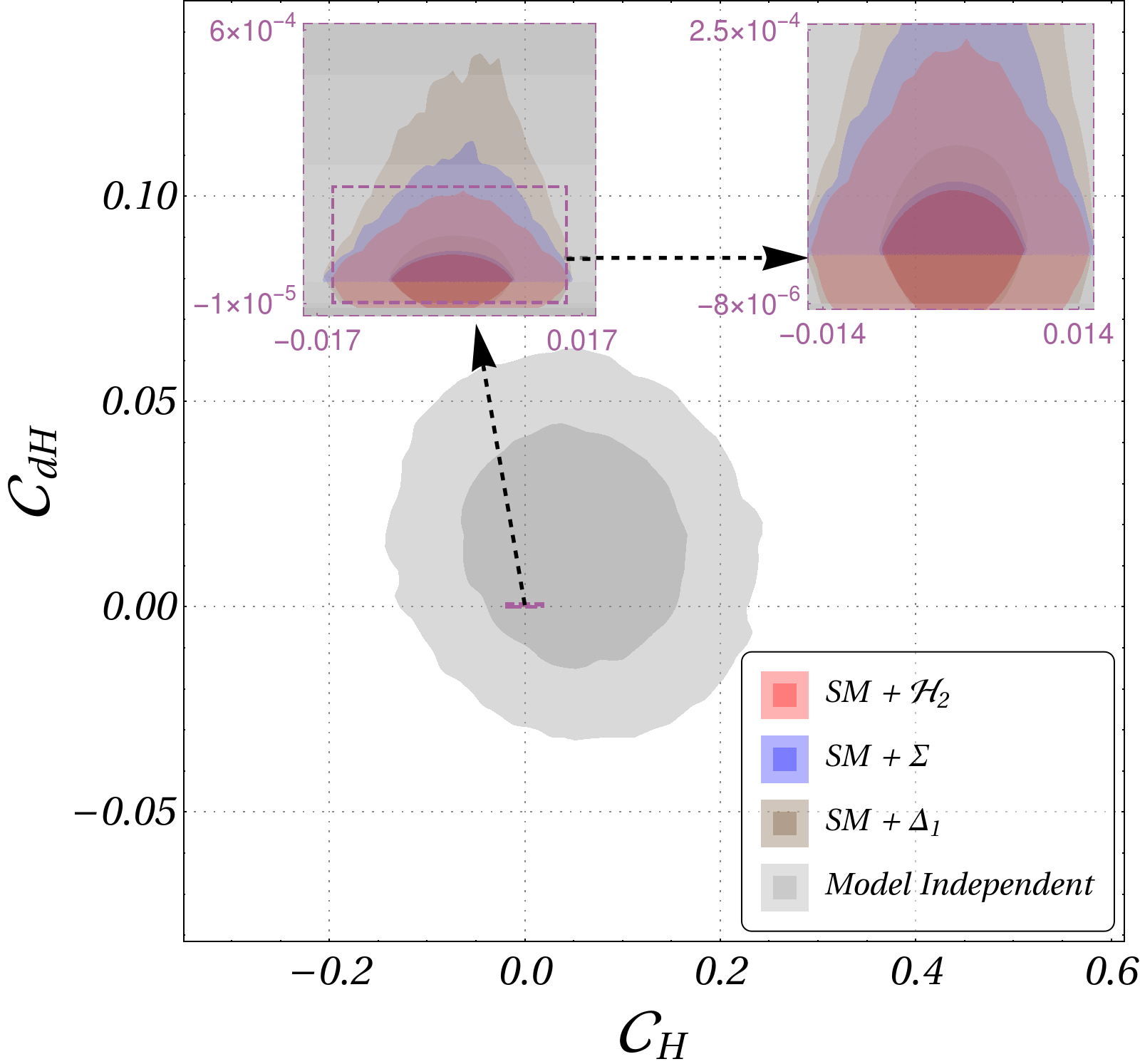}}
	\caption{Continued from figure \ref{fig:scalar_multiplet_wc1}.}
	\label{fig:scalar_multiplet_wc2}
\end{figure}

\begin{figure}[t]
	\centering
	\subfloat[$\mathcal{C}_{H}$ - $\mathcal{C}_{H\square}$]
	{\includegraphics[width=0.32\textwidth, height=5cm]{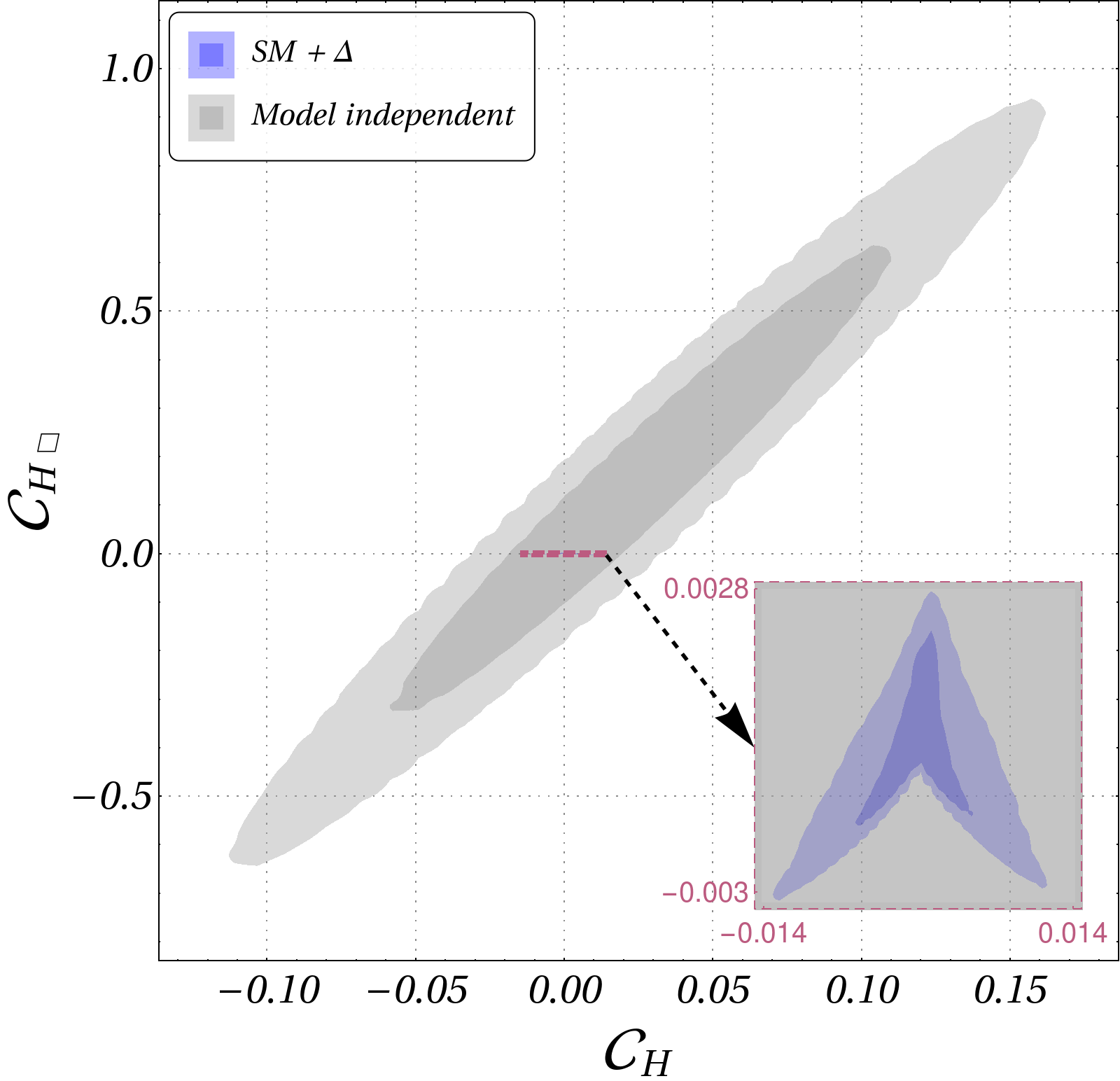}}~
	\subfloat[$\mathcal{C}_{H}$ - $\mathcal{C}_{HD}$]
	{\includegraphics[width=0.32\textwidth, height=5cm]{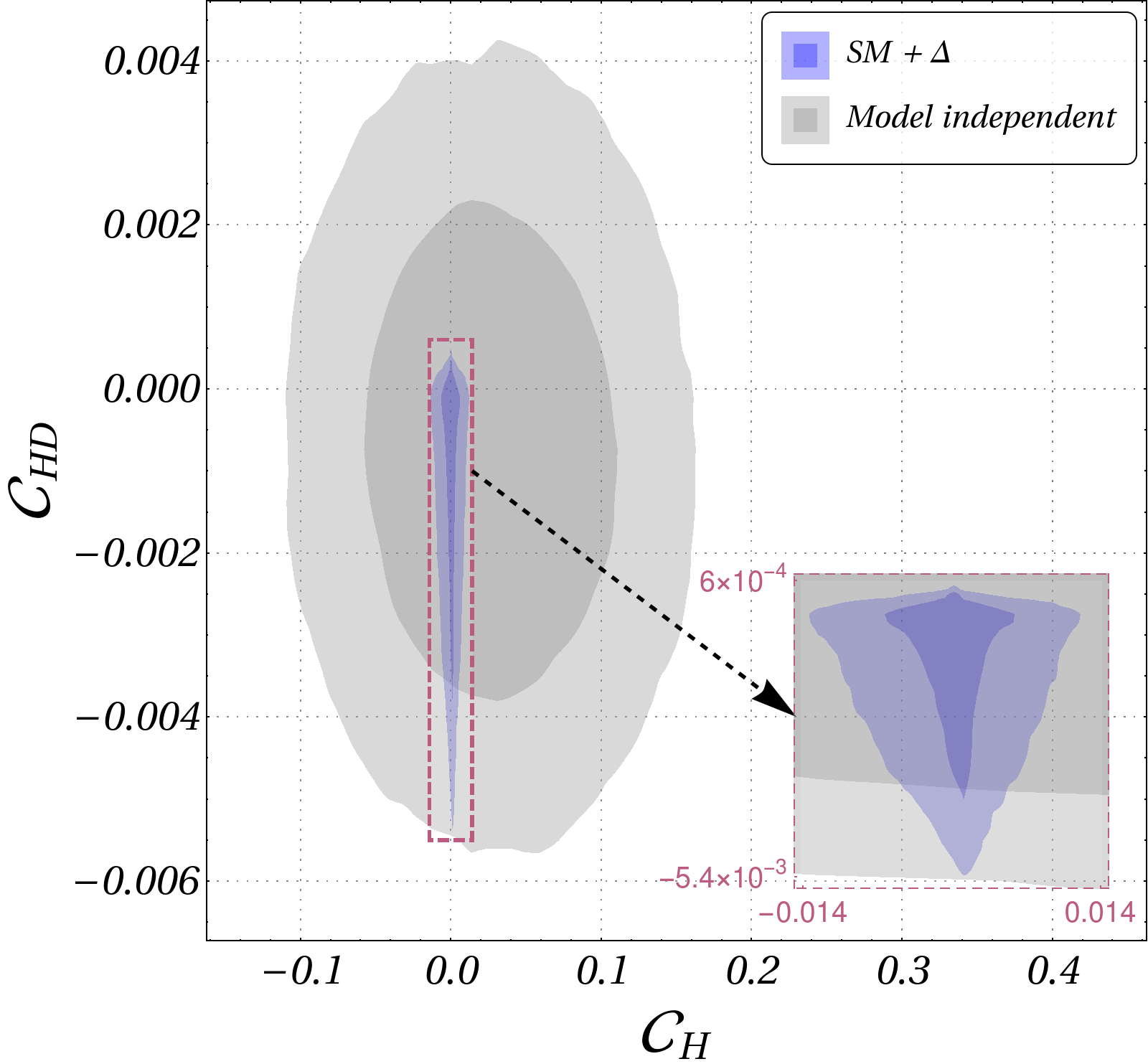}}~
	\subfloat[$\mathcal{C}_{H}$ - $\mathcal{C}_{HW}$]
	{\includegraphics[width=0.32\textwidth, height=5cm]{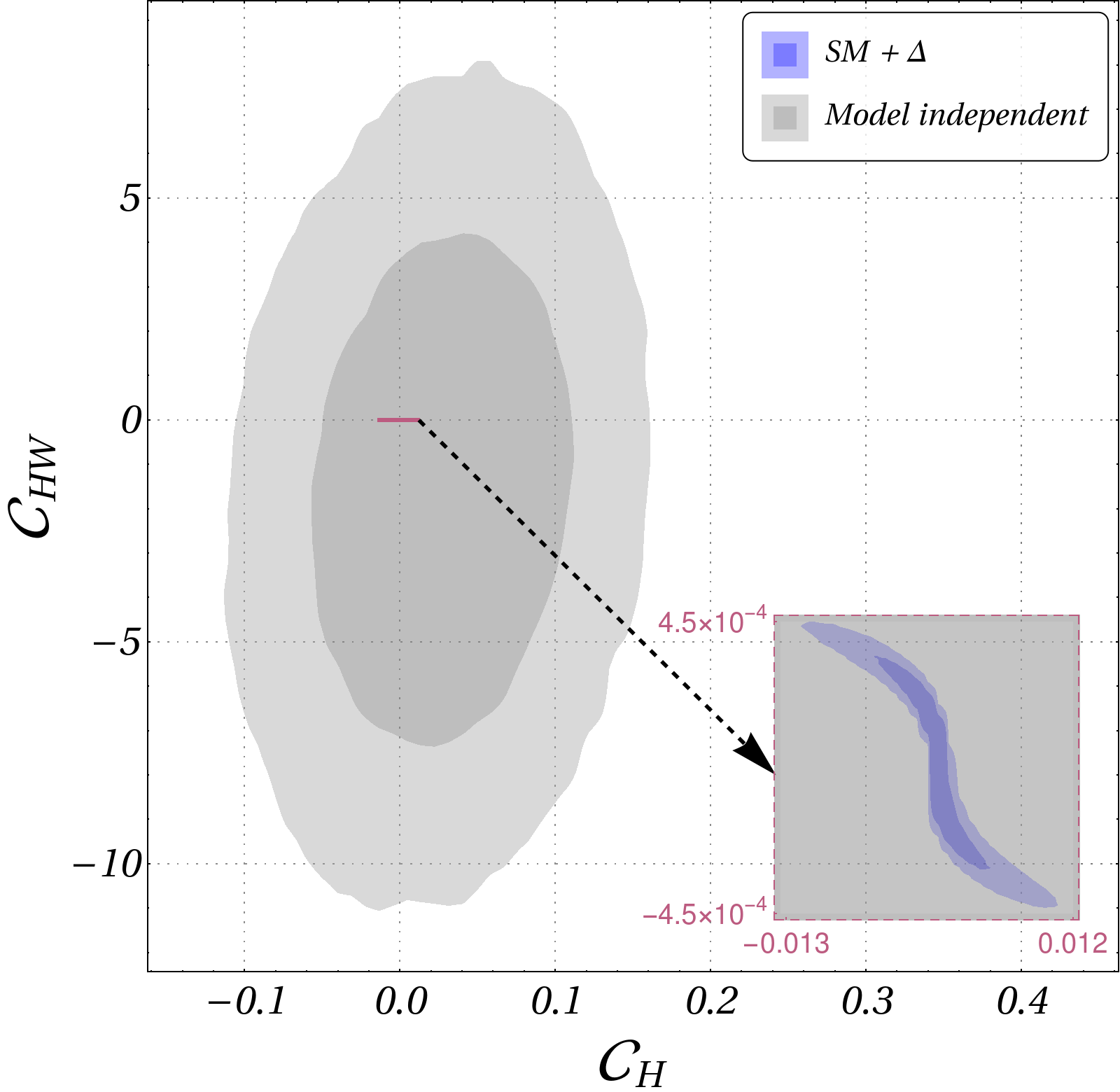}}\\
	\subfloat[$\mathcal{C}_{H}$ - $\mathcal{C}_{eH}$]
	{\includegraphics[width=0.32\textwidth, height=5cm]{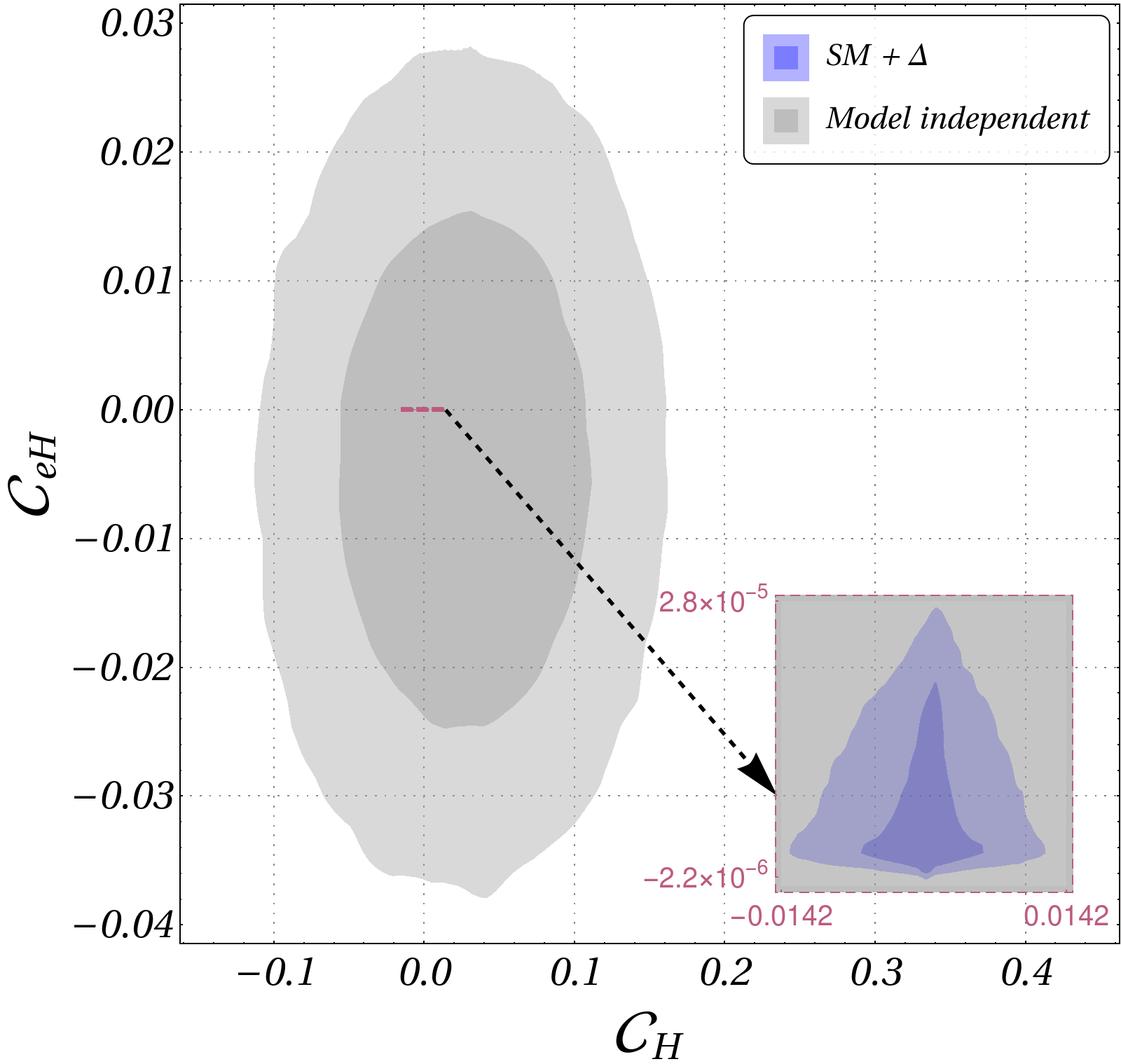}}~
	\subfloat[$\mathcal{C}_{H}$ - $\mathcal{C}_{uH}$]
	{\includegraphics[width=0.32\textwidth, height=5cm]{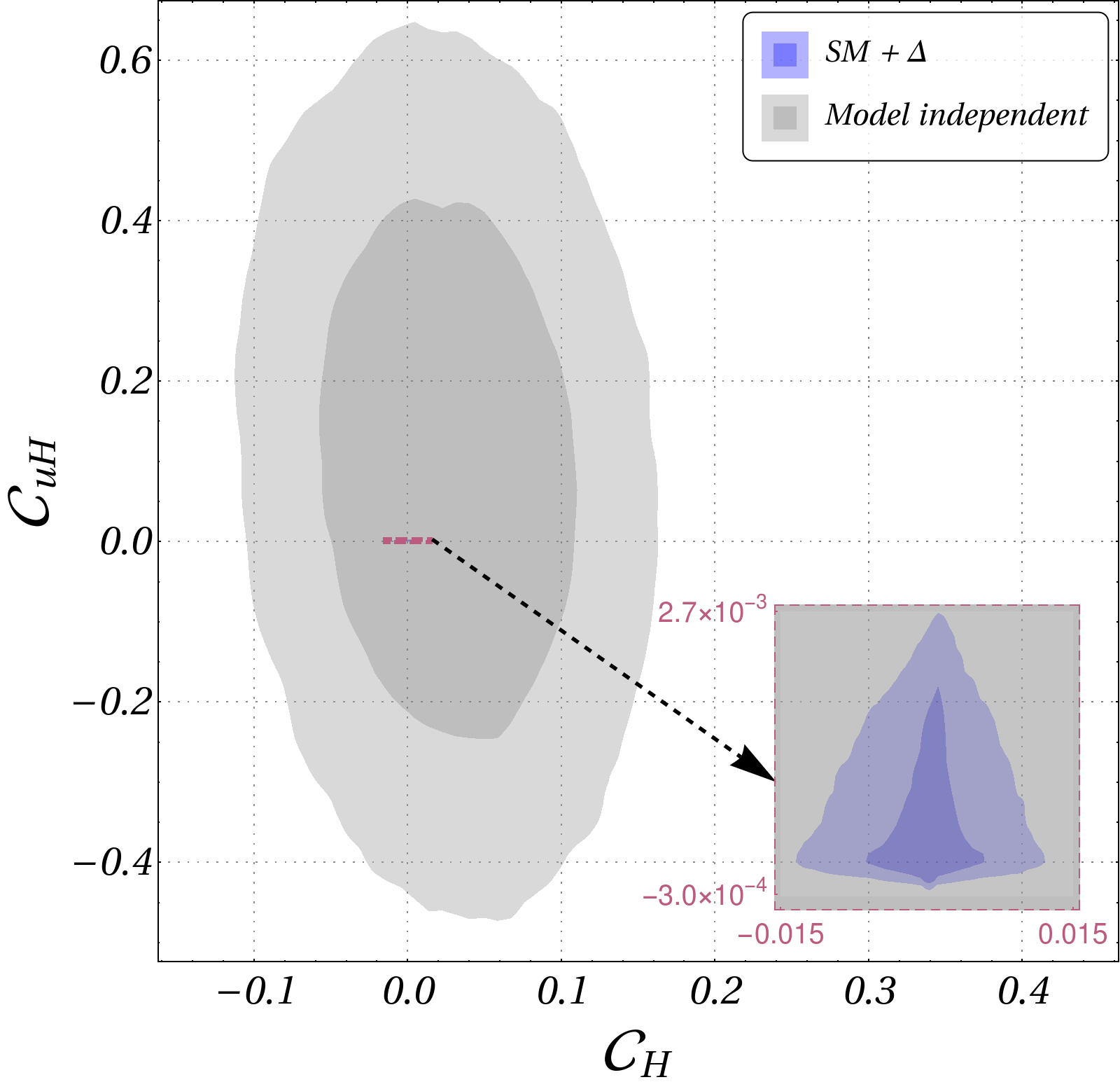}}~
	\subfloat[$\mathcal{C}_{H}$ - $\mathcal{C}_{dH}$]
	{\includegraphics[width=0.32\textwidth, height=5cm]{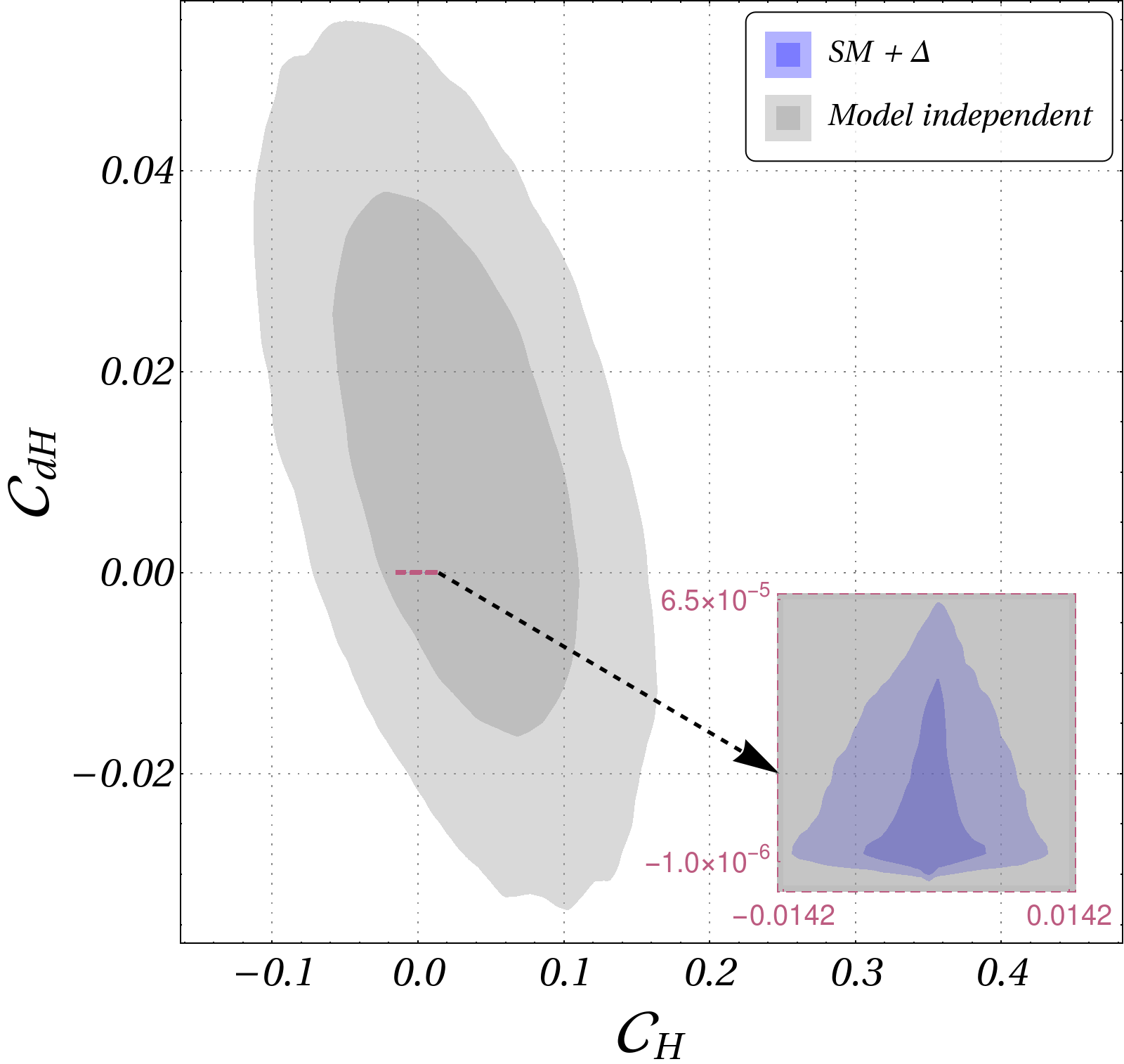}}
	\caption{\small Comparison of the two-dimensional posteriors of model-independent WCs (gray; 68\% (darker) and 95\% (lighter) credible intervals) with those generated from the class of isospin triplet real scalar (SM+$\Delta$). The blue region corresponds to the similar region obtained from the SM+$\Delta$. Zoomed-in spaces are shown inset.} 
	\label{fig:scalar_isospin_triplet_wc}
\end{figure}

	Comparing the span of the WCs for the fit with all effective operators present and fits with one WC at a time (second and third columns of table~\ref{table:wcfitval}) show us how the allowed space of each WC increases in presence of other independent WCs. This non-negligible variation of the WC-space is the reason for our including the 10 WCs fit result in the fourth column of the said table. This enunciates the fact that for a meaningful comparison of parameter-spaces between model-independent and model-dependent studies, we need to consider only those subsets of the WCs which come from some BSM model considered here. As can be seen from table~\ref{table:models}, some of the BSM scenarios considered in this analysis give rise to the same set of effective operators. This observation motivates us to club these 11 scenarios into 6 different classes of BSM theories. 
	
	Our thumb rule is: if two or more BSM scenarios lead to the exact same set of effective operators, they are declared degenerate and are clubbed in a single class. It is important to remark that though the degenerate models possess exactly the same operators, their respective WCs, which depend on model parameters, are entirely different, reflecting the intrinsic non-identical nature of those BSMs. This strategy of clubbing multiple models together in a class, using the power of EFT, enables us to bring down some of the BSMs in the same footing to be adjudged simultaneously, and thus to allow further comparative remarks. This methodology is  illustrated through the flowchart in Figure~\ref{fig:flowchart}.
	
	For each of these 6 classes, we first want to constrain the WCs from the data in a model-independent manner. To this end, we obtain the Bayesian inference on these classes of WCs without any model information and by varying them as free and independent parameters in each case. The operators that are not generated in a given model are taken to be zero. These results, from now on, will be considered to be the `model-independent' results. The remaining columns of table~\ref{table:wcfitval} list the best-fit values and uncertainties for each class. As can be seen from the second row (from top) in that table, the last three columns are results for classes with WCs, which can be connected to multiple BSM scenarios. The results clearly show the necessity of considering these individual classes, as the parameter-space of each WC varies largely from both the fit with all WCs present and the fit with only that WC, as well as fits for other classes. As an example, one can follow the row of $\mathcal{C}_{H\square }$ in table~\ref{table:wcfitval} and see how even the `model-independent' results vary.
	
\begin{figure}[t]
	\centering
	\subfloat[$\mathcal{C}_{H}$ - $\mathcal{C}_{H\square}$]
	{\includegraphics[width=0.32\textwidth, height=5cm]{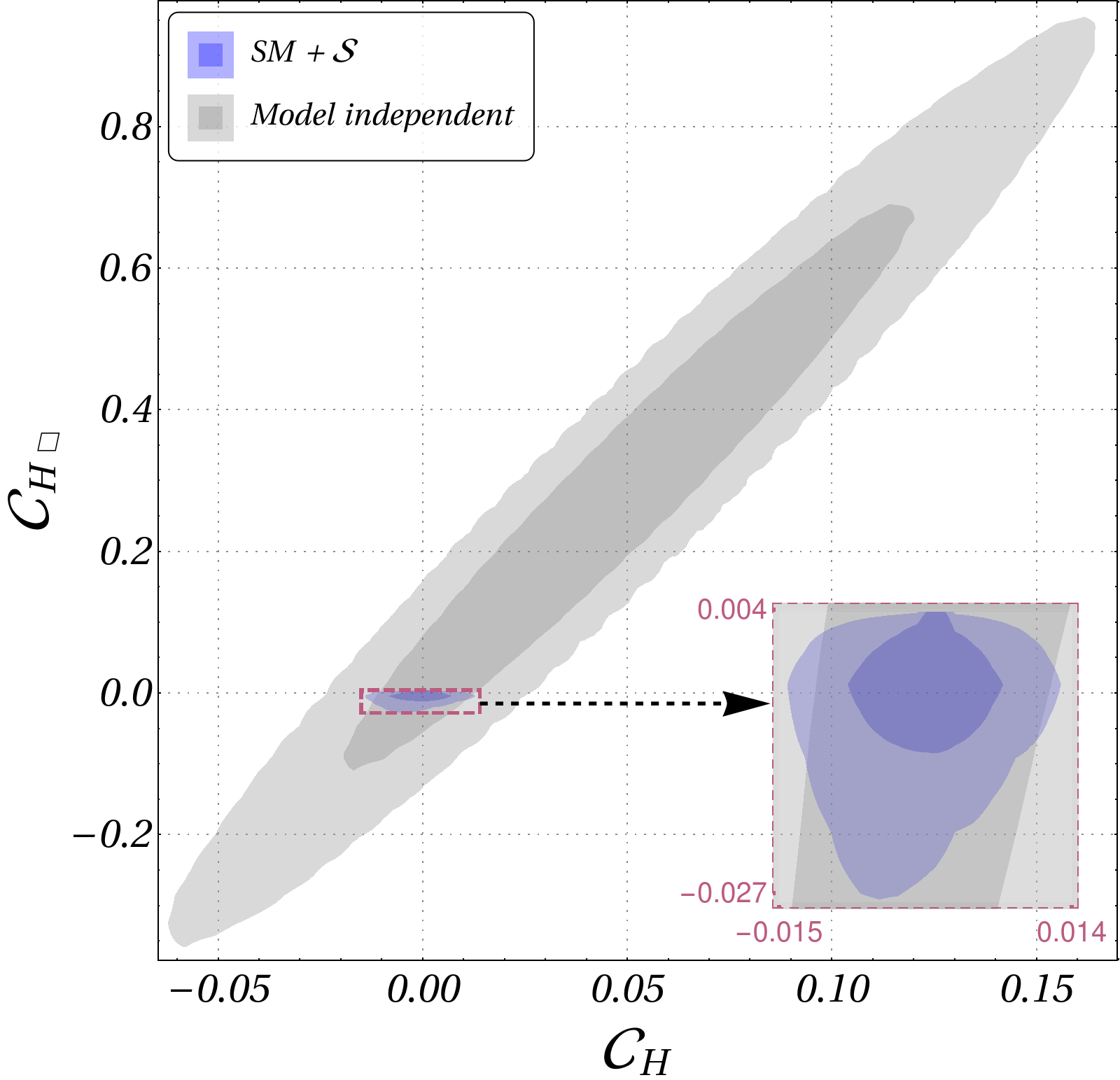}}
	\subfloat[$\mathcal{C}_{H}$ - $\mathcal{C}_{H\square}$]
	{\includegraphics[width=0.32\textwidth, height=5cm]{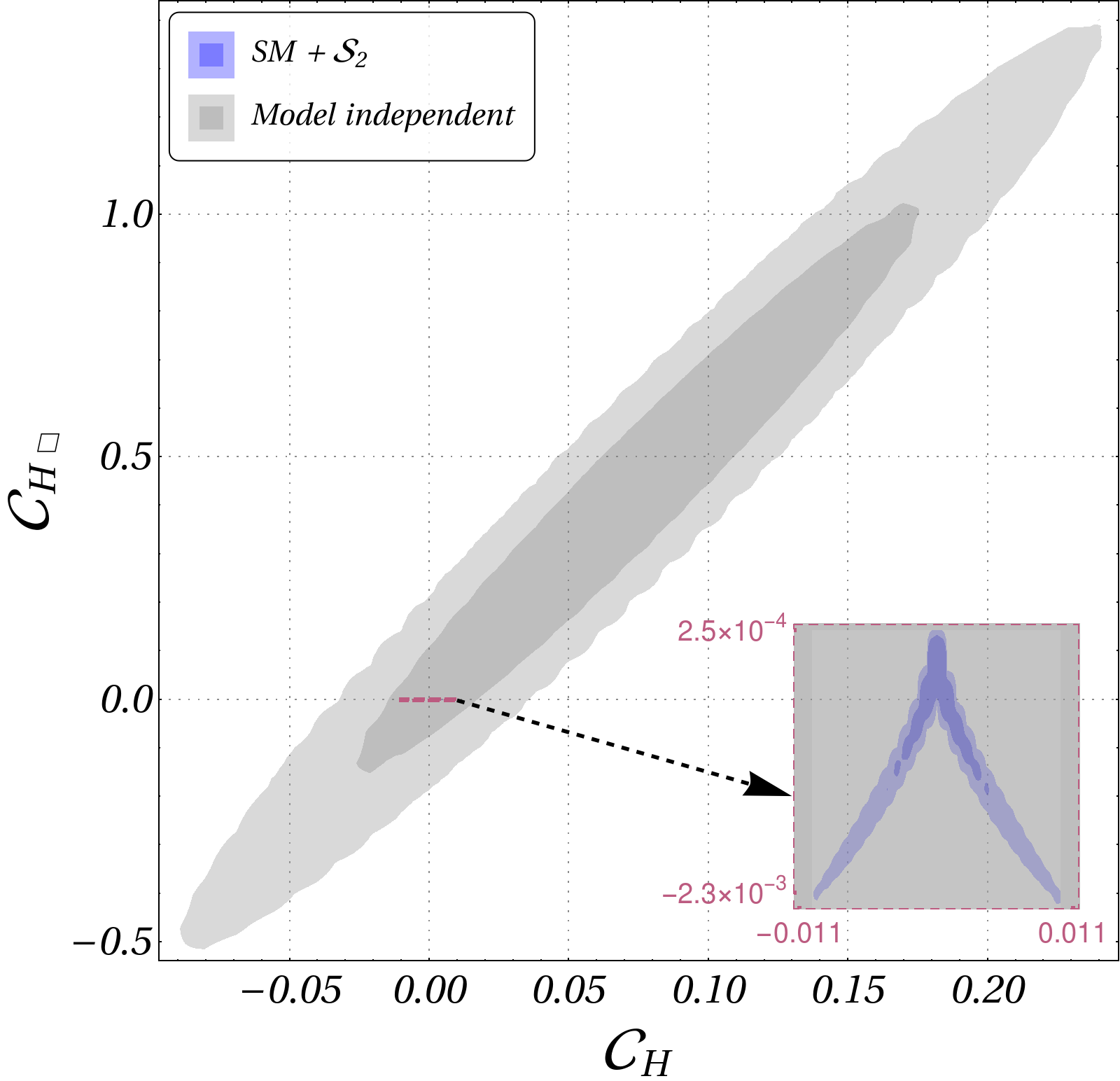}}
	\subfloat[$\mathcal{C}_{H}$ - $\mathcal{C}_{HB}$]
	{\includegraphics[width=0.32\textwidth, height=5cm]{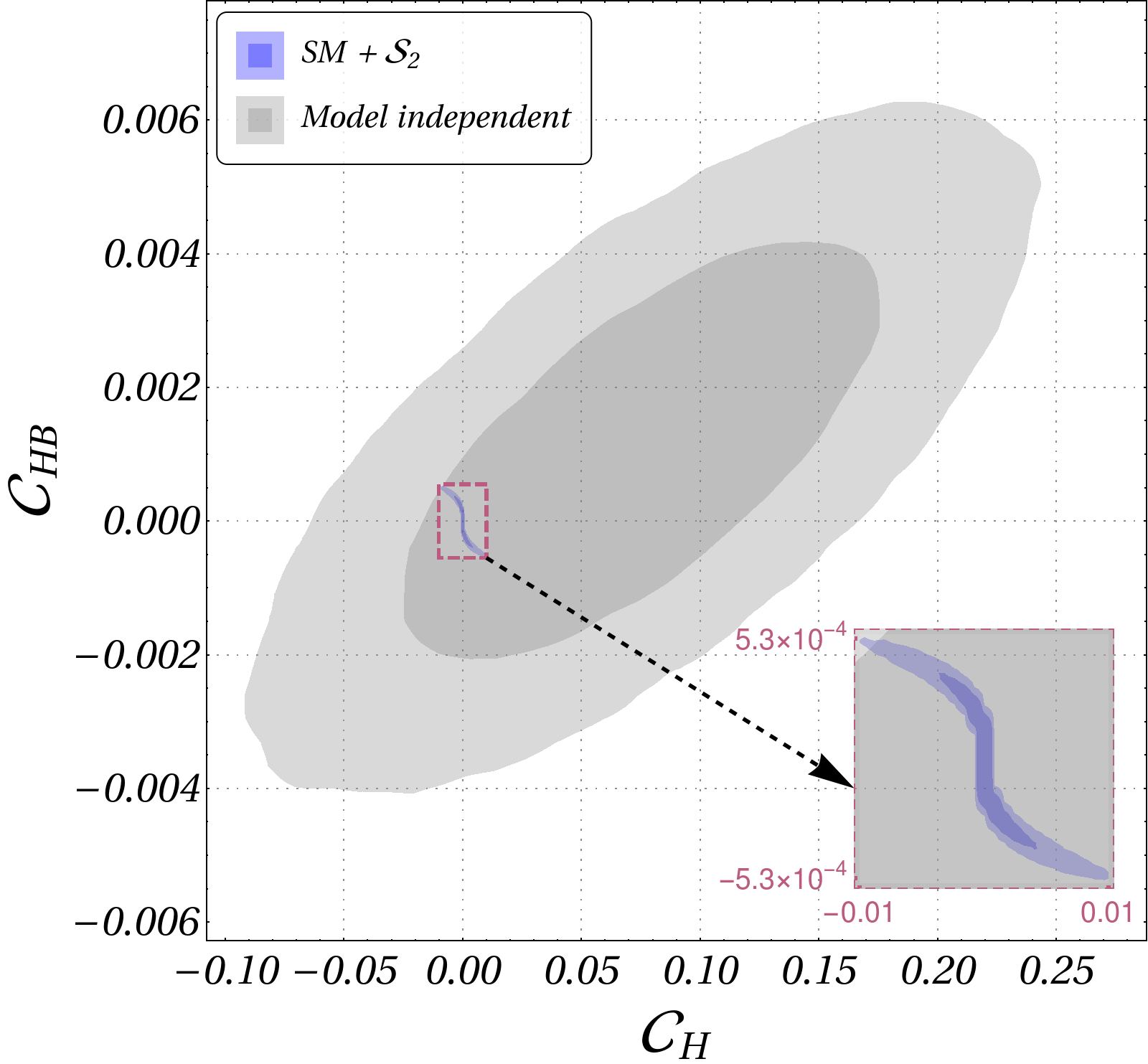}}
	\caption{\small Plots similar to Figure~\ref{fig:scalar_isospin_triplet_wc}, corresponding to the classes of SM+$\mathcal{S}$ and SM+$\mathcal{S}_{2}$ with two and three mapped effective operators respectively.}
	\label{fig:scalar_singlet_wc}
\end{figure}
	
	In the next step,using the data-driven BSM parameter posteriors of each model of a given class (Section~\ref{sec:moddep_stat}) and the matching results relating the WCs with the BSM parameters obtained using CoDEx (tables~\ref{tab:class1WCs}~-~\ref{tab:complexsingletWCs}), we proceed to constrain the same set of WCs again. In this case, we expect to obtain a different result for each BSM model, as the WCs are functions of BSM parameters and thus related to each other. We use the large samples generated in the MCMC processes for each model-fit to obtain the multi-variate distributions of corresponding WCs. Where the model-independent results for each class do not contain any model information, the WC-distributions generated in this way are naturally highly constrained by the structure of the specific BSM models. Comparing the WC-distributions obtained in these two ways is effectively one definitive way of comparing the model-independent and dependent results.

	To visualize these distributions, we use marginal posteriors for the WC-distributions, taken two at a time. For a specific class, the model-independent two-dimensional marginal posterior is the same, while those generated directly from several BSM scenarios from the same class are different. Whenever possible, we compare them together.

	Figure~\ref{fig:leptoquark_singlet_wc} showcases these results for the class of  models SM+$\varphi_{1}$, and SM+$\varphi_{2}$ with four associated effective SMEFT operators. The large gray regions correspond to the model-independent 68\% and 95\% credible intervals (darker to lighter) of corresponding WCs. Similarly, the red-dashed-bounded and blue regions correspond to the WC-regions obtained from models SM+$\varphi_{1}$ and SM+$\varphi_{2}$ respectively. As the latter ones occupy comparatively tiny regions in the main figure, we show their blown-up versions inset. As we are mainly interested in the allowed parameter-spaces for individual WCs from model-dependent or independent analyses, we show only the smallest subset of possible two-dimensional marginal distributions here onward, containing all the WC-regions. These, among all other possible figures not shown here, are organized in the GitHub repository associated with this work \cite{Githubeffex}.

	Similar figures are obtained for the rest of the classes with multiple models. WC-spaces for  models  SM+$\Theta_{1}$, SM+$\Theta_{2}$, and SM+$\Omega$ with five effective SMEFT operators are shown in Figure~\ref{fig:leptoquark_multiplet_wc}, whereas Figures~\ref{fig:scalar_multiplet_wc1} and \ref{fig:scalar_multiplet_wc2} show those regions for the class of models involving $SU(2)_{L}$ doublet, triplet, and quartet scalars with nine mapped SMEFT operators.

	The WC-spaces for classes with a single model in each, namely SM+$\Delta$, SM+$\mathcal{S}$, and SM+$\mathcal{S}_{2}$ are shown in Figures~\ref{fig:scalar_isospin_triplet_wc} and \ref{fig:scalar_singlet_wc} respectively. There is a single connecting theme throughout all these figures: the minuscule size of the model-constrained WC-regions, compared to their model-independent counterparts. Though all model-independent WC-regions are consistent with SM, looking only at the model-independent WC-spaces, one erroneously concludes that any model giving rise to these WCs would probably have quite a large parameter-space allowed, whereas, in reality, the allowed WC-space strictly coming from a single model is constrained in a tiny region around zero (SM). Instead of this being a quirk of one or a few models, we find this fact true for all BSM scenarios considered in this work.

	The situation turns even worse if we consider the fact that we only are using a partial set of WCs, relevant to the models in question, for the `model-independent' results. As is the norm in the community, model-independent inferences are generally obtained with all WCs present simultaneously. As we have seen from the first column of table~\ref{table:wcfitval}, the WC-spaces become considerably larger in that case. Using such results makes our `model-independent' inferences overwhelmingly conservative and in essence, inaccurate.

	This points us to the ominous realization that depending solely on the model-independent SMEFT fit results to infer parameter-spaces of individual BSM scenarios, is in fact, far from ideal. The results of this analysis motivate us to propose that during any consequential data-driven analysis of BSM theories in view of the low-energy observables, the bottom-up approach of expressing the observables in terms of SMEFT WCs should go hand-in-hand with the top-down way of calculating those WCs in terms of the BSM model parameters, to avoid erroneous, conservative, and in effect, too hopeful statistical inferences.

\section{Role of Theoretical Constraints}	
\begin{figure}[h!]
	\centering
	{\includegraphics[width=7cm, scale=0.4]{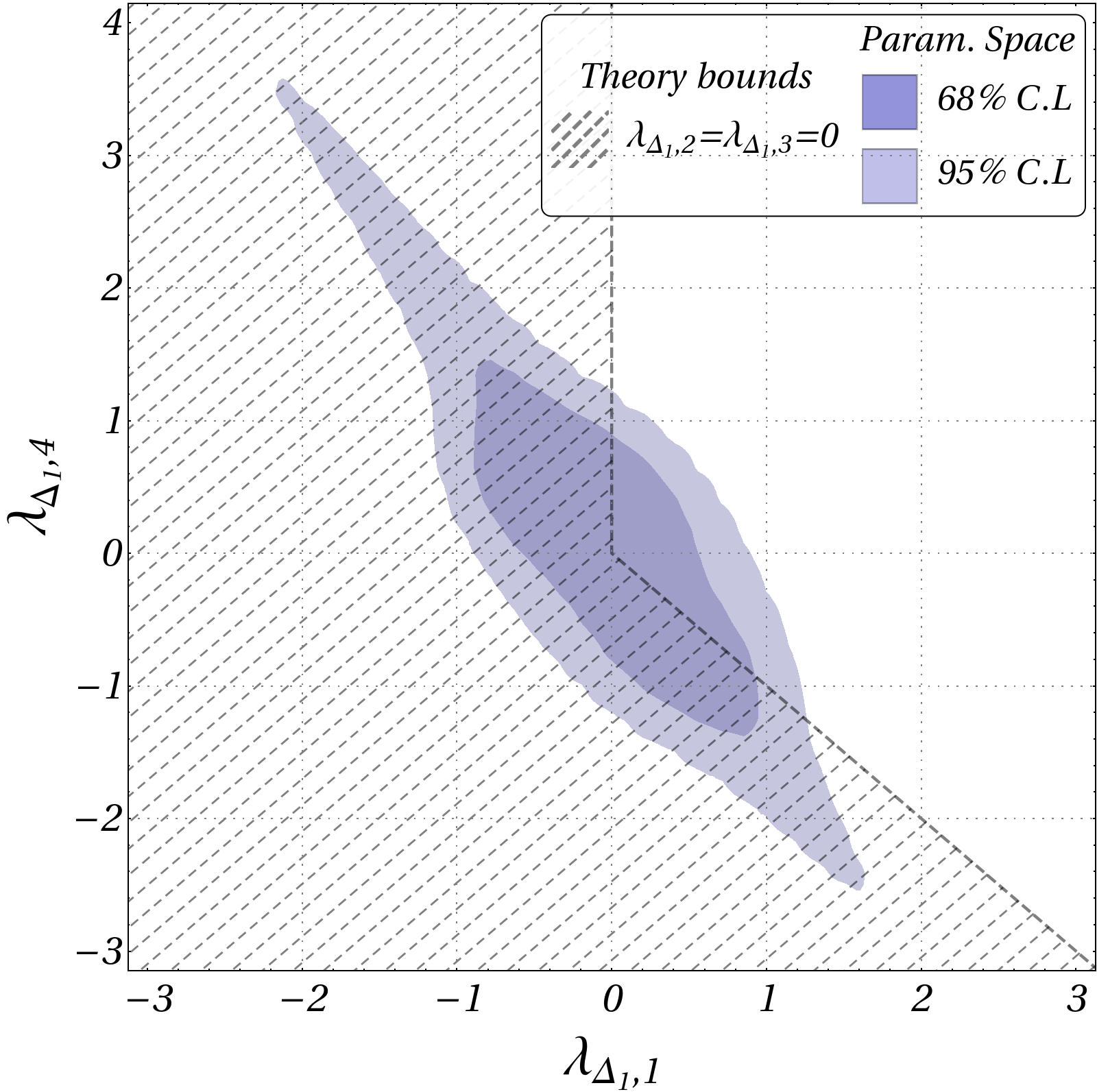}}
	\caption{\small The marginalized 2D posterior for the $\lambda_{\Delta_{1},1}$ and $\lambda_{\Delta_{1},4}$ for the SM+$\Delta_{1}$ extension. The cross hatched region denotes the parameter space disallowed by the theoretical bounds for $\lambda_{\Delta_{1},2}$ = $\lambda_{\Delta_{1},3}$ = 0.}
	\label{fig:ct2Dmodelparam}
\end{figure}

	In this section, we will show how the theoretical constraints can affect the model parameters derived in Section~\ref{sec:moddep_stat}. For the sake of demonstration, we consider the SM extension with $\Delta_{1}$ as an example model. We have noted the vacuum stability and unitarity bounds, see refs.~\cite{Arhrib:2011uy,Chun:2012jw,Haba:2016zbu}, as:\\
	\underline{\em Vacuum stability constraints}
\begin{align} \label{eq:ct_vs}
	&\lambda_{\Delta_{1},2}+\lambda_{\Delta_{1},3}\geq 0, ~~~\lambda_{\Delta_{1},2}+\frac{\lambda_{\Delta_{1},3}}{2}\geq 0, ~~~\lambda_{\Delta_{1},1}+\sqrt{4\lambda_{H}(\lambda_{\Delta_{1},2}+\lambda_{\Delta_{1},3})}\geq 0, \nonumber \\ &\lambda_{\Delta_{1},1}+\sqrt{4\lambda_{H}(\lambda_{\Delta_{1},2}+\frac{\lambda_{\Delta_{1},3}}{2})}\geq 0,~~~
	\lambda_{\Delta_{1},1}+\lambda_{\Delta_{1},4}+\sqrt{4\lambda_{H}(\lambda_{\Delta_{1},2}+\lambda_{\Delta_{1},3})}\geq 0,\nonumber \\ &\lambda_{\Delta_{1},1}+\lambda_{\Delta_{1},4}+\sqrt{4\lambda_{H}(\lambda_{\Delta_{1},2}+\frac{\lambda_{\Delta_{1},3}}{2})}\geq 0\,.
\end{align}
	\underline{\em Unitarity constraints}
\begin{align} \label{eq:ct_unitarity}
	&\lambda_{\Delta_{1},2}+2\lambda_{\Delta_{1},3} \leq 4\pi, ~~4\lambda_{\Delta_{1},2}+3\lambda_{\Delta_{1},3}\leq 4\pi,  ~~2\lambda_{\Delta_{1},2}-\lambda_{\Delta_{1},3} \leq 8 \pi,\nonumber\\ 
	&|\lambda_{\Delta_{1},1}+\lambda_{\Delta_{1},4}|\leq 8 \pi, ~~|\lambda_{\Delta_{1},1}|\leq 8 \pi, ~~|2 \lambda_{\Delta_{1},1}+3 \lambda_{\Delta_{1},4}|\leq 16 \pi, \nonumber\\ &|2\lambda_{\Delta_{1},1}-\lambda_{\Delta_{1},4}|\leq 8 \pi , ~~~|\lambda_{\Delta_{1},4}|\leq \text{min} \sqrt{(4\lambda_{H} \pm 16 \pi)(\lambda_{\Delta_{1},2}+2 \lambda_{\Delta_{1},3} \pm 4\pi)}, \nonumber\\
	&|2 \lambda_{\Delta_{1},1}+\lambda_{\Delta_{1},4}|\leq \sqrt{2(4\lambda_{H} - \frac{16}{3} \pi)(4\lambda_{\Delta_{1},2}+3 \lambda_{\Delta_{1},3} - 4\pi)}.
\end{align}
	Here, $\lambda_{H} =\frac{ m_{H}^2}{2 v^2}$ is the SM Higgs self-quartic coupling.
	
	The marginalized two-dimensional posterior distribution from the Bayesian fit for parameters $\lambda_{\Delta_{1},1}$ and $\lambda_{\Delta_{1},4}$ enclosing the 68\% and 95\% probability regions is shown in Figure~\ref{fig:ct2Dmodelparam}. The unitary and vacuum stability bounds stated in eqs.~\ref{eq:ct_vs} and \ref{eq:ct_unitarity} are also shown in the figure. The cross-hatched region denotes the parameter space disallowed by the theoretical bounds with $\lambda_{\Delta_{1},2}$=$\lambda_{\Delta_{1},3}$=0.  This shows that the allowed parameter space consistent with experimental data may not be compatible with the theoretical constraints for any arbitrary choice of other parameters of the model. Thus before performing the phenomenological analysis while choosing the benchmark values for the parameters, the inferred parameter space must be checked against the theoretical constraints.

\section{Conclusion and Remarks}
For long, experimental observations have persuaded us to propose numerous theoretically consistent models beyond the SM (BSM), with widely varying underlying symmetries and particle content. The unsettling aspect of the phenomenological landscape is that, on one hand, this diverse group of BSM scenarios, often proposed to address similar queries, are not viable for scrutiny under the same microscope. On the other hand, to understand the correct nature of new physics, it is necessary to find a common ground for multiple BSM scenarios, from where we can start making comparative remarks about them. The resolution of this apparent conflict has been our chief motivation for the present work. 

To achieve this, we proceed in this work to re-realize the minimal extensions of the Standard Model (SM) with the help of Effective Field Theory (EFT). As the experimental touchstone, we have defined our set of observables using the electroweak precision observables (EWPO) and the Higgs signal strengths from Run-I \& -II CMS and ATLAS data. Noting that the complete set of our adopted observables can be recast in terms of 18 SMEFT dimension-6 operators, we first estimate the Bayesian posteriors of the respective 18 WCs (both taken together and individually), assuming them to be independent. To estimate the WC-spaces and their correlations, we have used the marginalized one and two-dimensional posteriors of those WCs.

On the model side, we have considered 11 BSM scenarios, each of them an extension of the SM by a single heavy scalar multiplet. We have integrated out these heavy fields and computed the effective operators and associated WCs up to one-loop level, thus making the WCs correlated through the BSM parameters. While computing the WCs, we have ignored the heavy-light mixing in the loop. 

Further noticing that only 10 of the 18 operators can be generated from the said BSM scenarios, we recreated the statistical analysis considering the relevant 10 WCs independent. Comparing the marginal posteriors of each WC present in both 10 and 18 WCs-fits, we observe that simultaneous fits of a larger number of independent WCs non-trivially increases the allowed parameter-space of each WC. Based on this variation of the allowed WC-space, together with the observation that some of the scenarios lead to the exactly same set of effective dimension-6 operators, we have clubbed such BSMs to form 6 classes encapsulating 11 models, with each class containing a distinct set of WCs. Estimates of these groups of WCs constitute the `model-independent' part of our analysis.

Next we have performed a Bayesian analysis to estimate the ranges of the parameters appearing in the Lagrangians of all the 11 BSM scenarios. Using samples from these posteriors, we have then reconstructed the WCs of the class, to which those models belong. This enabled us to display and compare the ranges and correlations of the WCs coming from different models of the same class to their respective `model-independent' estimates, through two-dimensional marginal distributions in the WC-space. We have also shown, with an example model, how the theoretical constraints, e.g., vacuum stability, unitarity can further play a crucial role to rule out some of the BSM parameter space which is consistent with the experimental data. Numerical results of the entire analysis, along with all figures (including those not added in the draft) are available in the GitHub repository \cite{Githubeffex} \href{https://github.com/effExTeam/SMEFT-EWPO-Higgs}{\faGithub} associated with this work.

This method of employing EFT and using the common WC-spaces shared by BSM scenarios, provides a platform to compare apparently disconnected UV-theories, described by the same IR DOFs respecting the same symmetry, and paves the way towards a complete data-driven way of addressing the intractable inverse problem. Though EFT cannot replace the full theoretical computation, it can help us sniff out the correct nature of NP. It can be further used to understand the underlying degeneracy in model space, with a clue to break the same degeneracy including more observables. 
This approach can be replicated even in the event of the discovery of a new BSM particle. In that case, we need to compute the complete set of effective operators for the new theory (BSMEFT \cite{Banerjee:2020jun,Anisha:2019nzx}) and recast the full observable set in terms of these new operators. With the help of GrIP \cite{Banerjee:2020bym}, CoDEx \cite{Bakshi:2018ics}, and an increasing number of observables from different sectors, a suitable statistical inference process could, hopefully, unveil the correct nature of new theories.



\section{Acknowledgement}
The works of A, SDB, and JC are supported by the Science and Engineering Research Board,
Government of India, under the agreements SERB/PHY/2016348 (Early Career Research
Award) and SERB/PHY/2019501 (MATRICS) and Initiation Research Grant, agreement
number IITK/PHY/2015077, by IIT Kanpur.

 \clearpage

	
	\appendix
	
	\section{Appendix}
	\subsection{SM fit of electroweak precision observables}\label{appendix:SMfit}
\begin{table}[h!]
	\centering
	\caption{\small Results of the SM fit of EWPO. }
	\renewcommand{\arraystretch}{2.0}
	\begin{tabular}{|c|c|ccccc|}
		\hline 
		Parameters 		& Fit Values 		& \multicolumn{5}{c|}{Correlations} \\ 
		\hline 
		$m_{Z}$ [GeV]			& $91.188\pm0.002$  & $1$ & $0.002$ & $-0.097$ & $-0.007$ & $0.040$ \\
		$ m_{H}$ [GeV]	& $125.1\pm0.2$ & & $1$& $0.002$ &$-0.001$ & $-0.001$  \\
		$m_{t}$ [GeV]			& $173.554\pm0.843$ & & & $1$ & $0.045$ & $0.098$ \\
		$\alpha_{s}$	& $0.118\pm0.003$  	& & & & 1 & 0.010 \\
		$\Delta\alpha^{(5)}_{had}(m_{Z}^2)$ & $ 0.0276\pm0.0001$ & & & & & 1 \\ 
		\hline
	\end{tabular}
	\label{table:SM_EWPOfit}
\end{table}

\begin{figure}[h!]
	\centering
	\includegraphics[ width=0.6\textwidth]{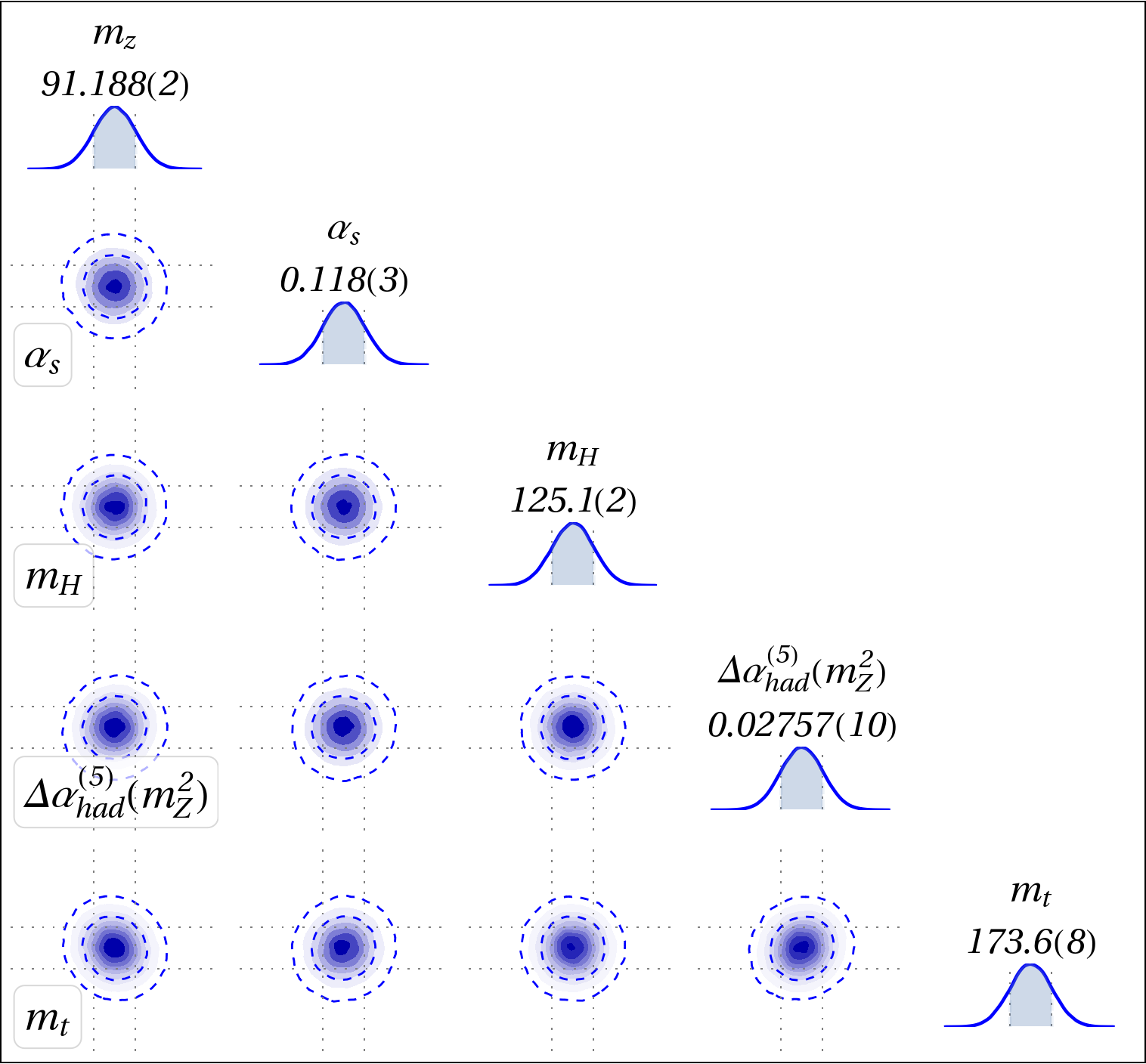}
	\caption{\small The one- and two-dimensional posterior distributions for the SM parameters showing the correlations among them.}
	\label{fig:smtriplot}
\end{figure}

	Using the experimental inputs and theoretical predictions of the electroweak precision observables mentioned in Section~\ref{subsection:EWPO_SM}, we perform the SM electroweak fit in terms of five parameters using the Bayesian framework.  The uniform priors are chosen with ranges of $\{90,92\}$, $\{120,130\}$, $\{170,180\}$, $\{0,0.2\}$ and $\{.02,0.03\}$ for SM parameters : mass of $Z$ boson ($m_{Z}$), mass of Higgs boson ($ m_{H}$), mass of top quark ($m_{t}$), strong coupling constant ($\alpha_{s}(m^{2}_{Z})$), and hadronic contribution to the running of $\alpha$ ($\Delta\alpha^{(5)}_{had}(m^{2}_{Z})$) respectively. The central tendencies and dispersions of the parameters obtained after performing a Bayesian fit in terms of 19 observables are given in the second and third columns of table~\ref{table:SM_EWPOfit}. The one- and two-dimensional posterior probability distributions of the fit parameters are shown in Figure~\ref{fig:smtriplot}.

	The obtained results are cross-checked and found to agree with the global electroweak fit performed by GFitter group \cite{Gfitter:2018}.
	These fitted SM parameters are considered as nuisance parameters in the main part of our analysis and these results are fed into the SMEFT fits as multi-normal priors.

		\subsection{Corrections to the EWPO from dimension-6 effective operators}\label{sec:dim6correction}
	The electroweak precision observables (EWPO) receive additional contributions from the dimension-6 effective operators through the re-definitions of the fields and the couplings.
	These modifications to the EWPO are captured through the corrections in $\alpha$, $m_{Z}$ and $G_{F}$ \cite{Berthier:2015oma,Brivio:2017vri}. In the process of estimating the corrections, the input values of  $\alpha$, $m_{Z}$, $G_{F}$, $m_{H}$, and the mass of light fermions (except top-quark) are not being varied. 
	\begin{table}[h]\label{tab:SMparam}
		\caption{\small The ``basis" inputs  that are used to define other parameters of the SM.}
		\centering
		\renewcommand{\arraystretch}{2.5}
		\begin{tabular}{|c|c|c|c|}
			\hline
			$\sin^2 2\theta_{_W}=\frac{4\pi\alpha}{\sqrt{2}G_{F} m^2_{Z}}$&$g_{_Y}=\frac{\sqrt{4\pi\alpha}}{\cos\theta_{_{W}}}$&$g_{_{W}}=\frac{\sqrt{4\pi\alpha}}{\sin\theta_{_{W}}}$&$g^{SM}_{_L}=T_{3}- Q_{\text{em}} \sin^2\theta_{_{W}}$\\
			\hline
			$m^2_{W}=m^2_{Z}\cos^2 {\theta_{_{W}}}$&$v^2=\frac{1}{\sqrt{2}G_{F}}$&$g_{_{Z}}=-\frac{g_{_W}}{\cos\theta_{_{W}}}$&$g^{SM}_{_{R}}=- Q_{\text{em}} \sin^2\theta_{_{W}}$\\
			\hline
		\end{tabular}
	\end{table}
	
   Here, we list the corrections to the SM in put parameters and the EWPO in presence of the SMEFT operators following the guidelines in refs.~\cite{Alonso:2013hga,Berthier:2015oma, Brivio:2017vri,Dawson:2019clf}. The notable corrections to the:
	\begin{itemize}
		\item ``basis" input parameters \cite{Alonso:2013hga, Berthier:2015oma, Brivio:2017vri}:
		\begin{eqnarray}\label{eq:basis-inputs}
		\delta G_{F}&=&\frac{G_{F}}{\Lambda^2}\left(2 v^2 \mathcal{C}^{3}_{Hl}- v^2 \mathcal{C}_{ll}\right),\\
		\delta \alpha&=&\frac{ 2 \, \alpha \, g_{_Y} g_{_W} v^2 }{(g_{_Y}^2+g_{_W}^2)}\frac{\mathcal{C}_{HWB}}{\Lambda^2},\\
		\delta m^2_{Z}&=&\frac{1}{2\sqrt{2}}\frac{m^2_{Z}}{G_{F}}\frac{\mathcal{C}_{HD}}{\Lambda^2}+\frac{2^{1/4}\sqrt{\pi \alpha} m_{Z}}{G^{3/2}_{F}}\frac{\mathcal{C}_{HWB}}{\Lambda^2}.
		\end{eqnarray}
	\end{itemize}
	\begin{itemize}
		\item  Higgs mass \cite{Alonso:2013hga, Brivio:2017vri}:
		\begin{eqnarray}
		\delta m^2_{H}&=&\frac{m^2_{H}}{\sqrt{2}G_{F}\Lambda^2}\left(-\frac{3\mathcal{C}_{H}}{2\lambda_{H}}+2\mathcal{C}_{H\square}-\frac{\mathcal{C}_{HD}}{2}\right).
		\end{eqnarray}
		\item  Weinberg angle \cite{Berthier:2015oma, Brivio:2017vri}:
		\begin{eqnarray}
		\delta(\sin^2{\theta_{_{W}}})&=&\frac{2\sin 2\theta_{_{W}}}{8\sqrt{2}\cos 2\theta_{_{W}}  G_{F}\Lambda^2}\left(2\sin\theta_{_{W}}(\mathcal{C}_{HD}+4\mathcal{C}^{(3)}_{Hl}-2\mathcal{C}_{ll})+4 \mathcal{C}_{HWB}\right).
		\end{eqnarray} 
		\item  gauge coupling $g_{_W}$ is \cite{Alonso:2013hga,Berthier:2015oma}:
		\begin{eqnarray}
		\delta g_{_W}&=&\frac{g_{_W}}{2}\left( \frac{\delta \alpha}{\alpha}-\frac{ \delta(\sin^2{\theta_{_{W}}})}{\sin^2{\theta_{_{W}}}}\right).
		\end{eqnarray}
		
		\item  W boson mass \cite{Berthier:2015oma, Dawson:2019clf} and decay width \cite{Brivio:2017vri}:
		\begin{eqnarray}
		\delta m_{W}&=&\frac{m_{Z} \cos \theta_{_{W}}}{2} \left(\frac{2\delta g_{_W}}{g_{_W}} +  \frac{\delta G_{F}}{G_{F}}\right),\\
		\delta\Gamma_{W}&=&\Gamma_{W}\left(\frac{4}{3}\delta g^l_{W}+\frac{8}{3}\delta g^q_{W}+\frac{\delta m^2_{W}}{2 m^2_{W}}\right).
		\end{eqnarray}
		
		\item  left and right handed couplings of the fermions to $Z$ boson \cite{Berthier:2015oma,Brivio:2017vri, Dawson:2019clf}:
		\begin{eqnarray}\label{eq:fermigaugevert_neutral}
		\frac{\delta g_{Z}}{g_{Z}}&=&-\frac{\delta G_{F}}{G_{F}}-\frac{\delta m^2_{Z}}{2 m^2_{Z}}+\frac{\sin\theta_{_{W}} \cos\theta_{_{W}}}{\sqrt{2}G_{F}\Lambda^2}\mathcal{C}_{HWB},\nonumber \\
		\delta g^l_{L}&=&\delta(g_{Z}) g^{l}_{L}-\frac{g_{Z}(\mathcal{C}_{He}+\mathcal{C}^{(1)}_{Hl}+\mathcal{C}^{(3)}_{Hl})}{4\sqrt{2}G_{F}\Lambda^2} + g_{Z}\,\delta(\sin^{2}\theta_{_{W}}), \nonumber\\
		\delta g^{\nu}_{L} &=&\delta(g_{Z}) g^{\nu}_{L}-\frac{(\mathcal{C}^{(1)}_{Hl}+\mathcal{C}^{(3)}_{Hl})}{4\sqrt{2}G_{F}\Lambda^2},\\
		\delta g^l_{R}&=&\delta(g_{Z})
		g^{l}_{R}+\frac{g_{Z}(\mathcal{C}_{He}-\mathcal{C}^{(1)}_{Hl}-\mathcal{C}^{(3)}_{Hl})}{4\sqrt{2}G_{F}\Lambda^2}, \nonumber\\
		\delta g^{\nu}_{R} &=&	0,\nonumber\\
		\delta g^u_{L} &=&\delta(g_{Z}) g^{u}_{L}+\frac{(-\mathcal{C}_{Hq}+\mathcal{C}^{(3)}_{Hq}-\mathcal{C}_{Hu})}{4\sqrt{2}G_{F}\Lambda^2}+\frac{2}{3}g_{Z} \,\delta(\sin^{2}\theta_{_{W}}), \nonumber\\
		\delta g^u_{R} &=&\delta(g_{Z})
		g^{u}_{R}+\frac{(\mathcal{C}^{(1)}_{Hq}+\mathcal{C}^{(3)}_{Hq}-\mathcal{C}_{Hu})}{4\sqrt{2}G_{F}\Lambda^2}+\frac{2}{3}g_{Z}\,\delta(\sin^{2}\theta_{_{W}}),\nonumber\\
		\delta g^d_{L} &=&\delta(g_{Z})
		g^{d}_{L}-\frac{(\mathcal{C}^{(1)}_{Hq}+\mathcal{C}^{(3)}_{Hq}+\mathcal{C}_{Hd})}{4\sqrt{2}G_{F}\Lambda^2}+\frac{1}{3}g_{Z}\,\delta(\sin^{2}\theta_{_{W}}),\nonumber\\
		\delta g^d_{R} &=&\delta(g_{Z})
		g^{d}_{R}+\frac{(-\mathcal{C}^{(1)}_{Hq}-\mathcal{C}^{(3)}_{Hq}+\mathcal{C}_{Hd})}{4\sqrt{2}G_{F}\Lambda^2}.\nonumber
		\end{eqnarray}
		\item  couplings of fermions to charged gauge bosons \cite{Berthier:2015oma,Brivio:2017vri}:
		\begin{eqnarray}\label{eq:fermigaugevert_charged}
		\delta(g^{l}_{W})&=&\frac{g_{_W}}{2\sqrt{2}G_{F}\Lambda^2}\mathcal{C}^{(3)}_{Hl}+\delta g_{_W}, \\
		\delta(g^{q}_{W})&=&\frac{g_{_W}}{2\sqrt{2}G_{F}\Lambda^2}\mathcal{C}^{(3)}_{Hq}+\delta g_{_W}. \nonumber
		\end{eqnarray}
		\item partial decay width of Z boson 
		\begin{eqnarray}
		\Gamma_{f}=N_{c}\frac{m_{Z}}{12\pi}\sqrt{1-4\frac{m_{f}^2}{m_{Z}^2}}\left(\frac{1}{2}(g_{L}^2+g_{R}^2)+\frac{2m_{f}^2}{m_{Z}^2}\Big(-\frac{g_{L}^2}{4}-\frac{g_{R}^2}{4}-\frac{3}{2}g_{L} g_{R}\Big)\right),
		\end{eqnarray}
		using eqs.~\ref{eq:fermigaugevert_neutral} and \ref{eq:basis-inputs}.
		Here, $N_{C}$ and $m_f$ are the color charges and masses of the fermions. On top of that, we have also included the corrections to the other partial decay widths of $Z$, e.g., $\delta R_{l}$, $\delta R_{b}$ and $\delta R_{c}$, and then successively to the total decay width as well. Employing these corrections, we can further estimate the change in the total scattering cross section of Z using 
		\begin{eqnarray}
	\sigma^{0}_{had}=\frac{12\pi}{m_{Z}^2}\big(\frac{\Gamma_{e}\Gamma_{had}}{\Gamma_{Z}^2}\big).
		\end{eqnarray}

			\item  asymmetry parameters $(\delta A_{f})$ and  $(\delta A^{f}_{FB})$ using \ref{eq:fermigaugevert_neutral} \cite{Berthier:2015oma}.
			
	\end{itemize}
	

	\providecommand{\href}[2]{#2}
	\addcontentsline*{toc}{section}{}
	\bibliographystyle{JHEP}
	\bibliography{BSM_SMEFT}

	
\end{document}